\documentclass[aps,prd,twocolumn,amsmath,amssymb,amsfonts,nofootinbib,superscriptaddress,altaffilletter]{revtex4-1}
\usepackage{graphicx}
\usepackage{hyperref}
\hypersetup{
  bookmarksopen=true
}
\usepackage{amssymb}
\usepackage{amsmath}
\usepackage{color}
\usepackage{supertabular}
\usepackage{dcolumn}

\def\beq{\begin{equation}}
\def\eeq{\end{equation}}
\def\bea{\begin{eqnarray}}
\def\eea{\end{eqnarray}}

\def\Tobs{T_{\textrm{\mbox{\tiny{obs}}}}}
\def\Tcoh{T_{\textrm{\mbox{\tiny{coh}}}}}

\def\max{\textrm{\mbox{\tiny{max}}}}

\def\ec{\textrm{~,}}

\def\fdot{f^{(1)}}

\newcommand{\F}{\mathcal{F}}
\newcommand{\SNR}{\textrm{SNR}}
\newcommand{\Fstat}{$\mathcal{F}$-statistic}

\def\pf{{\it PowerFlux}}
\def\fh{{\it FrequencyHough}}
\def\sh{{\it SkyHough}}
\def\td{{\it Time-Domain \Fstat}}

\def\sci#1#2{#1\times10^{#2}}

\def\RAJ{\textrm{RA}_{\textrm J2000}}
\def\DECJ{\textrm{DEC}_{\textrm J2000}}

\begin{document}

\title{
All-sky Search for Periodic Gravitational Waves in the O1 LIGO Data
}



\author{%
B.~P.~Abbott,$^{1}$  
R.~Abbott,$^{1}$  
T.~D.~Abbott,$^{2}$  
F.~Acernese,$^{3,4}$ 
K.~Ackley,$^{5}$  
C.~Adams,$^{6}$  
T.~Adams,$^{7}$ 
P.~Addesso,$^{8}$  
R.~X.~Adhikari,$^{1}$  
V.~B.~Adya,$^{9}$
% ...
\\
\medskip
(LIGO Scientific Collaboration and Virgo Collaboration) 
\\
\medskip
{{}$^{*}$Deceased, March 2016. }%
{{}$^{\sharp}$Deceased, March 2017. }%
{${}^{\dag}$Deceased, February 2017. }%
{${}^{\ddag}$Deceased, December 2016. }%
}\noaffiliation
\affiliation {LIGO, California Institute of Technology, Pasadena, CA 91125, USA }
\affiliation {Louisiana State University, Baton Rouge, LA 70803, USA }
\affiliation {Universit\`a di Salerno, Fisciano, I-84084 Salerno, Italy }
\affiliation {INFN, Sezione di Napoli, Complesso Universitario di Monte S.Angelo, I-80126 Napoli, Italy }
\affiliation {University of Florida, Gainesville, FL 32611, USA }
\affiliation {LIGO Livingston Observatory, Livingston, LA 70754, USA }
\affiliation {Laboratoire d'Annecy-le-Vieux de Physique des Particules (LAPP), Universit\'e Savoie Mont Blanc, CNRS/IN2P3, F-74941 Annecy, France }
\affiliation {University of Sannio at Benevento, I-82100 Benevento, Italy and INFN, Sezione di Napoli, I-80100 Napoli, Italy }
\affiliation {Albert-Einstein-Institut, Max-Planck-Institut f\"ur Gravi\-ta\-tions\-physik, D-30167 Hannover, Germany }
\affiliation {The University of Mississippi, University, MS 38677, USA }
\affiliation {NCSA, University of Illinois at Urbana-Champaign, Urbana, IL 61801, USA }
\affiliation {Nikhef, Science Park, 1098 XG Amsterdam, The Netherlands }
\affiliation {LIGO, Massachusetts Institute of Technology, Cambridge, MA 02139, USA }
\affiliation {Instituto Nacional de Pesquisas Espaciais, 12227-010 S\~{a}o Jos\'{e} dos Campos, S\~{a}o Paulo, Brazil }
\affiliation {Gran Sasso Science Institute (GSSI), I-67100 L'Aquila, Italy }
\affiliation {INFN, Sezione di Roma Tor Vergata, I-00133 Roma, Italy }
\affiliation {Inter-University Centre for Astronomy and Astrophysics, Pune 411007, India }
\affiliation {International Centre for Theoretical Sciences, Tata Institute of Fundamental Research, Bengaluru 560089, India }
\affiliation {University of Wisconsin-Milwaukee, Milwaukee, WI 53201, USA }
\affiliation {Leibniz Universit\"at Hannover, D-30167 Hannover, Germany }
\affiliation {Universit\`a di Pisa, I-56127 Pisa, Italy }
\affiliation {INFN, Sezione di Pisa, I-56127 Pisa, Italy }
\affiliation {OzGrav, Australian National University, Canberra, Australian Capital Territory 0200, Australia }
\affiliation {Laboratoire des Mat\'eriaux Avanc\'es (LMA), CNRS/IN2P3, F-69622 Villeurbanne, France }
\affiliation {LAL, Univ. Paris-Sud, CNRS/IN2P3, Universit\'e Paris-Saclay, F-91898 Orsay, France }
\affiliation {California State University Fullerton, Fullerton, CA 92831, USA }
\affiliation {European Gravitational Observatory (EGO), I-56021 Cascina, Pisa, Italy }
\affiliation {Universit\`a di Roma Tor Vergata, I-00133 Roma, Italy }
\affiliation {Universit\"at Hamburg, D-22761 Hamburg, Germany }
\affiliation {INFN, Sezione di Roma, I-00185 Roma, Italy }
\affiliation {Embry-Riddle Aeronautical University, Prescott, AZ 86301, USA }
\affiliation {Albert-Einstein-Institut, Max-Planck-Institut f\"ur Gravitations\-physik, D-14476 Potsdam-Golm, Germany }
\affiliation {APC, AstroParticule et Cosmologie, Universit\'e Paris Diderot, CNRS/IN2P3, CEA/Irfu, Observatoire de Paris, Sorbonne Paris Cit\'e, F-75205 Paris Cedex 13, France }
\affiliation {Korea Institute of Science and Technology Information, Daejeon 34141, Korea }
\affiliation {West Virginia University, Morgantown, WV 26506, USA }
\affiliation {Center for Gravitational Waves and Cosmology, West Virginia University, Morgantown, WV 26505, USA }
\affiliation {Universit\`a di Perugia, I-06123 Perugia, Italy }
\affiliation {INFN, Sezione di Perugia, I-06123 Perugia, Italy }
\affiliation {Syracuse University, Syracuse, NY 13244, USA }
\affiliation {University of Minnesota, Minneapolis, MN 55455, USA }
\affiliation {SUPA, University of Glasgow, Glasgow G12 8QQ, United Kingdom }
\affiliation {LIGO Hanford Observatory, Richland, WA 99352, USA }
\affiliation {Wigner RCP, RMKI, H-1121 Budapest, Konkoly Thege Mikl\'os \'ut 29-33, Hungary }
\affiliation {Columbia University, New York, NY 10027, USA }
\affiliation {Stanford University, Stanford, CA 94305, USA }
\affiliation {Universit\`a di Camerino, Dipartimento di Fisica, I-62032 Camerino, Italy }
\affiliation {Universit\`a di Padova, Dipartimento di Fisica e Astronomia, I-35131 Padova, Italy }
\affiliation {INFN, Sezione di Padova, I-35131 Padova, Italy }
\affiliation {MTA E\"otv\"os University, ``Lendulet'' Astrophysics Research Group, Budapest 1117, Hungary }
\affiliation {Nicolaus Copernicus Astronomical Center, Polish Academy of Sciences, 00-716, Warsaw, Poland }
\affiliation {University of Birmingham, Birmingham B15 2TT, United Kingdom }
\affiliation {Universit\`a degli Studi di Genova, I-16146 Genova, Italy }
\affiliation {INFN, Sezione di Genova, I-16146 Genova, Italy }
\affiliation {RRCAT, Indore MP 452013, India }
\affiliation {Faculty of Physics, Lomonosov Moscow State University, Moscow 119991, Russia }
\affiliation {SUPA, University of the West of Scotland, Paisley PA1 2BE, United Kingdom }
\affiliation {Caltech CaRT, Pasadena, CA 91125, USA }
\affiliation {OzGrav, University of Western Australia, Crawley, Western Australia 6009, Australia }
\affiliation {Department of Astrophysics/IMAPP, Radboud University Nijmegen, P.O. Box 9010, 6500 GL Nijmegen, The Netherlands }
\affiliation {Artemis, Universit\'e C\^ote d'Azur, Observatoire C\^ote d'Azur, CNRS, CS 34229, F-06304 Nice Cedex 4, France }
\affiliation {Institut de Physique de Rennes, CNRS, Universit\'e de Rennes 1, F-35042 Rennes, France }
\affiliation {Washington State University, Pullman, WA 99164, USA }
\affiliation {Universit\`a degli Studi di Urbino 'Carlo Bo', I-61029 Urbino, Italy }
\affiliation {INFN, Sezione di Firenze, I-50019 Sesto Fiorentino, Firenze, Italy }
\affiliation {University of Oregon, Eugene, OR 97403, USA }
\affiliation {Laboratoire Kastler Brossel, UPMC-Sorbonne Universit\'es, CNRS, ENS-PSL Research University, Coll\`ege de France, F-75005 Paris, France }
\affiliation {Carleton College, Northfield, MN 55057, USA }
\affiliation {Astronomical Observatory Warsaw University, 00-478 Warsaw, Poland }
\affiliation {VU University Amsterdam, 1081 HV Amsterdam, The Netherlands }
\affiliation {University of Maryland, College Park, MD 20742, USA }
\affiliation {Center for Relativistic Astrophysics and School of Physics, Georgia Institute of Technology, Atlanta, GA 30332, USA }
\affiliation {Universit\'e Claude Bernard Lyon 1, F-69622 Villeurbanne, France }
\affiliation {Universit\`a di Napoli 'Federico II', Complesso Universitario di Monte S.Angelo, I-80126 Napoli, Italy }
\affiliation {NASA Goddard Space Flight Center, Greenbelt, MD 20771, USA }
\affiliation {RESCEU, University of Tokyo, Tokyo, 113-0033, Japan. }
\affiliation {OzGrav, University of Adelaide, Adelaide, South Australia 5005, Australia }
\affiliation {Tsinghua University, Beijing 100084, China }
\affiliation {Texas Tech University, Lubbock, TX 79409, USA }
\affiliation {Kenyon College, Gambier, OH 43022, USA }
\affiliation {The Pennsylvania State University, University Park, PA 16802, USA }
\affiliation {National Tsing Hua University, Hsinchu City, 30013 Taiwan, Republic of China }
\affiliation {Charles Sturt University, Wagga Wagga, New South Wales 2678, Australia }
\affiliation {Canadian Institute for Theoretical Astrophysics, University of Toronto, Toronto, Ontario M5S 3H8, Canada }
\affiliation {University of Chicago, Chicago, IL 60637, USA }
\affiliation {Pusan National University, Busan 46241, Korea }
\affiliation {University of Cambridge, Cambridge CB2 1TN, United Kingdom }
\affiliation {The Chinese University of Hong Kong, Shatin, NT, Hong Kong }
\affiliation {INAF, Osservatorio Astronomico di Padova, Vicolo dell'Osservatorio 5, I-35122 Padova, Italy }
\affiliation {INFN, Trento Institute for Fundamental Physics and Applications, I-38123 Povo, Trento, Italy }
\affiliation {Universit\`a di Roma 'La Sapienza', I-00185 Roma, Italy }
\affiliation {Universit\'e Libre de Bruxelles, Brussels 1050, Belgium }
\affiliation {Sonoma State University, Rohnert Park, CA 94928, USA }
\affiliation {Montana State University, Bozeman, MT 59717, USA }
\affiliation {Center for Interdisciplinary Exploration \& Research in Astrophysics (CIERA), Northwestern University, Evanston, IL 60208, USA }
\affiliation {Cardiff University, Cardiff CF24 3AA, United Kingdom }
\affiliation {Universitat de les Illes Balears, IAC3---IEEC, E-07122 Palma de Mallorca, Spain }
\affiliation {The University of Texas Rio Grande Valley, Brownsville, TX 78520, USA }
\affiliation {Bellevue College, Bellevue, WA 98007, USA }
\affiliation {Institute for Plasma Research, Bhat, Gandhinagar 382428, India }
\affiliation {The University of Sheffield, Sheffield S10 2TN, United Kingdom }
\affiliation {University of Southern California Information Sciences Institute, Marina Del Rey, CA 90292, USA }
\affiliation {California State University, Los Angeles, 5151 State University Dr, Los Angeles, CA 90032, USA }
\affiliation {Universit\`a di Trento, Dipartimento di Fisica, I-38123 Povo, Trento, Italy }
\affiliation {Montclair State University, Montclair, NJ 07043, USA }
\affiliation {National Astronomical Observatory of Japan, 2-21-1 Osawa, Mitaka, Tokyo 181-8588, Japan }
\affiliation {Whitman College, 345 Boyer Avenue, Walla Walla, WA 99362 USA }
\affiliation {School of Mathematics, University of Edinburgh, Edinburgh EH9 3FD, United Kingdom }
\affiliation {University and Institute of Advanced Research, Gandhinagar Gujarat 382007, India }
\affiliation {IISER-TVM, CET Campus, Trivandrum Kerala 695016, India }
\affiliation {University of Szeged, D\'om t\'er 9, Szeged 6720, Hungary }
\affiliation {Tata Institute of Fundamental Research, Mumbai 400005, India }
\affiliation {INAF, Osservatorio Astronomico di Capodimonte, I-80131, Napoli, Italy }
\affiliation {University of Michigan, Ann Arbor, MI 48109, USA }
\affiliation {American University, Washington, D.C. 20016, USA }
\affiliation {Rochester Institute of Technology, Rochester, NY 14623, USA }
\affiliation {University of Bia{\l }ystok, 15-424 Bia{\l }ystok, Poland }
\affiliation {SUPA, University of Strathclyde, Glasgow G1 1XQ, United Kingdom }
\affiliation {University of Southampton, Southampton SO17 1BJ, United Kingdom }
\affiliation {University of Washington Bothell, 18115 Campus Way NE, Bothell, WA 98011, USA }
\affiliation {Institute of Applied Physics, Nizhny Novgorod, 603950, Russia }
\affiliation {Seoul National University, Seoul 08826, Korea }
\affiliation {Inje University Gimhae, South Gyeongsang 50834, Korea }
\affiliation {National Institute for Mathematical Sciences, Daejeon 34047, Korea }
\affiliation {NCBJ, 05-400 \'Swierk-Otwock, Poland }
\affiliation {Institute of Mathematics, Polish Academy of Sciences, 00656 Warsaw, Poland }
\affiliation {OzGrav, School of Physics \& Astronomy, Monash University, Clayton 3800, Victoria, Australia }
\affiliation {Hanyang University, Seoul 04763, Korea }
\affiliation {NASA Marshall Space Flight Center, Huntsville, AL 35811, USA }
\affiliation {ESPCI, CNRS, F-75005 Paris, France }
\affiliation {Southern University and A\&M College, Baton Rouge, LA 70813, USA }
\affiliation {OzGrav, University of Melbourne, Parkville, Victoria 3010, Australia }
\affiliation {College of William and Mary, Williamsburg, VA 23187, USA }
\affiliation {Indian Institute of Technology Madras, Chennai 600036, India }
\affiliation {IISER-Kolkata, Mohanpur, West Bengal 741252, India }
\affiliation {Scuola Normale Superiore, Piazza dei Cavalieri 7, I-56126 Pisa, Italy }
\affiliation {Universit\'e de Lyon, F-69361 Lyon, France }
\affiliation {Hobart and William Smith Colleges, Geneva, NY 14456, USA }
\affiliation {Janusz Gil Institute of Astronomy, University of Zielona G\'ora, 65-265 Zielona G\'ora, Poland }
\affiliation {University of Washington, Seattle, WA 98195, USA }
\affiliation {King's College London, University of London, London WC2R 2LS, United Kingdom }
\affiliation {Indian Institute of Technology, Gandhinagar Ahmedabad Gujarat 382424, India }
\affiliation {International Institute of Physics, Universidade Federal do Rio Grande do Norte, Natal RN 59078-970, Brazil }
\affiliation {Andrews University, Berrien Springs, MI 49104, USA }
\affiliation {Universit\`a di Siena, I-53100 Siena, Italy }
\affiliation {Trinity University, San Antonio, TX 78212, USA }
\affiliation {Abilene Christian University, Abilene, TX 79699, USA }


\begin{abstract}
  We report on an all-sky search for periodic gravitational waves in the frequency band 20-475\,Hz
  and with a frequency time derivative in the range of $\sci{[-1.0, +0.1]}{-8}$\,Hz/s. Such a signal could be produced by
  a nearby spinning and slightly non-axisymmetric isolated neutron star in our galaxy.
  This search uses the data from Advanced LIGO's first observational run, O1.
  No periodic gravitational wave signals were observed, and upper limits were placed on their strengths.
  The lowest upper limits on worst-case (linearly polarized) strain amplitude $h_0$ are  $\sim\sci{4}{-25}$ near 170\,Hz.
  For a circularly polarized source (most favorable orientation), the smallest
  upper limits obtained are  $\sim\sci{1.5}{-25}$.
  These upper limits refer to all sky locations and the entire range of frequency derivative values.
  For a population-averaged ensemble of sky locations and stellar orientations,
  the lowest upper limits obtained for the strain amplitude are $\sim\sci{2.5}{-25}$.
  
\end{abstract}

%
%
\maketitle

\section{Introduction}
\label{sec:introduction}

We report the results of an all-sky, multi-pipeline search for continuous, nearly monochromatic gravitational waves from rapidly rotating isolated neutron stars using data from the first observing run (O1) of the Advanced Laser Interferometer Gravitational wave Observatory (Advanced LIGO \cite{aligo}).
Several different analysis algorithms are employed and cover frequencies from 20~Hz through 475~Hz and frequency derivatives over the range $\sci{[-1.0, +0.1]}{-8}$~Hz/s.

A number of previous searches for periodic gravitational waves from isolated neutron stars
have been carried out in initial
LIGO and Virgo data~\cite{S1Paper, S2TDPaper, S3S4TDPaper, Crab, S5TDPaper, CasA, S6SNRPaper, S6GlobularCluster, S2HoughPaper, S2FstatPaper, S4IncoherentPaper, S4EH, EarlyS5Paper, S5EH, FullS5Semicoherent, FullS5EH, S6PowerFlux, S6BucketEH, ref:VSRFH, orionspur, S5Hough, VSR1TDFstat}.
These searches have included coherent searches for gravitational radiation from known radio and X-ray pulsars,
directed searches for known stars or locations having unknown signal frequencies, and spotlight or all-sky
searches for stars with unknown signal frequency and sky location.

Here, we apply four different all-sky search programs (pipelines) used in previous searches. In summary,
\begin{itemize}
\item The \pf\ pipeline has been used in previous searches of
  initial LIGO data from the S4, S5 and S6 Science Runs~\cite{S4IncoherentPaper,EarlyS5Paper,FullS5Semicoherent,S6PowerFlux}.
  The program uses a {\em Loosely Coherent} method for following up outliers~\cite{loosely_coherent}, 
  and also a new universal statistic that provides correct upper limits
  regardless of the noise distribution of the underlying data, but which yields near-optimal
  performance for Gaussian data~\cite{universal_statistics}.

\item The \fh\ hierarchical pipeline has been used in the previous all-sky search of Virgo VSR2 and VSR4 Science Runs~\cite{ref:VSRFH}. It consists of an
initial multi-stage phase, in which candidates are produced, and a follow-up phase in which the candidates are confirmed or discarded. Frequentist upper limits are computed with a computationally cheap
procedure, based on the injection of a large number of simulated signals into the data.

\item The \sh\ pipeline has been used in previous searches of
  initial LIGO data from the  S2, S4  and S5  Science Runs~\cite{S2HoughPaper, S4IncoherentPaper,S5Hough}, as well as in the second stage of Einstein@Home searches~\cite{S4EH,S5EH}. An improved pipeline, applying a clustering algorithm to coincident candidates was employed in~\cite{AllSkyMDC}. Frequentist upper limits are derived based on a number of simulated software signal injections into the data.

\item The \td\ pipeline has been used in the all-sky search of the Virgo VSR1 data~\cite{VSR1TDFstat}.
This program performs a coherent analysis of
narrow-band time-domain sequences (each a few days long) with the {\Fstat}
method~\cite{jks}, followed by a search for coincidences among candidates found
in different time sequences, for a given band. In order to estimate the
sensitivity, frequentist upper limits are obtained by injecting simulated
signals into the data.

\end{itemize}

These different analysis programs employ a variety of algorithmic and parameter choices, in order to reduce the possibility of discarding a gravitational wave signal due to sub-optimal treatment of detector artifacts or by adhering to an overly restrictive signal model.
The coherence times used in first-stage data processing range from 1800~s to 6~days, and the treatment of narrow spectral artifacts (``lines'') differs substantially among the different search programs. The latter is an especially important consideration for the O1 data set, because lines are, unfortunately, especially prevalent.

After following up the first-stage outliers, none of the different search pipelines found evidence for continuous gravitational waves in the O1 data over the range of frequencies and frequency derivatives searched.
Upper limits are derived for each analysis, with some variation in techniques among the different programs.

This article is organized as follows: Sec.~\ref{sec:LIGO_O1} describes the Advanced LIGO interferometers and the first observing run. 
Sec.~\ref{sec:PipelineOverview} provides an overview of the four pipelines, discussing common and differing features. The individual pipelines are described in more detail in sec.~\ref{sec:searches}.
In sec~\ref{sec:results}, the results of these four searches are presented; describing both the outliers and their follow up, and the derived upper limits.
Finally, we conclude in sec.~\ref{sec:conclusions}.

\section{LIGO interferometers and the O1 observing run}
\label{sec:LIGO_O1}

Advanced LIGO consists of two detectors, one in Hanford, Washington, and the other in Livingston, Louisiana, separated by a $\sim$3000-km baseline~\cite{ALIGODescription}.
Each site hosts one, 4-km-long interferometer inside a vacuum envelope with the primary interferometer optics suspended by a cascaded, quadruple suspension system in order to isolate them from external disturbances.
The interferometer mirrors act as test masses, and the passage of a gravitational wave induces a differential-arm length change which is proportional to the gravitational-wave strain amplitude.
The Advanced LIGO detectors began data collecting in September 2015 after a major upgrade targeting a 10-fold improvement in sensitivity over the initial LIGO detectors.
While not yet operating at design sensitivity, both detectors reached an instrument noise 3 to 4 times lower than the previous best with the initial-generation detectors in their most sensitive frequency band between 100 Hz and 300 Hz~\cite{DetectorPaper}.

Advanced LIGO's first observing run occurred between September 12, 2015 and
January 19, 2016, for which approximately 77 days and 66 days of analyzable
data was produced by the Hanford (H1) and Livingston (L1) interferometers, respectively.
Notable instrumental contaminations affecting the searches described here
included spectral combs of narrow lines in both interferometers, many of
which were identified after the run ended and mitigated for future running.
These artifacts included an 8-Hz comb in H1 with the even harmonics (16-Hz comb)
being especially strong, which was later tracked down to digitization roundoff
error in a high-frequency excitation applied to servo-control the cavity length
of the Output Mode Cleaner (OMC). Similarly, a set of lines found to be linear
combinations of 22.7 Hz and 25.6 Hz in the L1 data was tracked down to OMC
excitation at a still higher frequency, for which digitization error occurred.

In addition, the low-frequency band of the H1 and L1 data (below $\sim$140 Hz)
was heavily contaminated by combs with spacings of 1 Hz, near-1-Hz and 0.5-Hz
and a variety of non-zero offsets from harmonicity. Many of these lines originated
from the observatory timing system, which includes both GPS-locked
clocks and free-running local oscillators. The couplings into the interferometer
appeared to come primarily through common current draws among power supplies in
electronics racks. These couplings were reduced following O1 via isolation of
power supplies, and in some cases, reduction of periodic current draws in
the timing system itself (blinking LEDs). A subset of these lines with common
origins at the two observatories contaminated the O1 search for a stochastic
background of gravitational waves, which relies upon cross-correlation of
H1 and L1 data, requiring excision of affected bands~\cite{O1StochasticPaper}.

Although most of these strong and narrow lines are stationary in frequency
and hence do not exhibit the Doppler modulations due to the Earth's motion
expected for a continuous wave (CW) signal from most sky locations, the lines pollute the spectrum
for such sources. In sky locations near the ecliptic poles, the lines contribute
extreme contamination for certain signal frequencies. For a run like O1 that spans
only a modest fraction of a full year, there are also other regions of the sky and spindown
parameter space for which the Earth's average acceleration toward the Sun
largely cancels a non-zero source frequency derivative, leading to signal templates
with substantial contamination from stationary instrumental lines~\cite{S4IncoherentPaper}.
The search programs used here have chosen a variety of methods to cope with this contamination, as
described below.

\section{Overview of Search Pipelines}
\label{sec:PipelineOverview}

The four search pipelines have many features in 
common, but also have important differences, both major and minor. 
In this section we provide a broad overview of similarities
and differences before describing the individual pipelines
in more detail in the following section. 

\subsection{Signal Model}

All four search methods presented here assume the same signal model, based on
a classical model of a spinning neutron star with a time-varying quadrupole moment that produces circularly polarized gravitational radiation along the rotation axis and linearly polarized radiation in the directions perpendicular to the rotation axis.
The linear polarization is the most unfavorable case because the gravitational wave flux impinging on the detectors is smallest compared to the flux from circularly polarized waves.

The assumed strain signal model for a periodic source is given as
\begin{equation}
\begin{array}{l}
h(t)=h_0\left(F_+(t, \alpha_0, \delta_0, \psi)\frac{1+\cos^2(\iota)}{2}\cos(\Phi(t))+\right.\\
\quad\quad\quad \left.\vphantom{\frac{1+\cos^2(\iota)}{2}}+F_\times(t, \alpha_0, \delta_0, \psi)\cos(\iota)\sin(\Phi(t))\right)\ec
\end{array}
\end{equation}

\noindent where $F_+$ and $F_\times$ characterize the detector responses to signals with ``$+$'' and ``$\times$''
quadrupolar polarizations \cite{S4IncoherentPaper, EarlyS5Paper, FullS5Semicoherent}, the sky location is described by right ascension $\alpha_0$ and declination $\delta_0$, $\psi$ is the polarization angle of the projected source rotation axis in the sky plane, and the inclination of the source rotation axis to the detector line-of-sight is $\iota$. The phase evolution of the signal is given by the formula
\begin{equation}
\label{eqn:phase_evolution}
\Phi(t)=2\pi\left(f_\textrm{source}\cdot (t-t_0)+\fdot\cdot (t-t_0)^2/2\right)+\phi\ec
\end{equation}
where $f_\textrm{source}$ is the source frequency, $\fdot$ is the first frequency derivative (which, when negative, is termed the {\em spindown}), time in the Solar System barycenter is $t$, and the initial phase $\phi$ is computed relative to reference time $t_0$.
When expressed as a function of local time of ground-based detectors, Eq.~\ref{eqn:phase_evolution} acquires sky-position-dependent Doppler shift terms.

Most natural sources are expected to have a negative first frequency derivative, as the energy lost in gravitational or electromagnetic waves would make the source spin more slowly. The frequency derivative can be positive when the source is affected by a strong slowly-varying Doppler shift, such as due to a long-period orbit with a companion.

\subsection{Detection Statistics}

All four methods look for excess detected strain power that follows a time
evolution of peak frequency consistent with the signal model.
Each program begins with sets of ``short Fourier transforms'' (SFTs)
that span the observation period, with coherence times ranging from 1800 to 7200 s. 
The first three pipelines (\pf, \fh\ and \sh) compute measures of
strain power directly from the SFTs and create detection statistics
by stacking those powers or stacking weights for powers exceeding threshold,
with corrections for frequency evolution applied in the semi-coherent power stacking. 
The fourth pipeline (\td) uses a much longer coherence time (6 d) and applies
frequency evolution corrections coherently in band-limited time-domain data
recreated from the SFT sets, to obtain the \Fstat~\cite{jks}.
Coincidences are then required among multiple data segments with no stacking.

The \pf\ method includes explicit searches over different signal polarizations,
while the other three methods use a detection statistic that performs well on
average over an ensemble of polarizations.

All methods search for initial frequency in the
range $20\,$--$\,2000$~Hz, but with template grid spacings that depend inversely
upon the effective coherence time used. 

The range of $\dot{f}$ values searched is $[-\sci{1}{-8},+\sci{1}{-9}]~\mathrm{Hz}~\mathrm{s}^{-1}$.
All known isolated pulsars spin down more slowly
than the two values of $|\dot{f}|_\max$ used here, and as seen in the results
section, the ellipticity required for higher $|\dot{f}|$ is improbably high
for a source losing rotational energy primarily via gravitational radiation at low
frequencies.
A small number of isolated pulsars in globular clusters exhibit slight spin-up, 
believed to arise from acceleration in the Earth's direction;  such spin-up values have
magnitudes small enough to be detectable with the zero-spin-down templates
used in these searches, given a strong enough signal. Another potential source
of apparent spin-up is Dopper modulation from an unseen, long-period binary companion.

\subsection{Upper limits}
\label{subsubsec:themethodandul}

While the parameter space searched is the same for the three methods,
there are important differences in the way upper limits are
determined. The \pf\ pipeline sets strict frequentist upper limits
on detected strain power in circular and linear polarizations that
apply everywhere on the sky except for small regions near the ecliptic
poles, where signals with small Doppler modulations can be masked by
stationary instrumental spectral lines. The other three pipelines set
population-averaged upper limits over the parameter search volume,
relying upon Monte Carlo simulations. 

\subsection{Outlier follow-up}

The \pf\ and \fh\ pipelines have hierarchical structures
that permit systematic follow-up of loud outliers in the
initial stage to improve intrinsic strain sensitivity by
increasing effective coherence time while dramatically
reducing the parameter space volume over which the follow-up
is pursued. The \pf\ pipeline uses ``loose coherence''~\cite{loosely_coherent}
with stages of improving refinement via steadily increasing effective
coherence times, while the \fh\ pipeline
increases the effective coherence time by a factor of 10 and
recomputes strain power ``peakmaps.'' Any outliers that survive
all stages of any of the four pipelines are examined manually for
contamination from known instrumental artifacts and for evidence of
contamination from a previously unknown single-interferometer artifact.
Those for which no artifacts are found are subjected to
additional systematic follow-up used for Einstein@Home searches~\cite{ref:ehfu,ref:eho1},
which includes a final stage with full coherence across the entire data run.

\section{Details of Search Methods}
\label{sec:searches}

\subsection{\pf\ Search Method}

The \pf\ search pipeline has two principal stages. First, the main \pf\ algorithm \cite{S4IncoherentPaper, EarlyS5Paper, FullS5Semicoherent, PowerFluxTechNote, PowerFlux2TechNote, PowerFluxPolarizationNote} is run to establish upper limits and produce lists of outliers with signal-to-noise ratio (SNR) greater than a threshold of 5. These outliers are then followed up with the {\em Loosely Coherent} detection pipeline \cite{loosely_coherent, loosely_coherent2, FullS5Semicoherent}, which is used to reject or confirm the outliers.

Both algorithms calculate power for a bank of signal model templates.  The upper limits and signal-to-noise ratios for each template are computed by  comparison to templates with nearby frequencies and the same sky location and spindown \cite{PowerFluxTechNote, PowerFlux2TechNote, universal_statistics}.  The calibrated detector output time series, $h(t)$, for each detector, is broken into $50$\%-overlapping $7200$~s-long segments which are Hann-windowed and Fourier-transformed. The resulting Short Fourier Transforms (SFTs) are arranged into an input matrix with time and frequency dimensions. The power calculation of the data can be expressed as a bilinear form of the input matrix $\left\{a_{t,f}\right\}$:

\begin{equation}
P[f] = \sum_{t_1, t_2} a_{t_1, f+\Delta f(t_1)} a_{t_2, f+\Delta f(t_2)}^* K_{t_1, t_2, f}\ec
\end{equation}
where $\Delta f(t)$ is the detector frame frequency drift due to the effects from both Doppler shifts and the first frequency derivative. The sum is taken over all times $t$ corresponding to the midpoint of the short Fourier transform time interval. The kernel $K_{t_1, t_2, f}$ includes the contribution of time dependent SFT noise weights, antenna response, signal polarization parameters, and relative phase terms~\cite{loosely_coherent, loosely_coherent2}.

The first-stage, \pf\ algorithm uses a kernel with main diagonal terms only and is very fast. The second-stage, {\em Loosely Coherent} algorithm increases coherence time while still allowing for controlled deviation in phase~\cite{loosely_coherent}. This is done by more complicated kernels that increase effective coherence length.

The effective coherence length is captured in a parameter $\delta$,
which describes the amount of phase drift that the kernel allows between SFTs. A value of $\delta=0$ corresponds to a fully coherent case, and $\delta=2\pi$ corresponds to incoherent power sums.

Depending on the terms used, the data from different interferometers can be combined incoherently (such as in stages 0 and 1, see Table \ref{tab:PowerFlux_followup_parameters}) or coherently (as used in stages 2, 3 and 4). The coherent combination is more computationally expensive but provides much better parameter estimation.

The upper limits presented in section~\ref{sec:PowerFluxResults} (Figure~\ref{fig:powerflux_O1_upper_limits}) are reported in terms of the worst-case value of $h_0$ (which applies to linear polarizations with $\iota=\pi/2$) and for the most sensitive circular polarization ($\iota=0$~or~$\pi$).
As described previously~\cite{FullS5Semicoherent}, the pipeline retains some sensitivity, however, to non-general-relativity GW polarization models, including a longitudinal component, and to slow amplitude evolution.


The 95\% confidence level upper limits (see Figure~\ref{fig:powerflux_O1_upper_limits}) produced in the first stage are based on the overall noise level and largest outlier in strain found for every template in each $62.5$~mHz band in
the first stage of the pipeline. The $62.5$~mHz bands are analyzed by separate instances of \pf~\cite{FullS5Semicoherent}.
A followup search for detection is carried out for high-SNR outliers found in the first stage.


\subsubsection{Universal statistics}

As discussed above, a multitude of spectral combs contaminated the O1 low-frequency band, and, in contrast to the 23-month-long
S5 Science Run and 15-month-long S6 Science Runs of initial LIGO, the 4-month-long O1 run did not span the Earth's
full orbit. This means the Doppler shift magnitudes from the Earth's motion are reduced, on the whole,
in O1 compared to those of the other, earlier runs. In particular, for certain combinations of sky location, frequency
and spindown, a signal can appear relatively stationary in frequency in the detector frame of
reference. This effect is most pronounced for low signal frequencies, a pathology noted in searches of the 1-month-long S4 run~\cite{S4IncoherentPaper}.
At the same time, putative signals with low frequencies permit the use of 7200-s SFT spans, longer than the typical 1800-s SFTs used in the past,
which helps to resolve stationary instrumental lines from signals. One downside, though, of longer coherence length
is that there are far fewer SFTs in power sums compared with previous runs, which contributes to larger
deviations from ideal Gaussian behavior for power-sum statistics.

\begin{figure}[htbp]
\begin{center}
  \includegraphics[width=3.0in]{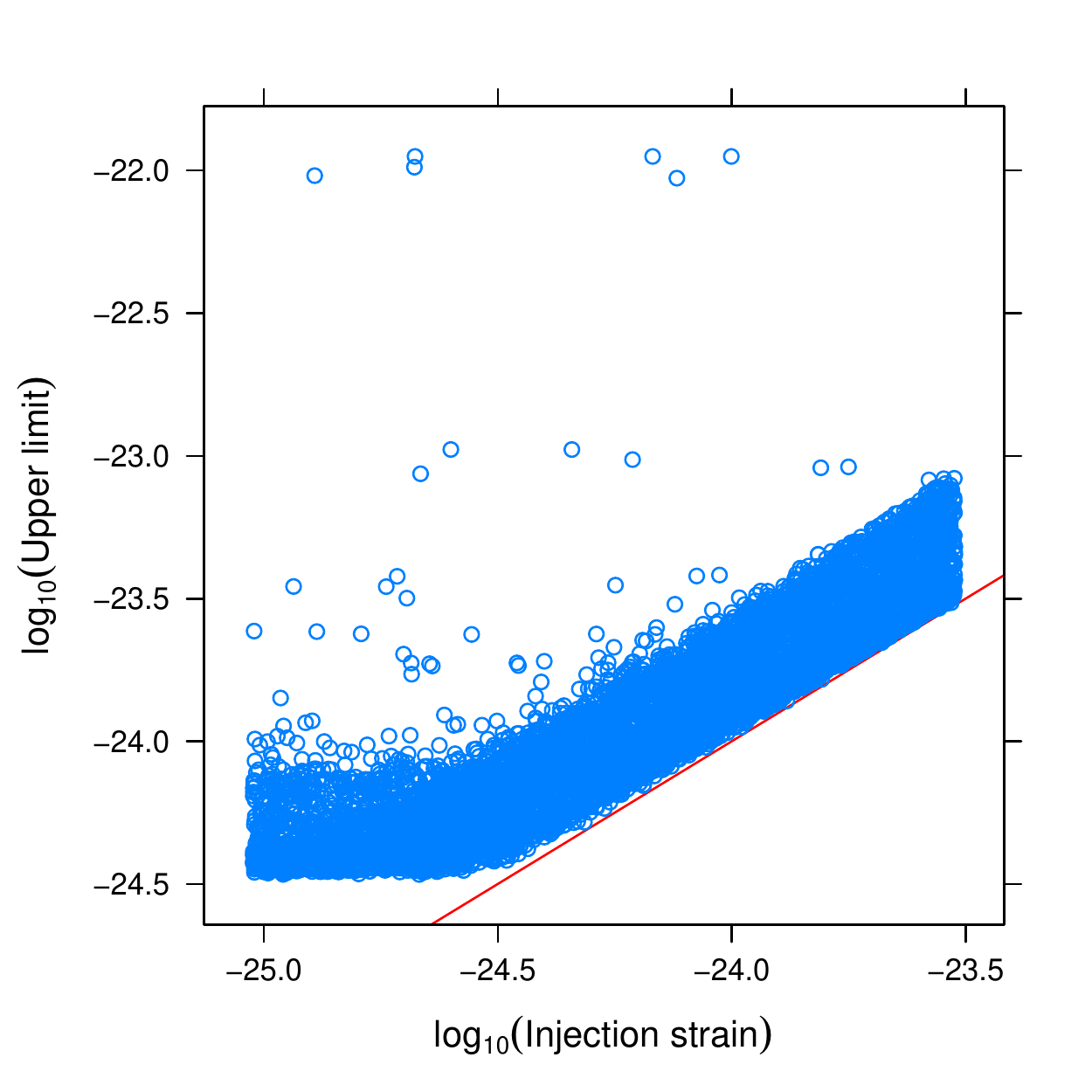}
 \caption[Upper limit versus injected strain]{\pf\ upper limit validation. Each point represents a separate injection in the 50-200\,Hz frequency range. Each established upper limit (vertical axis) is compared against the injected strain value (horizontal axis, red line) (color online).}
\label{fig:ul_vs_strain}
\end{center}
\end{figure}

To allow robust analysis of the entire spectrum, including the especially challenging lowest frequencies, the
{\em Universal statistic} algorithm~\cite{universal_statistics} is used for establishing upper limits. The algorithm is derived from the Markov inequality and shares its independence from the underlying noise distribution. It produces upper limits less than $5$\% above optimal in case of Gaussian noise. In non-Gaussian bands, it can report values larger than what would be obtained if the distribution were known, but the upper limits are always at least 95\% valid. Figure~\ref{fig:ul_vs_strain} shows results of an injection run performed as described in~\cite{FullS5Semicoherent}. Correctly established upper limits lie above the red line.

\begin{table*}[htbp]
\begin{center}
\begin{tabular}{llccccc}\hline
Stage & Instrument sum & {Phase coherence} & \multicolumn{1}{c}{Spindown step} & \multicolumn{1}{c}{Sky refinement} & \multicolumn{1}{c}{Frequency refinement} & \multicolumn{1}{c}{SNR increase} \\
 & & \multicolumn{1}{c}{rad} & \multicolumn{1}{c}{Hz/s} &  &  & \multicolumn{1}{c}{\%}\\
\hline \hline \\

  0  & Initial/upper limit semi-coherent & NA  & $\sci{1}{-10}$ & $1$ & $1/2$ & NA \\
  1 & incoherent & $\pi/2$ & $\sci{1.0}{-10}$ & $1/4$ & $1/8$ & 20 \\
  2 & coherent & $\pi/2$ & $\sci{5.0}{-11}$ & $1/4$ & $1/8$ & 10  \\
  3 & coherent & $\pi/4$ & $\sci{2.5}{-11}$ & $1/8$ & $1/16$ & 10 \\
  4 & coherent & $\pi/8$ & $\sci{5.0}{-12}$ & $1/16$ & $1/32$ & 7 \\
 \hline
\end{tabular}
\caption[Analysis pipeline parameters]{\pf\ analysis pipeline parameters. Starting with stage 1, all stages used the {\em Loosely Coherent} algorithm for demodulation. The sky and frequency refinement parameters are relative to values used in the semicoherent \pf\ search.}
\label{tab:PowerFlux_followup_parameters}
\end{center}
\end{table*}

\subsubsection{Outlier follow-up}

The initial stage (labeled 0) scans the entire sky with the semi-coherent \pf\ algorithm that computes weighted sums of powers of $7200$~s Hann-windowed SFTs. These power sums are then analyzed to identify high-SNR outliers. A separate algorithm uses the universal statistic \cite{universal_statistics} to establish upper limits.

The outlier follow-up procedure used in~\cite{FullS5Semicoherent, S6PowerFlux} has been extended with additional stages (see Table~\ref{tab:PowerFlux_followup_parameters}) to reduce the larger number of initial outliers, expected because of non-Gaussian artifacts and larger initial search space.

The entire dataset is partitioned into 3 stretches of equal length, and power sums are produced independently for any contiguous combinations of these stretches. As is done in~\cite{orionspur, S6PowerFlux}, the outlier identification is performed independently in each stretch.

High-SNR outliers are subject to a coincidence test.  For each
outlier with $\SNR>7$ in the combined H1 and L1 data, we require there
to be outliers in the individual detector data of the same sky area that had $\SNR>5$,
matching the parameters of the combined-detector outlier within 83~$\mu$Hz in frequency, and
$\sci{7}{-11}$\,Hz/s in spindown.  The combined-detector SNR is additionally required to be above both single-detector SNRs.

The identified outliers using combined data are then passed to the followup stage using the {\em Loosely Coherent} algorithm~\cite{loosely_coherent} with progressively tighter phase coherence parameters $\delta$, and improved determination of frequency, spindown, and sky location.

As the initial stage 0 sums only powers, it does not use the relative phase between interferometers, which results in some degeneracy among sky position, frequency, and spindown. The first {\em Loosely Coherent} followup stage also combines
interferometer powers incoherently, but demands greater temporal
coherence (smaller $\delta$) within each interferometer, which should
boost the SNR of viable outliers by at least 20\%. Subsequent stages use data coherently, providing tighter bounds on outlier location.

\begin{figure}[htbp]
\begin{center}
 \includegraphics[width=3.0in]{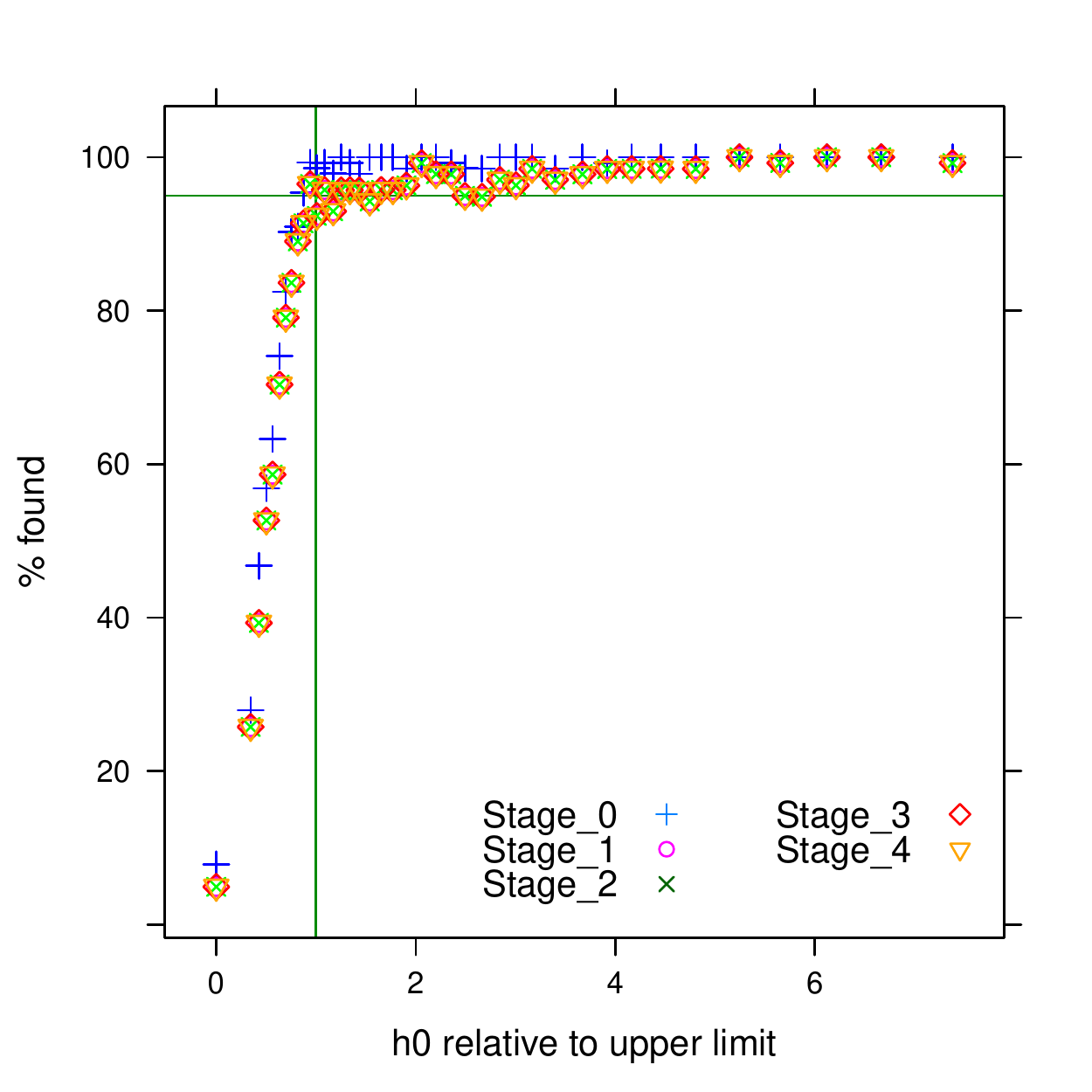}
 \caption[Injection recovery]{\pf\ injection recovery. The injections were performed in the 50-200\,Hz band. The injected strain divided by the upper limit in this band (before injection) is shown on the horizontal axis. The percentage of surviving injections is shown on the vertical axis, with a horizontal line drawn at the $95$\% level. Stage 0 is the output of the coincidence test after the initial semi-coherent search. (color online).}
\label{fig:powerflux_injection_recovery}
\end{center}
\end{figure}

Testing of the pipeline was performed for frequencies above 50~Hz. Injection recovery efficiencies from simulations covering the 50-200~Hz range are shown in Figure~\ref{fig:powerflux_injection_recovery}. The same followup parameters were applied to the 20-50~Hz region, but with stage 0 utilizing twice as dense spindown stepping.

Because the followup parameters were not tuned for the 20-50~Hz low frequency region and because of the highly disturbed spectrum, we do not expect a 95\% recovery rate.

Only a mild influence from parameter mismatch is expected, as the parameters are chosen to accommodate the worst few percent of injections. The followup procedure establishes very wide margins for outlier followup. For example, when transitioning from the semi-coherent stage 0 to the {\em Loosely Coherent} stage 1, the effective coherence length increases by a factor of 4. The average true signal SNR should then increase by more than $40$\%. But the threshold used in followup is only $20$\%, which accommodates unfavorable noise conditions, template mismatch, and detector artifacts.

The followup code was verified to recover 95\% of injections at or above the upper limit level for a uniform distribution of injection frequencies. (Figure~\ref{fig:powerflux_injection_recovery}).

The recovery criteria require that an outlier close to the true injection location (within 2~mHz in frequency $f$, $\sci{3}{-10}$~Hz/s in spindown and 12~rad$\cdot$Hz$/f$ in sky location) be found and successfully pass  through all stages of the detection pipeline. As each stage of the pipeline passes only outliers with an increase in SNR, this resulted in simulated outliers that stood out strongly above the background, with good estimates of the parameters of the underlying signals.

\subsection{\fh\ search method}
\label{sec:fh_method}
The \fh\ method is described in detail in~\cite{ref:FH0,ref:FH1,ref:VSRFH}. Calibrated detector data is used to create SFTs with coherence time depending on the frequency band being considered, see Table~\ref{tab:fh_fft}. Short time-domain disturbance are removed from the data before constructing the SFTs~\cite{peakmap}.
A time-frequency map, called a peakmap, is built by selecting the local maxima (called peaks) over a threshold of 1.58 on the square root of the equalized power $\mathcal R(i)=\sqrt{{\mathcal P(i)}/{S_{\textrm{AR}}(i)}}$, being $\mathcal P(i)$ the value of the periodogram of the data at the frequency index $i$, and $S_{\textrm{AR}}(i)$ an auto-regressive estimation of the average power spectrum at the same frequency index~\cite{peakmap} (then the index $i$ runs along the full frequency band being considered). The peakmap is cleaned using a line "persistency" veto, described in~\cite{ref:FH1}, which consists of projecting the peakmap onto the frequency axis and removing the frequency bins in which the projection is higher than a given threshold. 

\begin{table*}[htbp]
\begin{center}
\begin{tabular}{ccccc}\hline
Band [Hz] & Time duration [s] &  $\delta f$ [Hz] & $\delta \dot{f}$ [Hz/s]\\
\hline \hline \\
  $[20-128]$ & 8192 & $1.22\cdot 10^{-4}$ & $1.18\cdot 10^{-11}$ \\
  $[128-512]$ & 4096 &  $2.44\cdot 10^{-4}$ & $2.36\cdot 10^{-11}$ \\
  $[512-1024]$ & 2048 &  $4.88\cdot 10^{-4}$ & $4.71\cdot 10^{-11}$ \\
  $[1024-2048]$ & 1024 &  $9.76\cdot 10^{-4}$ & $9.42\cdot 10^{-11}$ \\
 \hline
\end{tabular}
\caption[FH FFT]{Properties of the FFTs used in the \fh\ pipeline. The time duration refers to the length in seconds of the data chunks on which the FFT is computed. The frequency bin width is the inverse of the time duration, while the spin-down bin width is computed as $\delta \dot{f}={\delta
f}/{T_{\textrm{obs}}}$. In the analysis described in this paper only the first two bands have been considered.}
\label{tab:fh_fft}
\end{center}
\end{table*}

After defining a grid in the sky, the peakmap for each sky position is corrected for the Doppler effect caused by the detector motion by shifting each peak to compensate for this effect. Shifted peaks are then fed to the
\fh\ algorithm, transforming each peak to the frequency/spin-down plane of the source. The \fh\ algorithm is a particular implementation of the Hough transform, which is a robust parameter estimator of patterns in images. The frequency and spin-down bins depend on the frequency band, as indicated in Tab.~\ref{tab:fh_fft}. The transformation properly weights any noise non-stationarity and the time-varying
detector response~\cite{ref:Hough_adap}.  

The computation of the \fh\ transform is the most computationally demanding part of the analysis and needs to be split into thousands of independent jobs, each computing a \fh\ transform covering a
small portion of the parameter space. In practice, each job covers 1 Hz, a small sky region--with a frequency-dependent size such that jobs at lower frequencies cover larger
regions---and a range of spin-down values, as detailed below. The output of a \fh\ transform is a 2-D histogram in the frequency/spin-down plane of the source.

For each \fh\ histogram, candidates for each sky location are selected by dividing the 1-Hz band into 20 intervals and taking the most or (in most cases) the two most significant
candidates for each interval, on the basis of the histogram number count. This is an effective procedure to avoid blinding by large disturbances in the data, which would contribute a large
number of candidates if we used a toplist, i.e. if only candidates globally corresponding to the highest number count were selected. All the steps described thus far are applied separately to  the data of each detector involved in the analysis.

Candidates from each detector are
clustered and coincidence tests are applied between the two sets of clusters, using a distance metric built in the four-dimensional
parameter space of position $(\lambda,~\beta)$ (in ecliptical coordinates), frequency $f$ and spin-down $\dot{f}$, defined as
\begin{equation}
d_{\rm FH}=\sqrt{\left(\frac{\Delta f}{\delta f}\right)^2+\left(\frac{\Delta \dot{f}}{\delta \dot{f}}\right)^2+\left(\frac{\Delta \lambda}{\delta \lambda}\right)^2+\left(\frac{\Delta 
\beta}{\delta \beta}\right)^2},
\label{eq:dfh}
\end{equation}
where $\Delta f$, $\Delta \dot{f}$, $\Delta \lambda$, and $\Delta \beta$ are the differences, for each parameter, among pairs of candidates of the two detectors, and $\delta f$, $\delta 
\dot{f}$, $\delta 
\lambda$, and $\delta \beta$ are the corresponding bins, that is the step width in a given parameter grid. Candidates within $d_{\rm FH}=3$ are considered coincident. Coincident candidates are
subject to a ranking procedure, based on the value of a statistic built using the distance and the \fh\ histogram weighted number count of the coincident candidates. More precisely, let us indicate with $N$ the total number of coincident candidates in each 0.1-Hz band. First, candidates are ordered in descending order of the number count, separately for each dataset, and a rank $r_{i,j}$ is assigned to each of them, from $1/N$ to the highest to 1 to the smallest, where $i=1,2$ identifies the dataset and $j$ runs over the coincident candidates of a given dataset in a given 0.1-Hz band. Then, coincident candidates are ordered in ascending order of their distance, and a rank $r_{d,j}$ is assigned to each pair, going from $1/N$ for the nearest to 1 for the farthest. A final rank $r_j=r_{d,j}\times \prod_{i=1}^{2} r_{i,j}$ is computed and will take smaller values for more significant candidates, i.e. having smaller distance and higher number counts.  A number of the most significant
candidates are selected for each 0.1-Hz band and are subject to a follow-up procedure in order to confirm or reject them. This number depends on the amount of computing power which can be devoted to the follow-up. For the analysis described in this paper 4 candidates have been selected in each 0.1-Hz band.

\subsubsection{Candidate follow-up}
\label{sec:fh_followup}
The follow-up consists of several steps (a detailed description is given in~\cite{ref:VSRFH}). First, for each candidate, a fully coherent search using data from
both detectors is performed assuming the parameters found for the candidate in this analysis. Although the
coherent search corrects exactly for the Doppler and spin-effect at a single particular point in the parameter space, corresponding to the candidate, the correction is extended, by
linear
interpolation, to the neighbors of the candidate. In practice, this means that from the resulting corrected and down-sampled time series, a new set of longer SFTs can be built, by a factor
of 10 in this analysis, as well as a set of new (Doppler corrected) peakmaps. The new peakmaps are valid even if the true signal parameters are slightly different from those of the candidate under consideration.

The joint corrected peakmaps (individually corrected for each detector's motion) are input to the \fh\ algorithm: overall, 1681 transforms are computed, covering $\pm 50$~mHz, $\pm
1$ spindown bins (of initial width) and $\pm 0.75$ bins (of initial width) for both $\lambda$ and $\beta$ around the candidate and the loudest candidate among the full set of \fh\ maps is selected (note that the bin widths are now 10 times smaller than those of the initial stage of the analysis). The starting
peakmap is then corrected using the parameters of the loudest candidate and projected on the frequency axis. We take the maximum of this projection in a range of $\pm 2$ bins (of initial width) around
the candidate frequency. We divide the rest of the 0.1-Hz band (which we consider the "off-source" region) into $\sim$200 intervals of the same width, take the maximum of the peakmap projection in each of these intervals and sort in decreasing order all these maxima. We tag the candidate as "interesting" if it ranks first or second in this list. 

Those candidates passing these tests are subject to further analysis: those candidates coincident with known noise lines (and that survived previous cleaning
steps) are discarded, candidates with multi-interferometer significance less than the single-interferometer significance are discarded, candidates with single-interferometer significances differing by more than a factor
of five are discarded, or candidates that have single-interferometer critical ratios ($\textrm{CR}={\left(A_p-\mu_p\right)}/{\sigma_p}$, being $A_p$ the candidate projection amplitude, $\mu_p$ and $\sigma_p$ the mean and standard deviation of the projection) differing by more than a factor of five are discarded. The choice of this factor is a conservative one, validated by simulations, such that a detectable signal would not be vetoed at this stage. The outliers passing also these steps are subject to additional, manual scrutiny, see Sec.~\ref{sec:fh_results} for more details concerning the O1 outliers.

\begin{figure}[htbp]
\begin{center}
 \includegraphics[width=3.0in]{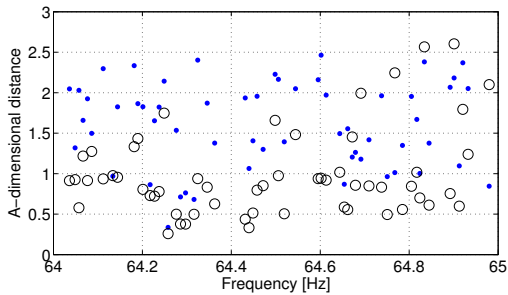}
\caption[]{A-dimensional distance between 56 injected signals in the band 64-65 Hz and the corresponding candidates (blue dots: before the follow-up; red circles: after the follow-up).}
\label{fig:fh_dist64}
\end{center}
\end{figure}

\begin{figure}[htbp]
\begin{center}
 \includegraphics[width=3.0in]{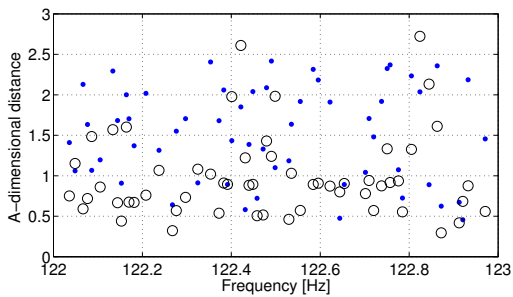}
\caption[Detection efficiency]{A-dimensional distance between 57 injected signals in the band 122-123 Hz and the corresponding candidates (blue dots: before the follow-up; red circles: after the follow-up).}
\label{fig:fh_dist122}
\end{center}
\end{figure}

As a validation of the follow-up we have made a study of software injection recovery. Specifically, we have generated about 110 signals which have been injected into representative 1-Hz bands, 64-65 Hz and 122-123 Hz, following the procedure described at the beginning of Sec.~\ref{sec:fh_upperlimits} and with amplitude equal to the upper limit computed in those bands. The data has been analyzed with the \fh\ pipeline and candidates selected, as discussed in Sec.~\ref{sec:fh_method}. These candidates have been subject to the follow-up and all the candidates due to injected signals, i.e. within the standard follow-up volume around the corresponding injection, have been confirmed, showing a CR$>$11 (to be compared with 7.57 which is the threshold used in the real analysis to select outliers, see Sec.~\ref{sec:fh_results}). Moreover, we have verified that in most cases (about 90\% of the cases in this test) the follow-up allows to improve parameter estimation. In Figs. \ref{fig:fh_dist64}, \ref{fig:fh_dist122} we show the a-dimensional distance of candidates associated to simulated signals from their injection, defined by Eq. \ref{eq:dfh}, both before and after the follow-up. The median of the distance reduces from 1.62 to 0.85 for the first band and from 1.55 to 0.88 for the second.  

\subsubsection{Upper limit computation}
\label{sec:fh_upperlimits}
Upper limits are computed in each 1-Hz band between 20~Hz and 475~Hz by injecting software simulated signals, with the same procedure used in~\cite{ref:VSRFH}. For each 1~Hz band 20 sets of 100 signals each are generated, with fixed amplitude within each set and random parameters (sky location, frequency, spin-down, and polarization parameters). These are complex signals generated in the time domain at a reduced sampling frequency of 1~Hz, and then added to the data of both detectors in the frequency domain. For each injected signal in a set of 100, an analysis is done using the \fh\ pipeline over a frequency band of 0.1 Hz around the injection frequency, the full spin-down range used in the real analysis, and nine sky points around the injection position~\cite{ref:VSRFH}. Candidates are selected exactly as in the real analysis, but no clustering is applied because it would be affected by the presence of too many signals. Then, coincidences are required directly among candidates (clustering has been used mainly to reduce computational cost). Coincident candidates that are within the follow-up volume around the injection parameters, and that have a critical ratio larger than the largest critical ratio found in the real analysis in the same band are considered as ``detections'' (excluding those that fall in a frequency bin vetoed by the persistency veto).

\begin{figure}[htbp]
\begin{center}
 \includegraphics[width=3.0in]{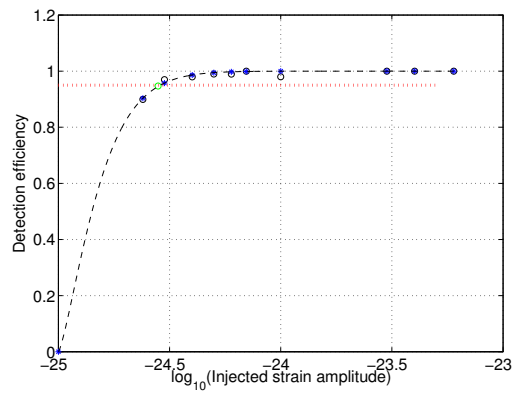}
\caption[Detection efficiency]{Measured detection efficiency values for the band 423-424 Hz (circles) and their fit done using Eq. \ref{eq:weifit} (asterisks). The dashed line 
represents the full fitted curve. The dotted horizontal line indicates the 95\% level of the detection efficiency.}
\label{fig:weigood}
\end{center}
\end{figure}

The upper limit in a given 1~Hz band is given by the signal amplitude such that 95\% of the injected signals are detected. In practice, typically, a fit is used to the measured detection efficiency values in order to interpolate the detection efficiency when injections do not cover densely enough the 95\% region. The fit has been done with the cumulative of a modified Weibull distribution function, given by
\begin{equation}
D(x)=K\left(1-e^{-A_1(x-x_{\textrm{min}})^{A_2}}\right),
\label{eq:weifit}
\end{equation}
where $x=\log_{10}(h_{\textrm{inj}})$, $h_{\textrm{inj}}$ is the injected amplitude, $x_{\textrm{min}}$ is the value such that $D(x_{\textrm{ min}})=0$, $K$ is a scaling factor such that the maximum of $D(x)$ is equal to the maximum measured detection efficiency, and $A_1$ and $A_2$ are the fit parameters. As an example, in Figure~\ref{fig:weigood} the measured detection efficiency values for the band 423-424~Hz are shown together with the fit. In cases like this, corresponding to artifact-free bands, the fit is accurate.

\begin{figure}[htbp]
\begin{center}
 \includegraphics[width=3.0in]{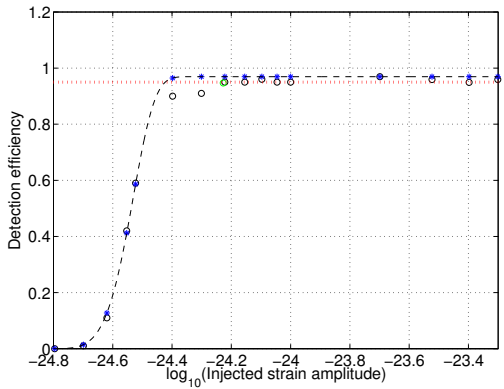}
\caption[Detection efficiency]{Measured detection efficiency values for the band 51-52~Hz (circles) and their fit done using Eq. \ref{eq:weifit} (asterisks). The dashed line represents 
the full fitted curve. The dotted horizontal line indicates the 95\% level of the detection efficiency.}
\label{fig:weibad}
\end{center}
\end{figure}

In more disturbed bands, the fit is not able to closely follow the measured values, as shown, for example, in Figure~\ref{fig:weibad}. In such cases, if an interpolation is needed, a linear interpolation is used between the two detection efficiency points nearest (one from below and one from above) to the 95\% level.



\subsection{\sh\ search method}
\label{sec:SH_method}

The \sh\ search method is described in detail in~\cite{SkyHough1, SkyHough2, Chi2Hough, S5Hough}.
Calibrated detector $h(t)$ data from the O1 run is used to create 1800-s Tukey-windowed SFTs, where each SFT is created from a segment of detector data that is at least 1800~s long. From this step, 3684 and 3007 SFTs are created for H1 and L1, respectively.
The data from the two LIGO interferometers are initially analyzed in separate all-sky searches for continuous gravitational wave signals, and then coincidence requirements on candidates are imposed.

SFT data from a single interferometer is analyzed by creating a peak-gram (a collection of zeros and ones) by setting a threshold of 1.6 on their normalized power.
This is similar to the \fh\ method, but, in this case, the averaged spectrum is determined via a running-median estimation~\cite{S4IncoherentPaper}.

An implementation of the weighted Hough transform~\cite{SkyHough2,S5Hough} is used to map
points from the time-frequency plane of the peak-grams into the space of the source parameters.
Similar to the methods described previously, the algorithm searches for signals whose frequency
evolution fits the pattern produced by the Doppler shift and spindown in the time-frequency plane of the data.
In this case, the Hough number count is the weighted sum of the ones and zeroes,
 $n_k^{(i)} $, of the different  peak-grams along the track corresponding to each point in parameter space.
This sum is computed as
\begin{equation}
  n = \sum_{i=0}^{N-1} w^{(i)}_{k} n^{(i)}_{k}\,,
\end{equation}
where the choice of weights is optimal~\cite{SkyHough2}. These weights are given by
\begin{equation}
\label{sh:w}
  w^{(i)}_{k} \propto  \frac{1}{S^{(i)}_{k}}\left\{
    \left(F_{+1/2}^{(i)}\right)^2 +
    \left(F_{\times 1/2}^{(i)}\right)^2\right\}\,,
\end{equation}
where $F_{+1/2}^{(i)}$ and $F_{\times 1/2}^{(i)}$ are the values of the antenna pattern functions at the
mid-point of the $i^{th}$ SFT for the sky location of interest and $S^{(i)}_{k}$ is the SFT noise level.
A particularly useful detection statistic is the \textit{significance} (or critical ratio), and is given by
\beq
 s= \frac{n-\langle n\rangle}{\sigma} \, ,
 \eeq
 where $\langle n\rangle$ and $\sigma$ are the expected mean and standard deviation of the Hough number count for pure noise.

The \sh\ search analyzes 0.1-Hz bands over the frequency interval 50-500~Hz, frequency time derivatives in the range $\sci{[-1.0, +0.1]}{-8}$~Hz/s, and covering the entire sky.
A uniform grid spacing, equal to the size of a SFT frequency bin,
$\delta f = 1/ \Tcoh= \sci{5.556}{-4}~\mathrm{Hz}$
is chosen, where $\Tcoh$ is the duration of an SFT.
The resolution $\delta \dot{f}$
is given by the smallest value of $\dot{f}$ for which
the intrinsic signal frequency does not drift by more than one frequency bin during
the total observation time $\Tobs$:
$\delta \dot{f} = \delta f / \Tobs \sim \sci{4.95}{-11}~\mathrm{Hz}~\mathrm{s}^{-1}$.
This yields 224 spin-down values for each frequency.
The sky resolution, $\delta \theta$
is frequency dependent, with the number of templates increasing with frequency,
as given by
Eq.(4.14) of Ref.~\cite{SkyHough1}, using a pixel-factor of $N_p=2$:
\beq
\delta \theta= \frac{10^4  \, \delta f}{f N_p} \, .
\eeq


For each frequency band, the parameter space is split further into 209 sub-regions of the sky.
For every sky region and frequency band the analysis program compiles a list of the 1000 most
significant candidates. A final list of the 1000 most significant candidates per frequency band is constructed, with no more than 300 candidates from a single sky region.
This procedure reduces the influence of instrumental spectral disturbances that affect specific sky regions.


The post-processing of the results for each 0.1-Hz band consists of the following steps:

(i) Apply a $\chi^2$ test, as described below, to eliminate candidates caused by detector artifacts.

(ii) Search for coincident candidates among the two data sets, using a coincidence window of $d_{\rm SH}<\sqrt{ 14}$. This dimensionless quantity, similar to the parameter $d_{\rm FH}$ used in the \fh\ pipeline, is defined as:
\beq
d_{\rm SH}= \sqrt{(\Delta f  /\delta f )^2 + (\Delta \dot{f}  / \delta \dot{f}  )^2 +( \Delta \theta /\delta \theta)^2}
\eeq
to take into account the distances in frequency, spin-down and sky location with respect to the grid resolution in parameter space. Here, $\Delta\theta$ is the sky angle separation.
Each coincidence pair is then characterized by its harmonic mean significance value and a center in parameter space: the mean weighted value of frequency, spin-down and sky-location obtained by using their corresponding individual significance values.
Subsequently, a list containing the 1000 most significant coincident pairs is produced for each 0.1-Hz band.

(iii) The surviving coincidence pairs are clustered, using the same coincidence window of  $d_{\rm SH}=\sqrt{14}$
applied to the coincidence centers. Each coincident candidate can belong to only a single cluster, and an element belongs to a cluster if there exists at least another element within that distance. Only the highest ranked cluster, if any,  will be selected for each 0.1-Hz band.
Clusters are ranked based on their mean significance value, but where all
clusters overlapping with a known instrumental line are ranked below any
cluster with no overlap.
A cluster is always selected for each of the 0.1-Hz  bands that had  coincidence candidates. In most cases the cluster with the largest mean significance value coincides also with the one containing the highest value.

Steps (ii) and (iii) take into account the possibility of coincidences  and formation of clusters across boundaries of consecutive 0.1-Hz frequency bands. The following tests are performed on those candidates remaining:

(iv) Based on previous studies~\cite{AllSkyMDC}, we require that interesting clusters must have a minimum population of 6 and that coincidence pairs should be generated from at least 2 different candidates per detector.

(v) For those candidates remaining, a multi-detector search is performed to verify the consistency of a possible signal. Any candidate that has a combined significance more than 1.6 below the expected value is discarded.

Outliers that pass these tests are manually examined.
In particular, outliers are also discarded if the frequency span of the cluster coincides with the list of narrow instrumental lines described in Sec.~\ref{sec:LIGO_O1}, or if there are obvious spectral disturbances associated with one of the detectors.

\subsubsection{The $\chi^2$ veto}

The $\chi^2$-test was first implemented in the \sh\ analysis of initial LIGO era S5 data~\cite{S5Hough}, and is used to reduce the number of candidates from single interferometer analysis before the coincidence step.
A veto threshold for the $\chi^2$-test is derived empirically from the O1 SFT data set.
A large number of simulated periodic gravitational wave signals are added to the SFTs, with randomly chosen amplitude, frequency, frequency derivative, sky location, polarization angle, inclination angle, and initial phase.
Then the data is analyzed separately for each detector, H1 and L1.

To determine the $\chi^2$ veto threshold (characterized by a ``veto curve''), 125 0.1-Hz bands are selected for H1 and 107 bands for L1, bands free of known large spectral disturbances.
In total  2,340,000
injections are analyzed.
The $\chi^2$ values are defined with respect to a split of the SFT data into $p=16$ segments.
The results are sorted with respect to the significance and grouped in sets containing 2000 points.
For each set the mean value of the significance, the mean of the $\chi^2$, and its standard deviation are computed.
With this reduced set of points, we fit two power laws $p-1 + A_1 s^{A_2}$  and $ \sqrt{2p-2} + B_1 s^{B_2}$ to the mean and standard deviation curve.

A detailed study of the calibration of the $\chi^2$ test using LIGO O1 data can be found in~\cite{CovasMaster}.
This study revealed a frequency-dependent behavior.
In particular, the results obtained from injections below 100 Hz differ from those between 100 and 200 Hz, while the characterization of the $\chi^2$-significance plane was similar for frequencies higher than 200 Hz.
For this reason, three different veto curves have been derived for the 50-100~Hz band, 100-200~Hz band, and for frequencies higher than 200~Hz.
In the corresponding frequency bands, the characterization was similar for both interferometers. Therefore, common veto curves are derived.

\begin{table}[htbp]
\begin{center}
\begin{tabular}{ccccc}\hline
$f$ [Hz] & $A_1$ & $A_2$& $B_1$ & $B_2$\\
\hline \hline \\
 50-100 & 0.4902   & 1.414  & 0.3581  & 1.481 \\
 100-200 & 0.2168   &  1.428  & 0.1902 & 1.499  \\
  $>$200 &  0.1187   &  1.470  &  0.0678 &  1.697 \\ \hline
\end{tabular}
\caption[chi2]{Parameters obtained for the O1 $\chi^2$ veto curve characterization in different frequency bands}
\label{tab:sh_chi2table}
\end{center}
\end{table}

\begin{figure}[htbp]
\begin{center}
\includegraphics[width=3.0in]{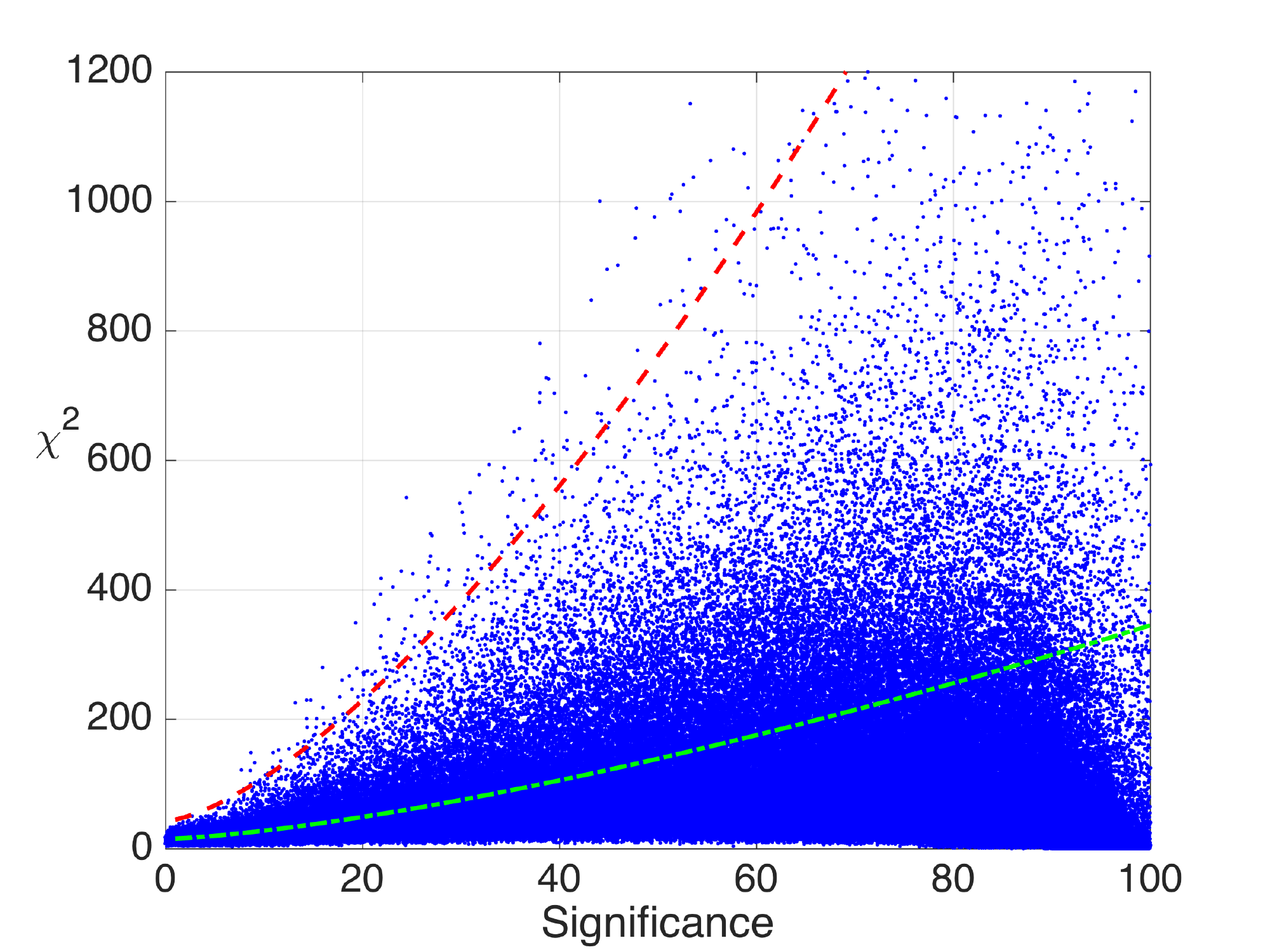}
\caption[chi2_50]{ Significance-$\chi^2$ plane for 180000 simulated injected signals in the 50 to 100 Hz band together with the fitted mean curve (dot-dashed line) and the veto curve (dashed line) corresponding to the mean $\chi^2$ plus five times its standard deviation for the H1-L1 combined data. The associated  false dismissal  rate (percentage of injections that are higher than the veto curve) is measured to be  0.12\%}
\label{fig:chi50}
\end{center}
\end{figure}

\begin{figure}[htbp]
\begin{center}
\includegraphics[width=3.0in]{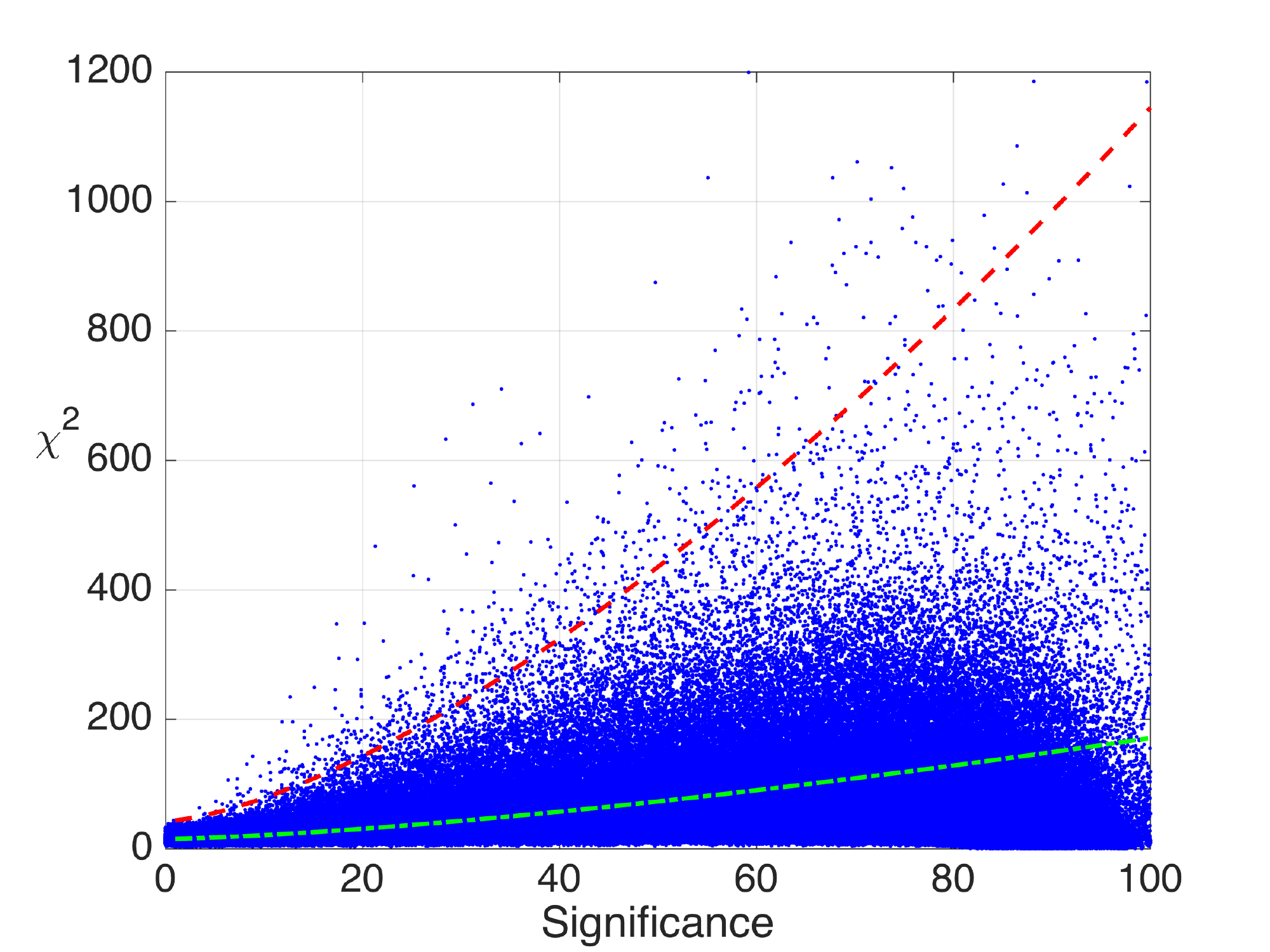}
\caption[chi2_100]{ Same as Fig.~\ref{fig:chi50}  for 320000 injections  in the 100 to 200 Hz band. The false dismissal rate being 0.21\%}
\label{fig:chi100}
\end{center}
\end{figure}

\begin{figure}[htbp]
\begin{center}
\includegraphics[width=3.0in]{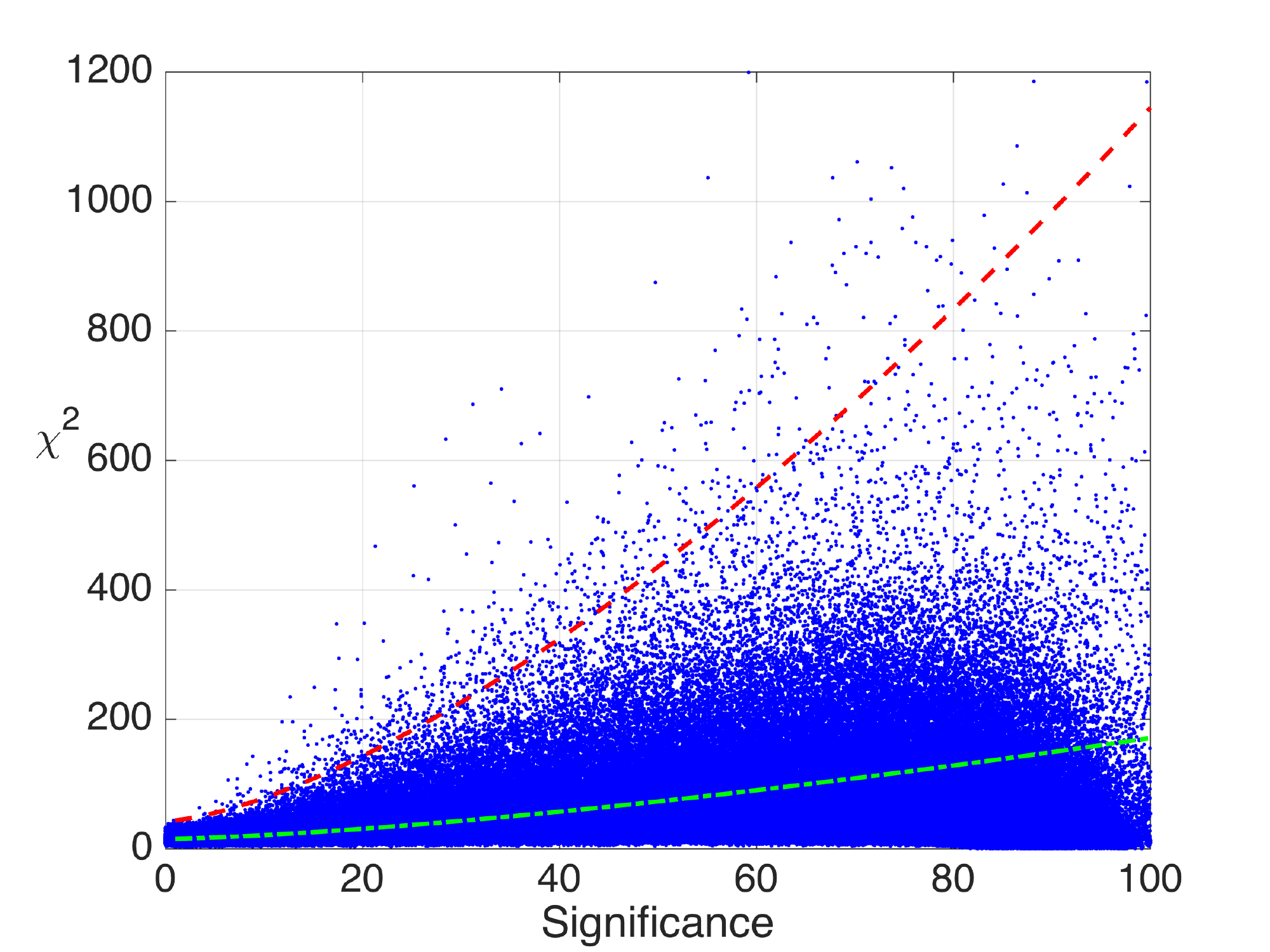}
\caption[chi2_200]{ Same as Fig.~\ref{fig:chi50} for 1840000 injections at frequencies higher than 200 Hz for the combined H1-L1 data.
The  false dismissal rate being 0.16\%}
\label{fig:chi200}
\end{center}
\end{figure}

The coefficients obtained for the proposed characterization can be found in Table~\ref{tab:sh_chi2table}. Figures~\ref{fig:chi50}, \ref{fig:chi100}, and \ref{fig:chi200} show the fitted curves and resulting veto curves corresponding veto for the mean $\chi^2$ plus five times its standard deviation for the H1-L1 combined data.
The associated false-dismissal rate for this veto is measured to be 0.12\% for the 50-100~Hz band, 0.21\% for the 100-200~Hz band, and 0.16\% for frequencies higher than 200~Hz.


\subsubsection{The multi-detector consistency veto}

Similar to the preceding $\chi^2$ test, a multi-detector consistency veto can be derived by comparing the significance results from a multi-detector search to those obtained by analyzing the data from the H1 and L1 detectors separately.

In particular, for each point in parameter space, we can derive the expected multi-detector significance from the significance obtained in the separate analysis of H1 and L1 data by using the weights defined by Eqn.~\ref{sh:w} and the SFT sets in use.
Since in this search the exact value of the weights is not stored, an approximation can be derived by ignoring the effect of the antenna pattern and considering only the influence of the varying noise levels of the different SFTs in a given frequency band.

The following expression can then be derived  for the multi-detector search
\beq
s_{\textrm{theo}} = \frac{s_{\textrm{L1}} \sqrt{\sum^{N_{\textrm{L1}}}_{i=0}\left({S^{(i)}_{\textrm{L1}}}\right)^{-2}} + s_{H1} \sqrt{\sum^{N_{\textrm{H1}}}_{i=0}\left({S^{(i)}_{\textrm{H1}}}\right)^{-2}}}
{\sqrt{\sum^{N_{\textrm{L1}}}_{i=0} \left({S^{(i)}_{\textrm{L1}}}\right)^{-2} + \sum^{N_{\textrm{H1}}}_{i=0}  \left({S^{(i)}_{\textrm{H1}}}\right)^{-2}}}
\label{eq:snrconsistency}
\eeq
where $N_{\textrm{H1}}$ and $N_{\textrm{L1}}$  are the number of SFTs of each detector, $S^{(i)}_{\textrm{H1}}$ and $S^{(i)}_{\textrm{L1}}$ are the one-sided PSDs of each detector averaged around a small frequency interval, and $s_{\textrm{H1}}$ and $s_{\textrm{L1}}$ are the significances of the separate single-detector searches.

Ideally, a coincidence pair from a periodic gravitational wave signal would have  $s_{\textrm{H1}}$, $s_{\textrm{L1}}$, and $s_{\textrm{theo}}$ values consistent with Eq.~\ref{eq:snrconsistency} within uncertainties arising from use of nearby--but not identical--templates and from noise fluctuations.
Furthermore, we are interested in characterizing its validity when considering the maximum significance values obtained in a small volume in parameter space.

\begin{figure}[htbp]
\begin{center}
\includegraphics[width=\columnwidth]{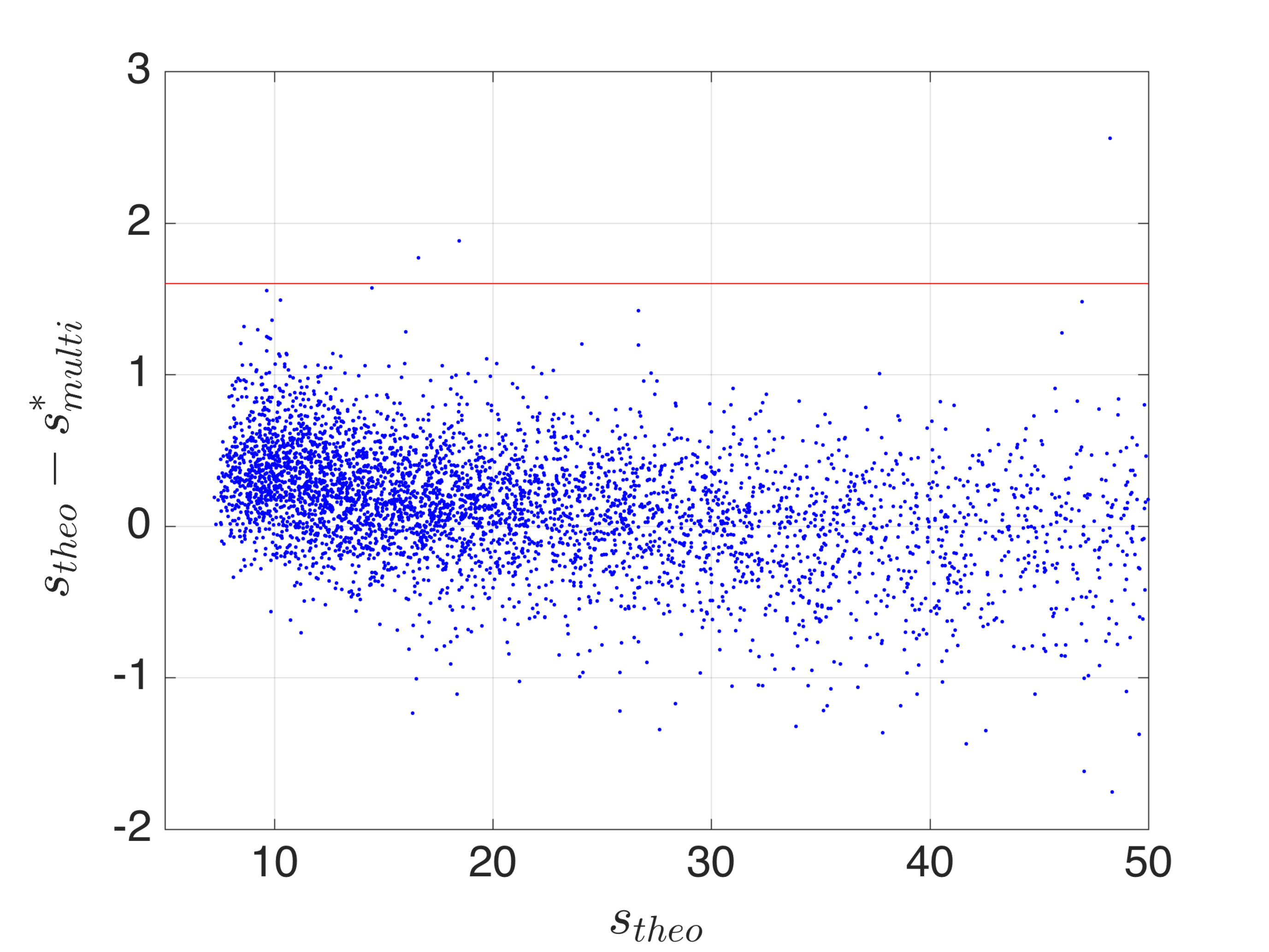}
\caption[MultiIfo]{ Characterization  of the multi-detector significance consistency veto using 4356 simulated injected signals  in the 50-500 Hz range. Each point represents a separate injection. The horizontal axis corresponds to the theoretical expected significance value, $s_{theo}$,  while the vertical axis is the difference between the theoretical and the measured value of significance, $s^*_{multi}$. The solid line is placed at  a difference
in significance of 1.6 that has only been exceeded by 3 injections. }
\label{fig:MixedSFTs_SigVeto}
\end{center}
\end{figure}

In order to test the validity of the consistency requirement, we have injected simulated signals in the 50-500 Hz range, randomly covering the same parameters of our search and for a variety of signal amplitudes. A full search, but covering only one sky patch, is performed on H1 and L1 data, as well as for the combined SFT data,
returning  a list of the most significant candidates for each of them. Of all the injections performed,  we considered only those with amplitudes strong enough that within a frequency  and spin-down window of 4 bins around the injected signal parameters, the maximum significance value would be at least $5$ for both individual single interferometer searches, and consequently a theoretical combined significance higher than $7$.
 A total of $4356$ injections with an expected theoretical combined significance between $7$ and $50$ were considered, and the results are presented in Figure~\ref{fig:MixedSFTs_SigVeto}.

In Figure~\ref{fig:MixedSFTs_SigVeto} we characterize the  difference in significance  obtained with respect to the theoretical expected value. From this plot, the multi-detector consistency veto for the O1 data used in this search can be determined: if the multi-detector combined significance has a value more than 1.6 below the nominal theoretically expected value, the candidate is vetoed.  This value of 1.6 yields a false dismissal rate of 0.07\%.

\subsubsection{Upper limit computation}

Upper limits are derived from the sensitivity depth for each 0.1-Hz band between 50 and 500 Hz.
The value of the depth corresponding to the  averaged 95\% confidence level upper limit is
obtained by means of simulated periodic gravitational wave signals added into the  SFT data of both detectors H1 and L1  in a limited number of frequency bands. In those bands, the detection efficiency, i.e., the fraction of signals that are considered detected (that have passed steps (i)-(iv) above), is computed as a function of signal strength, $h_0$ expressed by the sensitivity depth $\sqrt{S_n}/h_0$ ($1/\sqrt{\mathrm{Hz}}$). Here, $S_n$ is the maximum over both detectors of the power spectral density of the data, at the frequency of the signal, estimated
as the  power-2 mean value,
$\left( \sum_{i=1}^N  \left(S^{(i)}_{k} \right)^{-2}  /N \right)^{-2}$,   across the different noise level $S^{(i)}_{k}$  of   the different $N$ SFTs.

\begin{figure}[htbp]
\begin{center}
\includegraphics[width=\columnwidth]{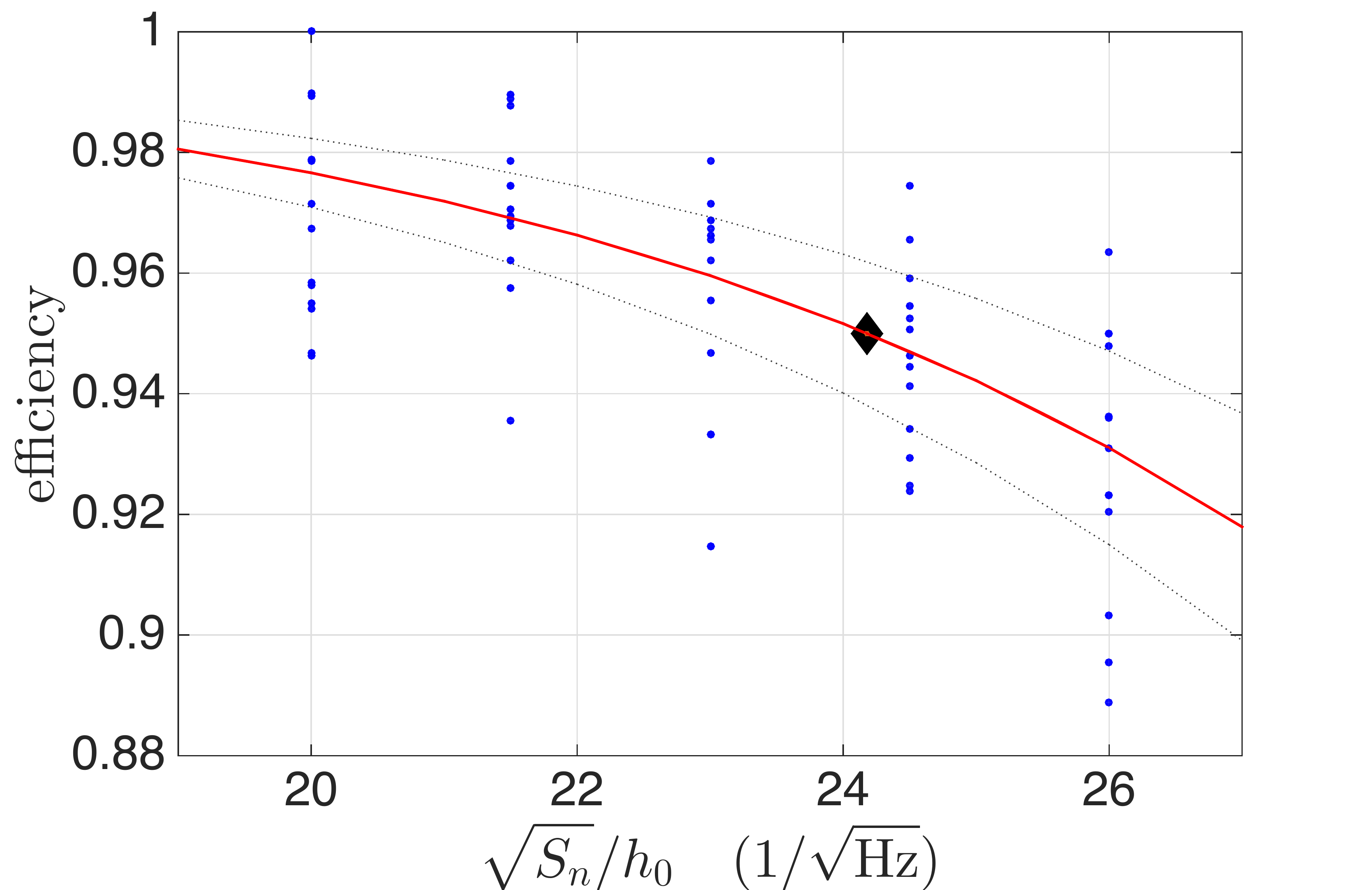}
\caption[MultiIfo]{ Detection efficiency as a function of depth obtained for 22  frequency bands.
Each dot corresponds to a set of 200 signal injections.
The solid line (red) correspond to the fitted sigmoid curve.
The diamond shows the depth value corresponding to an averaged all-sky 95\% detection efficiency,  $D^{95\%}=24.2$  ($1/\sqrt{\mathrm{Hz}}$). }
\label{fig:ULfit}
\end{center}
\end{figure}

\begin{figure*}[htbp]
\begin{center}
  \includegraphics[width=7.1in]{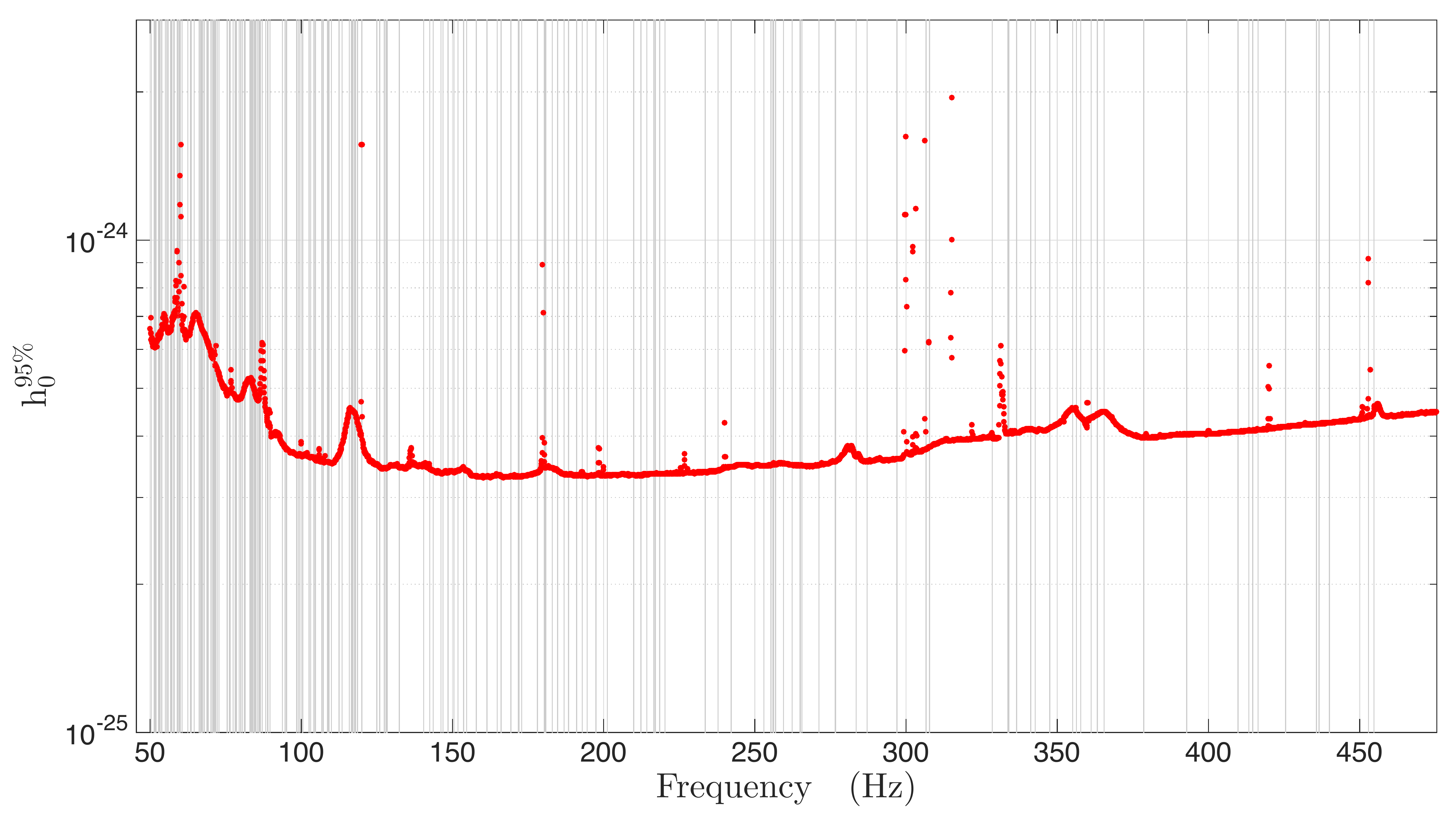}
 \caption{\sh\ O1 upper limits. The dotted (red) curve shows the averaged $95$\%~confidence level upper limits for every  analyzed 0.1-Hz band. The vertical (grey) lines indicate  194 0.1-Hz bands in which coincidence candidates were found and consequently no  upper limits are set.}
\label{fig:skyhough_O1_upper_limits}
\end{center}
\end{figure*}

Twenty-two different 0.1-Hz bands free of spectral disturbances  in both detectors were selected  with the following starting frequencies [73.6, 80.8, 98.3, 140.8, 170.2,
177.8, 201.1, 215.1, 240.7, 240.8, 250.7, 305.3, 320.3, 350.6, 381.6,
400.7, 402.1, 406.8, 416.2, 436.9, 446.9, 449.4].
In all these selected bands, we generated five sets of 200 signals each,
with fixed  sensitivity depth each set and random parameters  $(f, \alpha,\delta,\dot f, \varphi_0, \psi, \cos \iota)$.
For each injected signal added into the data of both detectors an analysis was
 done using the \sh\ search pipeline over a frequency  band of 0.1 Hz and the full spin-down range, but covering only one sky patch. For this sky-patch a list of 300 loudest candidates was produced.
 Then we imposed a threshold on significance, based on the minimum significance found in the all-sky search in the corresponding 0.1-Hz band before any injections.
 The post-processing is then done using the same parameters as in the search, including the population veto.

For each of those 22 frequency bands, a sigmoid curve,
\beq
\mathrm{Efficiency(Depth)} = 1 - \frac{1}{1 + \exp(b( \mathrm{Depth}-a))} \, ,
\eeq
was fitted by means of the least absolute residuals.
Then the 95\% confidence upper limit was deduced  from the corresponding value of the depth. With this procedure, the minimum and maximum values of the depth corresponding to the desired upper limit were 21.9 and 26.6  ($1/\sqrt{\mathrm{Hz}}$) respectively.
We also collected the results from all the frequency bands and, as shown in Figure~\ref{fig:ULfit},  performed a sigmoid fitting as before and obtained
the following fitted coefficients (with 95\% confidence bounds):
       $a$ =       39.83  (34.93, 44.73)   ($1/\sqrt{\mathrm{Hz}}$)  and
       $b$ =     $-$0.1882  ($-$0.2476, $-$0.1289)   ($\sqrt{\mathrm{Hz}}$),
that yields  the joint depth for corresponding to the 95\% upper limit  of  $D^{95\%}=24.2$  ($1/\sqrt{\mathrm{Hz}}$),  its uncertainty being smaller than 7\% for undisturbed bands, with the exception of the 98.3 Hz band for which the upper limit using this joint value would be underestimated by 10\% and for 406.8 Hz band for which the upper limit is overestimated by 9.5\%.

The 95\% confidence upper limit on $h_0$ for undisturbed bands can then be derived by simply scaling the power spectral density of the data, $h_0^{95\%}= \sqrt{S_n}/D^{95\%}$. The computed upper limits are shown in
Figure~\ref{fig:skyhough_O1_upper_limits}. No limits have been placed in 194 0.1-Hz bands in which coincidence candidates were detected, as this scaling  procedure can have larger errors in those bands due to the presence of spectral disturbances.

\subsection{\td\ search method}

The \td\ search method uses the algorithms described in~\cite{jks,AstoneBJPK2010,VSR1TDFstat,PisarskiJ2015} and has been applied to an all-sky search of VSR1 data~\cite{VSR1TDFstat}.

\begin{figure}[htbp]
  \includegraphics[width=0.9\columnwidth]{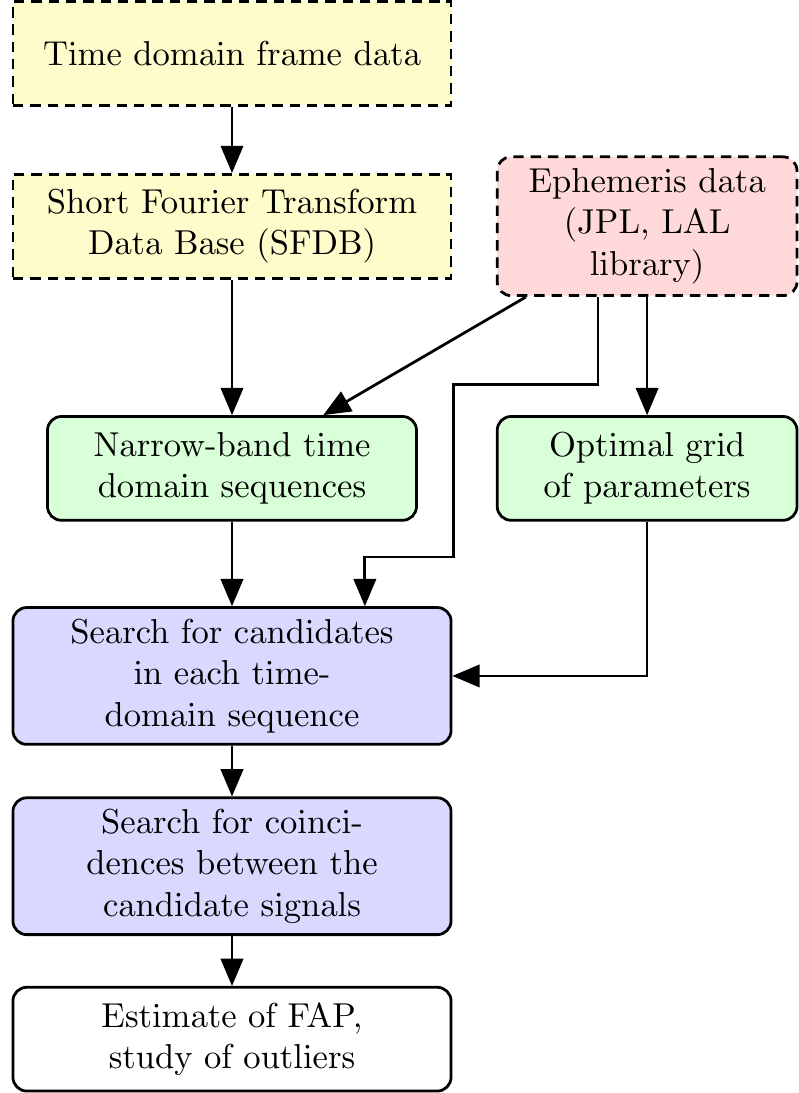}
  \caption{\td\ pipeline flowchart. Two main parts of the pipeline are
the coherent {\Fstat} search for candidate signals in time-domain segments, which is followed
by a search for coincidences between candidates from different time-domain segments.}
  \label{fig:tdfstat_flowchart}
\end{figure}

The search method consists primarily of two parts. The first part is  the
coherent search of narrowband, time-domain segments. The second part is the
search for coincidences among the candidates obtained from the coherent search
(see Figure~\ref{fig:tdfstat_flowchart}).

The time-domain segments of the data are extracted from the same set of SFTs
used by the \fh\ pipeline.  The data are split
into bands of 0.25~Hz long. The bands are overlapped by $2\times10^{-7}$~Hz. For each band, the data is inverse Fourier transformed to
extract a time series of O1 data from the SFTs. The time series is divided into
segments, called frames, of six sidereal days long each. Thus, the band $[10$-$475]$~Hz
has 1921 frequency bands. The band number $b$ is related to the
reference band frequency $f_b$ as follows:

\begin{equation}
f_b = 10~\textrm{Hz} + \frac{b(1-2^{-5})}{2 \Delta t},
  \label{eq:tdfstat_O1_band_freq}
\end{equation}
where the sampling time $\Delta t=2$~s. For O1 data, which is about 120 days long,
we obtain 20 time frames. Each 6-day narrowband segment contains $N = 258492$
data points. The O1 data has a number of non-science data segments. The
values of these bad data are set to zero. For analysis, we choose only segments
that have a fraction of bad data less than 1/3. This requirement results in eight 6-day-long
data segments for each band. Consequently, we have 15368 data segments to
analyze. These segments are analyzed coherently using the {\Fstat}. We
set a fixed threshold for the {\Fstat} of {$\mathcal{F}_0 = 14.5$} and record
parameters of all the threshold crossings together with the corresponding
values of the signal-to-noise ratio $\rho$

\begin{equation}
\rho = \sqrt{2(\F - 2)}.
  \label{eq:tdfstat_O1_fstat_snr}
\end{equation}

For the search we use a four-dimensional
(parametrized by frequency, spindown, and two more parameters
related to the position of the source in the sky)
grid of templates constructed in Sec.\ 4 of~\cite{PisarskiJ2015},
which belongs to the family $S_1$ of grids considered in~\cite{PisarskiJ2015}.
The grid has a minimal match $\textrm{MM}=\sqrt{3}/2$ and its thickness
equals 1.77959, what is only $\sim$0.8\% larger than the thickness of the
four-dimensional optimal lattice covering $A_4^*$ (equal to 1.76553).
We also veto candidates overlapping with lines
identified as instrumental artifacts.

In the second stage of the search we search for coincidences among the
candidates obtained in the coherent part of the search. We use exactly the same
coincidence search algorithm as in the analysis of VSR1 data and described in
detail in Section 8 of~\cite{VSR1TDFstat}. We  search for coincidences in each of the 1921
bands analyzed. To estimate the significance of a given coincidence, we use the
formula for the false alarm probability derived in the appendix of~\cite{VSR1TDFstat}.
Sufficiently significant coincidences are called outliers and subject
to further investigation.

\subsubsection{Sensitivity of the search}

The sensitivity of the search is taken to be the amplitude $h_0$ of the
gravitational wave signal that can be confidently detected. To estimate the
sensitivity we use a procedure developed in~\cite{S4EH}. We determine the
sensitivity of the search in each of the 1921 frequency bands that we have
searched. We perform Monte-Carlo simulations in
which, for a given amplitude $h_0$, we randomly select the other seven
parameters of the signal: $\omega_0, \omega_1, \alpha, \delta, \phi_0, \iota$
and $\psi$.  We choose frequency and spindown parameters uniformly over their
range, and source positions uniformly over the sky.  We choose angles $\phi_0$
and $\psi$ uniformly over the interval $[0, 2\pi]$ and $\cos\iota$
uniformly over the interval $[-1, 1]$.  For each band, the simulated signal is added to all
the data segments chosen for the analysis in that band. Then the data is processed through the pipeline.

\begin{figure}[htbp]
  \includegraphics[width=\columnwidth]{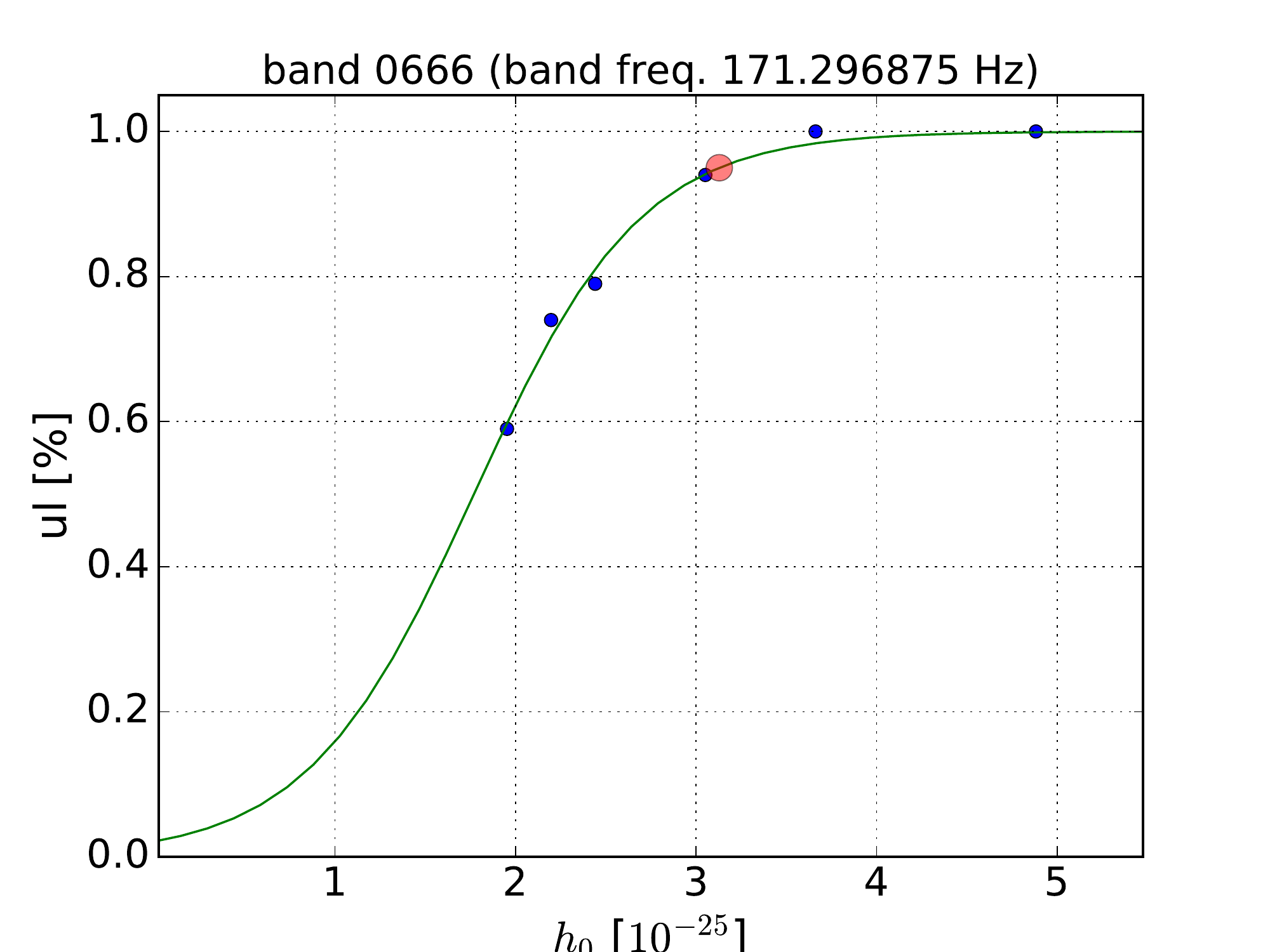}
  \caption{\td\ method of establishing upper limits from the
Monte-Carlo simulations with sigmoid fitting (an example for band no. $0666$). Blue
dots denote the results of the MC simulations for the range of amplitudes $h_0$
of injected signals with otherwise randomly chosen parameters. The green curve is
the best fit of Eq.~\ref{eq:tdfstat_O1_sigmoid}; the red dot corresponds to the
95\% upper limit (see Figure~\ref{fig:tdfstat_O1_upper_limits} for the summary of
results for the whole range of searched frequencies).}
\label{fig:tdfstat_O1_example_sigmoid}
\end{figure}

\begin{figure*}[htbp]
  \begin{center}
  \includegraphics[width=7.2in]{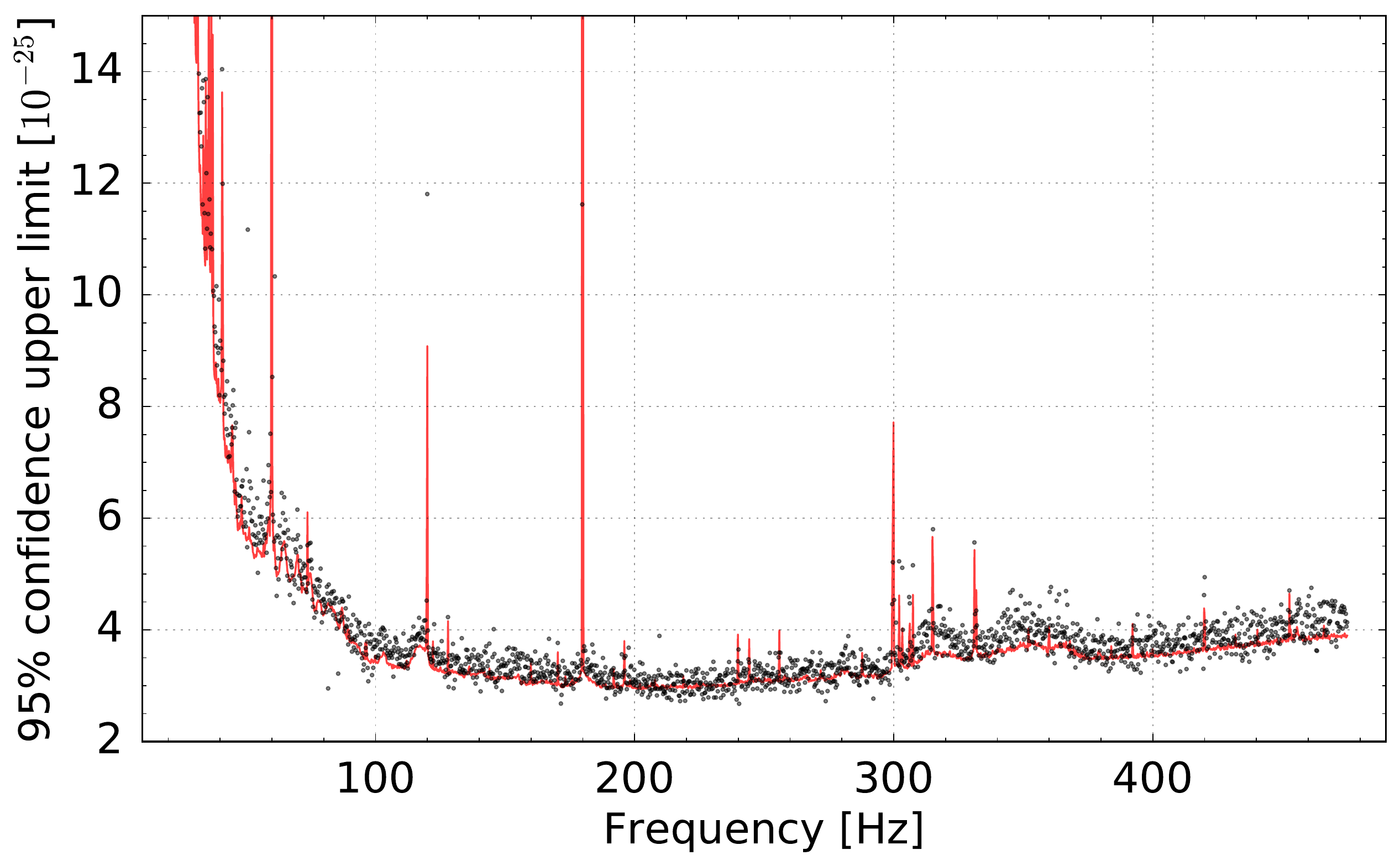}
  \caption{\td\ pipeline O1 upper limits. Black dots are the 95\% confidence upper limits for each frequency,
the red line denotes the H1 and L1 detectors' average noise curve rescaled by the factor $27.5/\sqrt{T_0}$,
where $T_0=516984$~s is the observational time of the 6-day time series segment (color online).}
  \label{fig:tdfstat_O1_upper_limits}
  \end{center}
\end{figure*}

First, we perform a coherent {\Fstat} search of each of the data segments where
the signal was added, and store all the candidates above a chosen {\Fstat}
threshold of 14.5. In this coherent analysis, to make the computation
manageable, we search over a limited parameter space consisting of $\pm 2$ grid
points around the nearest grid point where the signal was added. Then the
coincidence analysis of the candidates is performed. The signal is considered
to be detected, if it is coincident in more than 5 of the 8 time frames
analyzed for a given band. The ratio of numbers of cases in which the signal is
detected to the total number of simulations performed for a given $h_0$ determines
the frequentist sensitivity upper limits.  To obtain the 95\% confidence
sensitivity limit on $h_0$ we fit a sigmoid function,

\begin{equation}
S(x) = \frac{1}{1 + e^{k(x_0 - x)}},
\label{eq:tdfstat_O1_sigmoid}
\end{equation}
to these data, with $x_0$ and $k$ being the fitted parameters. An
example result, for a band frequency $171.296875$~Hz (corresponding to band
number $0666$) is presented in Figure~\ref{fig:tdfstat_O1_example_sigmoid}. The
95\% confidence upper limits for the whole range of frequencies are given in
Figure~\ref{fig:tdfstat_O1_upper_limits}; they follow very well the noise curves
of the O1 data that were analyzed.

\section{Search results}
\label{sec:results}

\subsection{Introduction}

Results from the four search pipelines are presented below.
In summary, no pipelines found a credible gravitational wave
signal after following up initial outlier candidates,
and each pipeline obtained a set of upper limits. In a number
of bands, particularly at low frequencies, instrumental artifacts
prevented setting of reliable upper limits. The sensitivities obtained
by the different pipelines are comparable and generally in line with
expectations from the previous mock data challenge that used data
from the Initial LIGO S6 run~\cite{AllSkyMDC}, but a greater density
of instrumental artifacts in the O1 data and refined algorithm parameter
choices led to additional small performance differences in this analysis.
In addition to the upper limits graphs presented below, numerical data
for these values can be obtained separately~\cite{data}. 

\subsection{\pf\ search results}

\label{sec:PowerFluxResults}

The \pf\ algorithm and {\em Loosely Coherent} method compute power estimates for continuous gravitational waves in a given frequency band for a fixed set of templates. The template parameters usually include frequency, first frequency derivative and sky location.

Since the search target is a rare monochromatic signal, it would contribute excess power to one of the frequency bins after demodulation. The upper limit on the maximum excess relative to the nearby power values can then be established. For this analysis we use a universal statistic \cite{universal_statistics} that places conservative 95\% confidence level upper limits for an arbitrary statistical distribution of noise power. The universal statistic has been designed to provide close to optimal values in the common case of Gaussian distribution.

\begin{figure*}[htbp]
\begin{center}
  \includegraphics[width=7.2in]{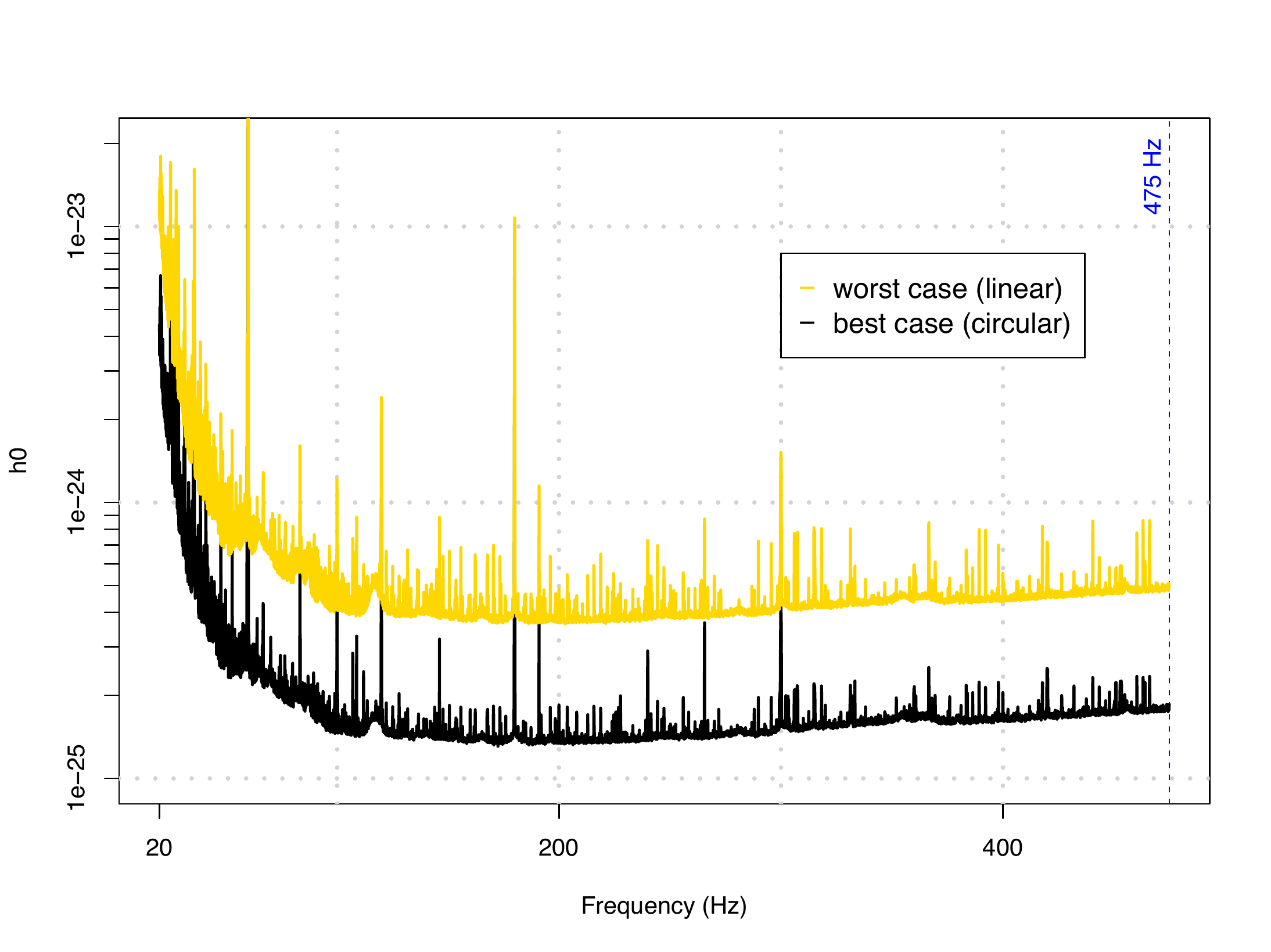}

 \caption{\pf\ O1 upper limits. The upper (yellow) curve shows worst-case (linearly polarized) $95$\%~CL upper limits in analyzed $62.5$\,mHz bands. The lower (grey) curve shows upper limits assuming a circularly polarized source.
 The data for this plot can be found in~\cite{data}. (color online)}
\label{fig:powerflux_O1_upper_limits}
\end{center}
\end{figure*}

The upper limits obtained in the search are shown in Figure~\ref{fig:powerflux_O1_upper_limits}.
The upper (yellow) curve shows the upper limits for a worst-case (linear) polarization when the smallest amount of gravitational energy is emitted toward the Earth. The lower curve shows upper limits for an optimally oriented source (circular polarization). Because of the day-night variability of the interferometer sensitivity due to anthropogenic noise, the linearly polarized sources are more susceptible to detector artifacts, as the detector response to such sources varies with the same period.

Each point in Figure~\ref{fig:powerflux_O1_upper_limits} represents a maximum over the sky, except for a small excluded portion of the sky near the ecliptic poles, which is highly susceptible to detector artifacts due to stationary frequency evolution produced by the combination of frequency derivative and Doppler shifts. The exclusion procedure is described in \cite{FullS5Semicoherent} and applied to $0.2$\%  of the sky over the entire run.

\begin{figure}[htbp]
\includegraphics[width=3in]{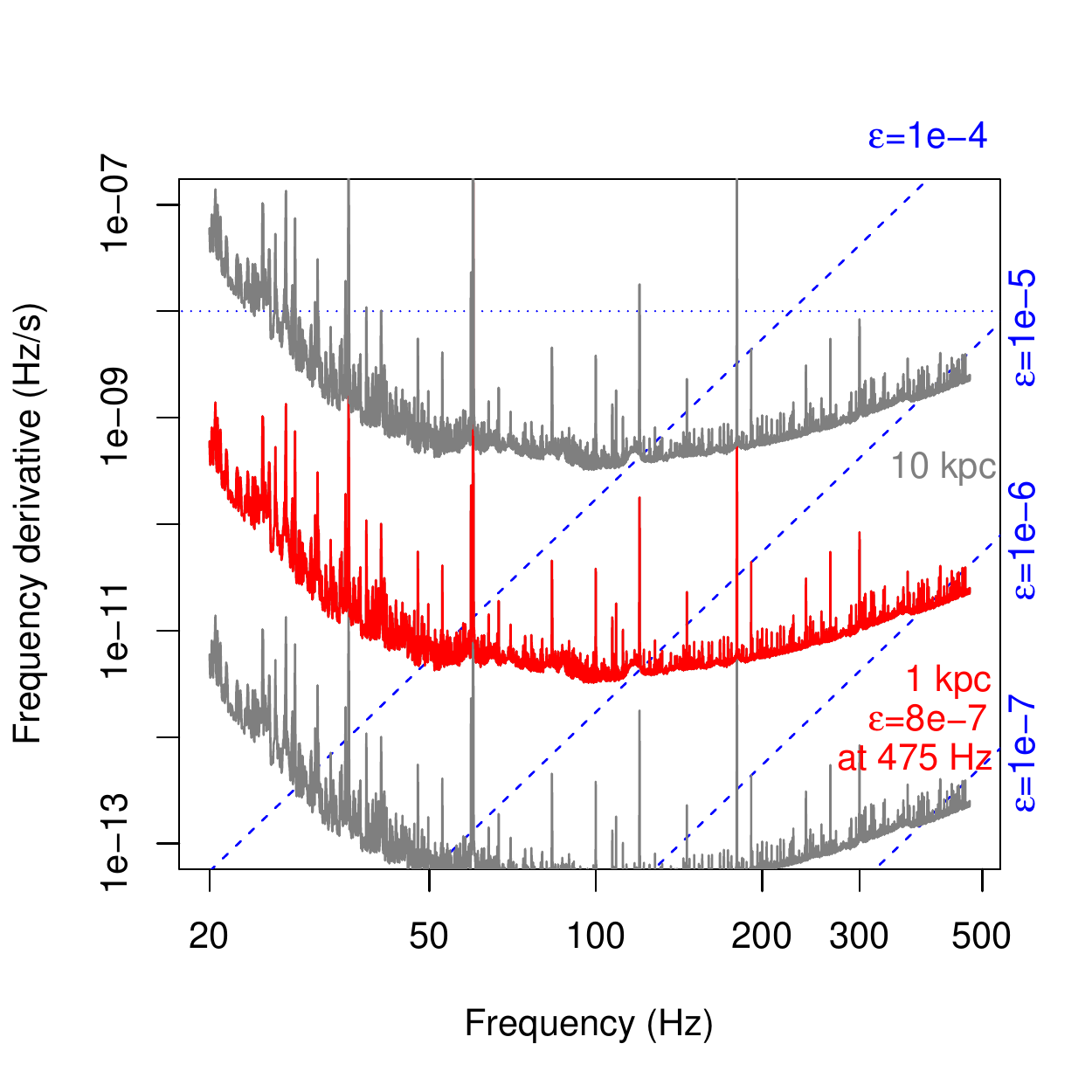}
\caption[Spindown range]{
\label{fig:spindown_range}
Range of the \pf\ search for neutron stars
spinning down solely due to gravitational radiation.  This is a
superposition of two contour plots.  The grey and red solid lines are contours of the maximum distance at which a neutron
star could be detected as a function of gravitational-wave frequency
$f$ and its derivative $\dot{f}$.  The dashed lines
are contours of the corresponding ellipticity
$\epsilon(f,\dot{f})$. The fine dotted line marks the maximum spindown searched. Together these quantities tell us the
maximum range of the search in terms of various populations (see text
for details) (color online). The other three search pipelines have similar ranges for circularly polarized sources.}
\end{figure}

If one assumes that the source spindown is solely due to emission of gravitational waves, then it is possible to recast upper limits on source amplitude as limits on source ellipticity. Figure~\ref{fig:spindown_range} shows the reach of the \pf\ search under different assumptions on source distance for circularly polarized signals. Superimposed are lines corresponding to sources of different ellipticities. Although not presented here, corresponding maximum ranges for circularly
polarized sources derived from the strain upper limits of the other three pipelines would be similar. 

The detection pipeline produced 62 outliers located near a 0.25-Hz comb of detector artifacts (table \ref{tab:PowerFluxOutliers25}), 74 outliers spanning only one data segment (about 1 month) which are particularly susceptible to detector artifacts (table \ref{tab:PowerFluxOutliersShort}) and  25 outliers (table \ref{tab:PowerFluxOutliers}) that do not fall into either of those two categories. Each outlier is identified by a numerical index. We report here SNR, frequency, spindown and sky location.

The ``Segment'' column describes the persistence of the outlier through
the data, and specifies which contiguous subset of the three equal
partitions of the timespan contributed most significantly to the
outlier: see \cite{orionspur} for details.  A true continuous signal
from an isolated source would normally have
[0,2] in this column (similar contribution from all 3 segments), or on
rare occasions [0,1] or [1,2].  Any other range strongly suggests a
statistical fluctuation, an artifact or a signal that does not conform
to the phase evolution of Equation~\ref{eqn:phase_evolution}.

During the O1 run several simulated pulsar signals were injected into the data by applying a small force to the interferometer mirrors with auxiliary lasers or via inductive forces from nearby electrodes~\cite{O1Injections}.
Several outliers were due to such hardware injections (Table \ref{tab:injections}). The hardware injection ip3 was exceptionally strong with a clear signature even in its non-Gaussian band. Note, however that these injections were not enabled for the H1 interferometer in the first part of the O1 run, leading to degraded efficiency for their detections.

\begin{table*}[htbp]
\begin{center}
\begin{tabular}{lD{.}{.}{6}rD{.}{.}{5}D{.}{.}{4}}
\hline
Label & \multicolumn{1}{c}{Frequency} & \multicolumn{1}{c}{Spindown} & \multicolumn{1}{c}{$\RAJ$} & \multicolumn{1}{c}{$\DECJ$} \\
 & \multicolumn{1}{c}{Hz} & \multicolumn{1}{c}{nHz/s} & \multicolumn{1}{c}{degrees} & \multicolumn{1}{c}{degrees} \\
\hline \hline
ip0   &  265.575533  & $\sci{-4.15}{-3}$   &   71.55193     &  -56.21749 \\
ip1   &  848.969641  & $\sci{-3.00}{-1}$   &   37.39385     &  -29.45246 \\
ip2   &  575.163521  & $\sci{-1.37}{-4}$   &  215.25617     &    3.44399 \\
ip3   &  108.857159  & $\sci{-1.46}{-8}$   &  178.37257     &  -33.4366  \\
ip4   & 1393.540559  & $\sci{-2.54}{-1}$   &  279.98768     &  -12.4666  \\
ip5   & 52.808324  & $\sci{-4.03}{-9}$     &  302.62664     &  -83.83914 \\
ip6   &  146.169370 & $\sci{-6.73}{0\mathrm{\;\;\;}}$     &  358.75095     &  -65.42262 \\
ip7   & 1220.555270 & $\sci{-1.12}{0\mathrm{\;\;\;}}$     &  223.42562     &  -20.45063 \\
ip8   &  191.031272 & $\sci{-8.65}{0\mathrm{\;\;\;}}$     &  351.38958     &  -33.41852 \\
ip9   &  763.847316 & $\sci{-1.45}{-8}$    &  198.88558     &   75.68959 \\
ip10  &   26.341917 & $\sci{-8.50}{-2}$    &  221.55565     &   42.87730 \\
ip11  &   31.424758 & $\sci{-5.07}{-4}$    &  285.09733     &  -58.27209 \\
ip12  &   38.477939 & $\sci{-6.25}{0\mathrm{\;\;\;}}$     &  331.85267     &  -16.97288 \\
ip13  &   12.428001 & $\sci{-1.00}{-2}$     &  14.32394     &  -14.32394 \\
ip14  & 1991.092401 & $\sci{-1.00}{-3}$     & 300.80284     &  -14.32394 \\
\hline
\end{tabular}
\caption[Parameters of hardware injections]{Parameters of the hardware-injected simulated continuous-wave signals during the O1 data run (epoch GPS $1130529362$). Because the interferometer configurations were largely frozen in a preliminary state after the first discovery of gravitational waves from a binary black hole merger, the hardware injections were not applied consistently. There were no injections in the H1 interferometer initially, and the initial injections in the L1 interferometer used an actuation method with significant inaccuracies at high frequencies.}
\label{tab:injections}
\end{center}
\end{table*}

The recovery of the hardware injections gives us additional confidence that no potential signals were missed.  Manual followup has shown non-injection outliers spanning all three segments to be caused by pronounced detector artifacts. Several outliers (numbers 47, 56, 70, 72, 134, 138, 154 in table \ref{tab:PowerFluxOutliers}) spanning 2 segments were also investigated with a fully coherent followup based on the Einstein@Home pipeline~\cite{ref:ehfu,ref:eho1}. None was found to be consistent with the astrophysical signal model. Tables with more details may be found in Appendix~\ref{app:pfout}.

\subsection{\fh\ search results}
\label{sec:fh_results}
In this section we report the main results of the O1 all-sky search using the \fh\ pipeline.
The number of initial candidates produced by the Hough transform stage was about $4.79\cdot 10^9$ (of which about $2.58\cdot 10^8$ belong to the band 20-128 Hz, and the rest to the band 128-512 Hz) for both Hanford and Livingston detectors.
As the number of coincident candidates remained too large, 231475 for the band 20-128 Hz and 
3109841 for the band 128-512 Hz, we reduced it with the ranking procedure described in Sec. 
\ref{sec:fh_method}. In 
practice, for computational efficiency reasons all the analysis was carried out separately for two different spin-down ranges: one from $+10^{-9}\textrm{Hz}/\textrm{s}$ to
$-2\cdot 10^{-9}\textrm{Hz}/\textrm{s}$ and the other from $-2\cdot 10^{-9}\textrm{Hz}/\textrm{s}$ to $-10^{-8}\textrm{Hz}/\textrm{s}$. As a consequence, the number of candidates selected after the ranking was 8640 for the band 20-128Hz, and 30720 for the band 128-512 Hz.
Each of these candidates was subject to the follow-up procedure, described in Sec. \ref{sec:fh_followup}. The number of candidates passing the first follow-up stage was 273 for the band 20-128 Hz and 1307 for the band 128-512 Hz and, after further vetoes, reduced to 64 for the band 20-128 Hz and 496 for the band 128-512 Hz.

From these surviving candidates we selected the outliers less consistent with noise fluctuations. In particular, we chose those for which the final peakmap projection has a critical ratio $\textrm{CR}>7.57$. This is the threshold corresponding to a false alarm probability of $1\%$ on the noise-only distribution after having taken into account the look-elsewhere effect (on the follow-up stage) \cite{narrowband}. The list of outliers is shown in Tab. \ref{tab:fh_outliers}. Each of these outliers has been manually examined, and for all of them a gravitational wave origin could be excluded, as discussed in Appendix~\ref{app:fhout}.

\begin{figure*}[htbp]
\begin{center}
\includegraphics[width=7.0in]{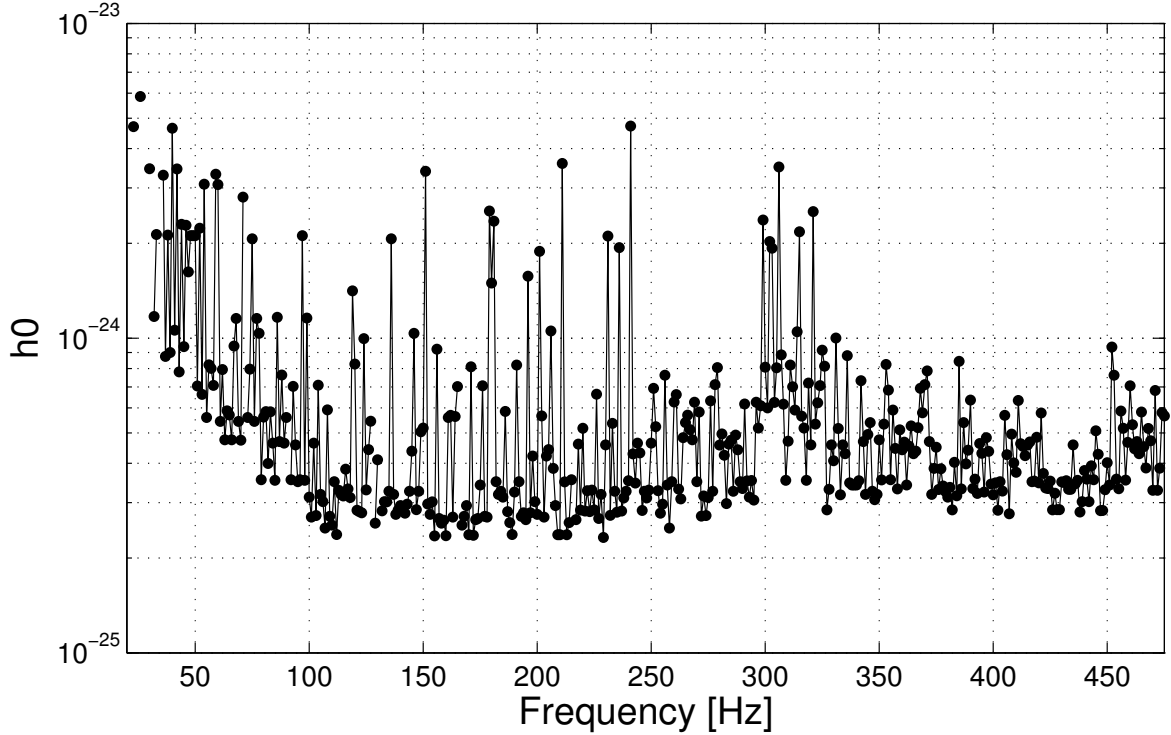}
\caption[Upper limit]{Upper limits for the \fh\ pipeline, in the range between 20 Hz and 475 Hz.}
\label{fig:fh_ul}
\end{center}
\end{figure*}

Upper limits have been computed in 1-Hz bands, as described in \ref{sec:fh_upperlimits} and are shown in Figure~\ref{fig:fh_ul}. The minimum value is about $2.5\cdot 10^{-25}$, reached in the range 150-200 Hz. In a number of frequency bands the upper limit value deviates from the smooth underlying distribution. This is a consequence of the typical behavior we see in disturbed bands and shown, as an example, in Figure~\ref{fig:weibad}: when the measured detection efficiency does not closely follow the Weibull fitting function, see Eq. \ref{eq:weifit}, in the interval around the 95\% level, the resulting upper limit can be significantly larger with respect to the value expected for a quiet band.
We have verified that such fluctuations could be significantly reduced by increasing the number of candidates selected at the ranking level: for instance going from 4 to 8 would yield much smoother results (at the price of doubling the number of follow-ups to be done).
There are some highly disturbed bands, especially below 40 Hz, for which we are unable to compute the upper limit because the detection efficiency never reaches the 95\% level.

As a further test of the capabilities of the pipeline to recover signals, in addition to those shown in Sec.~\ref{sec:fh_followup}, we report in Tab. \ref{tab:fh_hi} parameters of the recovered hardware injections, together with the error with respect to the injected signals. In general we find a good agreement (with the exception of ip5 and ip12, which are missed).
\begin{table*}[htbp]
\begin{center}
\begin{tabular}{cccccc}\hline
Label & CR & Frequency [Hz] & Spin-down [nHz/s] & $\alpha$ [deg] & $\delta$ [deg] \\
\hline \hline \\
ip10 & 37.9 & 26.34211 (0.00019) & $-$0.0671 (0.0179) & 219.6584 ($-$1.896) & 43.6464 (0.769)  \\
ip11 & 118.5 & 31.42512 (0.00027) & $-$0.0742 ($-$0.0736) & 274.4572 ($-$10.640) & $-$49.2362 (9.033) \\
ip3 & 87.8 & 108.85705 ($-$0.00011) & 0.01645 (0.0164) & 177.5770 ($-$0.795) & $-$34.1960 ($-$0.759) \\
ip6 & 42.2 & 146.16382 ($-$0.00045) & $-$6.7099 (0.0201) & 358.6904 ($-$0.061) & $-$65.2405 (0.182) \\
ip8 & 138.6 & 191.02390 ($-$0.00013) & $-$8.7161 ($-$0.0661) & 351.2037 ($-$0.186) & $-$35.0975 ($-$1.679) \\
ip0 & 180.7 & 265.57572 (0.00019) & $-$0.0483 ($-$0.0441) & 73.0276 (1.476) & $-$57.0156 ($-$0.798) \\
 \hline
\end{tabular}
\caption[]{Hardware injection recovery with the \fh\ pipeline. The reported values have been obtained at the end of the full analysis, including the follow-up. The values in parentheses are the absolute errors, that is the difference with respect to the injection parameters. The reference time is MJD=57328.59684. Two hardware injections are not accurately found (and not reported in the table): ip5 and ip12. In both cases the analysis detected ``children'' of the injected signal, with parameters significantly different from those of the injection. Injection ip11 has been discarded in the last stage of the follow-up because it did not pass the CR consistency veto among single detectors, although its parameters are recovered with good accuracy.}
\label{tab:fh_hi}
\end{center}
\end{table*}

\subsection{\sh\ search results}
\label{sec:SkyHoughResults}

In this section we report  the main results of the O1 all-sky search
 between 50 and 500 Hz using the \sh\
 pipeline, as described in section \ref{sec:SH_method}.
 In total, 194 0.1-Hz bands contained coincidence candidates,
and therefore 194 coincidence clusters were identified and further investigated.
The majority of these outliers corresponded to known spectral lines,
severe spectral disturbances or hardware injected signals.

\begin{table*}[htbp]
\begin{center}
\begin{tabular}{cccccc}\hline
Label & $s_{mean}$  & Frequency  & Spin-down & $\alpha$ & $\delta$  \\
 &  & [Hz] &  [nHz/s]&  [deg] &  [deg]    \\
\hline \hline \\
ip5 & 24.22 &  52.8084 (0.0001) & $-$0.0175 (0.0175) & 294.2376 (8.3890)& $-$83.1460 (0.6932)  \\
ip3 & 13.61 & 108.8573 (0.0002) & 0.0041  (0.0041) & 179.7435 (1.3709)& $-$32.7633 (0.6733)  \\
ip6 & 16.08 & 146.1994 (0.0006) & $-$6.6167 (0.1133) & 362.8627 (1.6137)& $-$63.7860 (1.6367) \\
ip8 & 22.83 & 191.0716 (0.0009) & $-$8.7553 (0.1053) & 348.0175 (3.3721)& $-$31.7070 (1.7115)  \\
ip0 & 21.16 & 265.5736 (0.0020) & 0.3441  (0.3482) & 68.7247 (2.8272)& $-$52.1531  (4.0643) \\
\hline
\end{tabular}
\caption[]{\sh\ hardware injection cluster information. The table provides the frequency, spin-down and sky location of the cluster center related to each of the hardware injections found by the \sh\ search. In parentheses the absolute errors with respect to the injected values are shown.
Frequencies are converted to epoch GPS 1125972653.}
\label{tab:sh_hi}
\end{center}
\end{table*}

This initial list was reduced to 59 after applying the cluster population veto
and to 26 after the multi-interferometer consistency veto. A detailed list of
these remaining  outliers is shown in Table~\ref{tab:sh_outliers}.
The multi-interferometer significance consistency
veto alone was able to reduce the initial list of 194 candidates to 33.

Of these 26 outliers, 5 corresponded to hardware injected pulsars
and 20 to  known line artifacts contaminating either H1 or L1 data.
The only unexplained outlier  around 452.89989 Hz is due to an unknown large spectral disturbance in the H1 detector. Table~\ref{tab:sh_hi} in Appendix~\ref{app:shout} provides details on these outliers.

Therefore this search did not find any evidence of a continuous
gravitational wave signal. Upper limits
have been computed in each 0.1-Hz band, except for the 194 bands
in which outliers were found.


\subsection{\td\ search results}
\label{sec:TDFstatResults}

\begin{figure}[htbp]
  \includegraphics[width=\columnwidth]{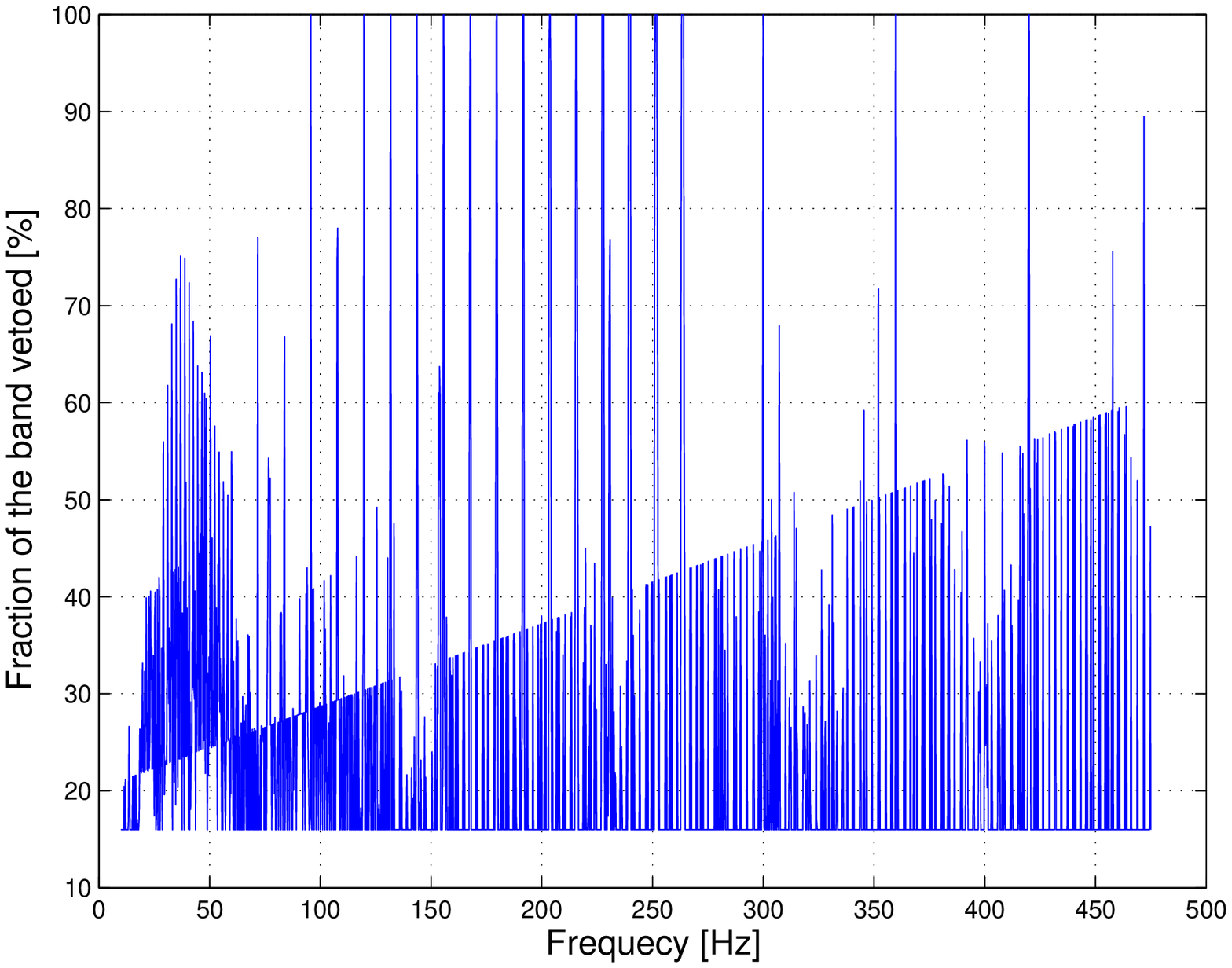}
  \caption{Fraction of the bandwidth vetoed for each band
searched by the \td\ pipeline.}
  \label{fig:tdfstat_O1_band_veto}
\end{figure}

In the bandwidth searched  $[10, 475]$ Hz, 1921
0.25-Hz long bands were defined (see Eq.~\ref{eq:tdfstat_O1_band_freq}).
As a result of vetoing candidates around the
known interference lines, a certain fraction of the bandwidth was not analyzed. In
Figure~\ref{fig:tdfstat_O1_band_veto} we show the fraction of the bandwidth
vetoed for each band. As a result 22\% of the $[10, 475]$ Hz band was vetoed, overall.

\begin{figure}[htbp]
  \includegraphics[width=\columnwidth]{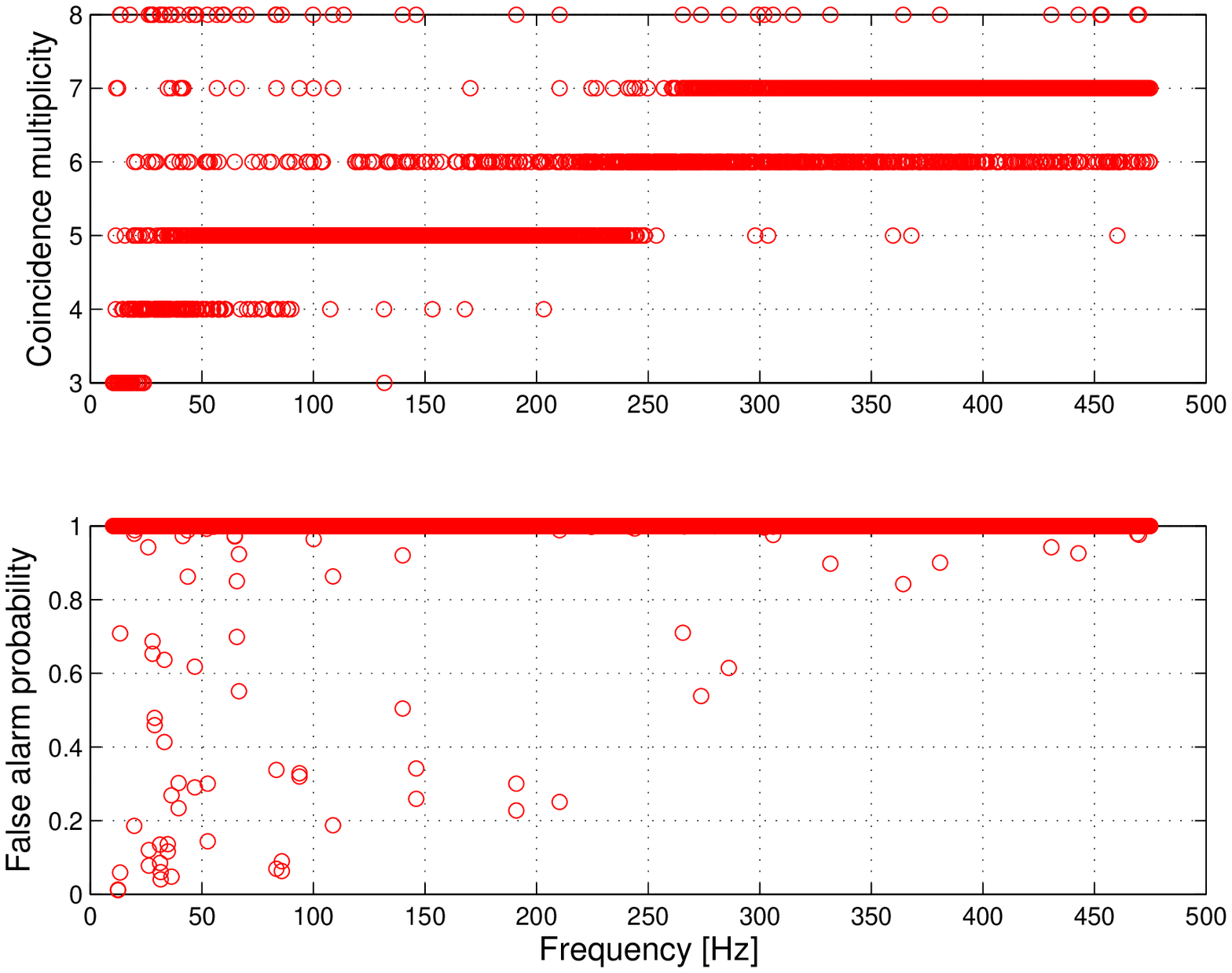}
  \caption{\td\ pipeline coincidences results as a function of the
band frequency. Top panel: maximum coincidence multiplicity. Bottom panel:
false alarm probability for the coincidence with the maximum multiplicity.}
  \label{fig:tdfstat_O1_Coinc_FA}
\end{figure}

Of 1921 bands analyzed, 38 bands were completely vetoed because of line artifacts.  As
a result of the coherent search in 15368 data segments, we obtained around $6.2
\times 10^{11}$ candidates. These candidates were subject to a search for initial
coincidences in a second stage of the \td\ analysis. The search
for coincidences was performed in all the bands except for the above-mentioned
38 that were completely vetoed.  Also, in addition to the 38 bands vetoed because of
line artifacts, there were 13 highly disturbed bands for which no coincidence
results were obtained because there were too many
candidates for the current coincidence program to handle properly.
In the coincidence analysis, for each band, the coincidences
among the candidates were searched in eight 6-day long time frames.  In
Figure~\ref{fig:tdfstat_O1_Coinc_FA} the results of the coincidence search are
presented. The top panel shows the maximum coincidence multiplicity for each of
the bands analyzed. The maximum multiplicity is an integer that varies from 3
to 8 because we require coincidence multiplicity of at least 3, 
and 8 is the number of time frames analyzed.

The bottom panel of Fig.~\ref{fig:tdfstat_O1_Coinc_FA} shows the results for
the false alarm probability of coincidence for the coincidence with the maximum
multiplicity. This false alarm probability is calculated using the formula from the
Appendix of \cite{VSR1TDFstat}.

For further analysis 49 coincidences with the lowest false alarm probability
were selected. The parameters of these coincidences are listed in Table
\ref{tab:TDFstat_outliers} in Appendix~\ref{app:tdout}: 
they are the outliers of the search. The
 parameters of a given coincidence are calculated as the mean values of the
parameters of the candidates that enter a given coincidence. Among these 49
outliers, 11 are identified with the hardware injections. 
Table \ref{tab:tdfstat_hi} presents the estimated parameters obtained for these
hardware injections, along with 
the absolute errors of the reconstructed parameters
(the differences with respect to the injected parameters).
The remaining 38 outliers include 6 associated with the 0.25 Hz comb,
15 seen in only one interferometer, 4 in only the first half of the
run, 1 transient disturbance, 8 corresponding to \pf\ outliers already
excluded, and 2 (numbers 10 and 11) requiring further, deep follow-up (although
inconsistent structures seen in run-averaged H1 and L1 spectra in that band
already cast doubt on an astrophysical origin). The deep follow-up used the
same method~\cite{ref:ehfu,ref:eho1} as for persistent outliers in the other
search pipelines. Again, no credible signals were found. 

\begin{table*}[htbp]
\begin{center}
\begin{tabular}{cccccc}\hline
Label & FA & Frequency [Hz] & Spin-down [nHz/s] & $\alpha$ [deg] & $\delta$ [deg] \\
\hline \hline \\
ip0 & 70 & 265.57565 ($-$0.00012) &   0.2582 ($-$0.2602) &  68.6196 (2.932)  & $-$53.7294 ($-$2.488) \\
ip3 & 19 & 108.85713 ($-$0.00003) & 0.0158 (0.0158) &  172.0773 (6.295) & $-$30.6495 ($-$2.787) \\
ip5 & 30 & 52.8085 ($-$0.00015) & $-$0.2168 (0.2168) &  273.6538 (28.972) & $-$63.6095 ($-$20.230) \\
ip6 & 34 & 146.16861 ($-$0.00064) & $-$6.8469 (0.1169) & 350.4083 (8.343) & $-$64.2301 ($-$1.192) \\
ip8 & 23 & 191.02942 (0.00004) & $-$8.2475 ($-$0.4024) & 340.2146 (11.175) & 8.6891 ($-$42.108) \\
ip10 & 55 & 26.34206 ($-$0.00016) & $-$0.0763 (0.0087) & 226.9401 ($-$5.384) & 41.1968 (1.680)  \\
ip11 & 89 & 31.42490 ($-$0.00014) & $-$0.0798 ($-$0.803) & 301.7315 ($-$16.634) & $-$53.2623 (5.010) \\
\hline
\end{tabular}
\caption[]{Hardware injection recovery with the \td\ pipeline. The values in parentheses are the absolute errors, that is, the difference with respect to the injection parameters. Frequencies are converted to epoch GPS 1130737464.}
\label{tab:tdfstat_hi}
\end{center}
\end{table*}

\begin{figure*}[htbp]
\includegraphics[width=7.2in]{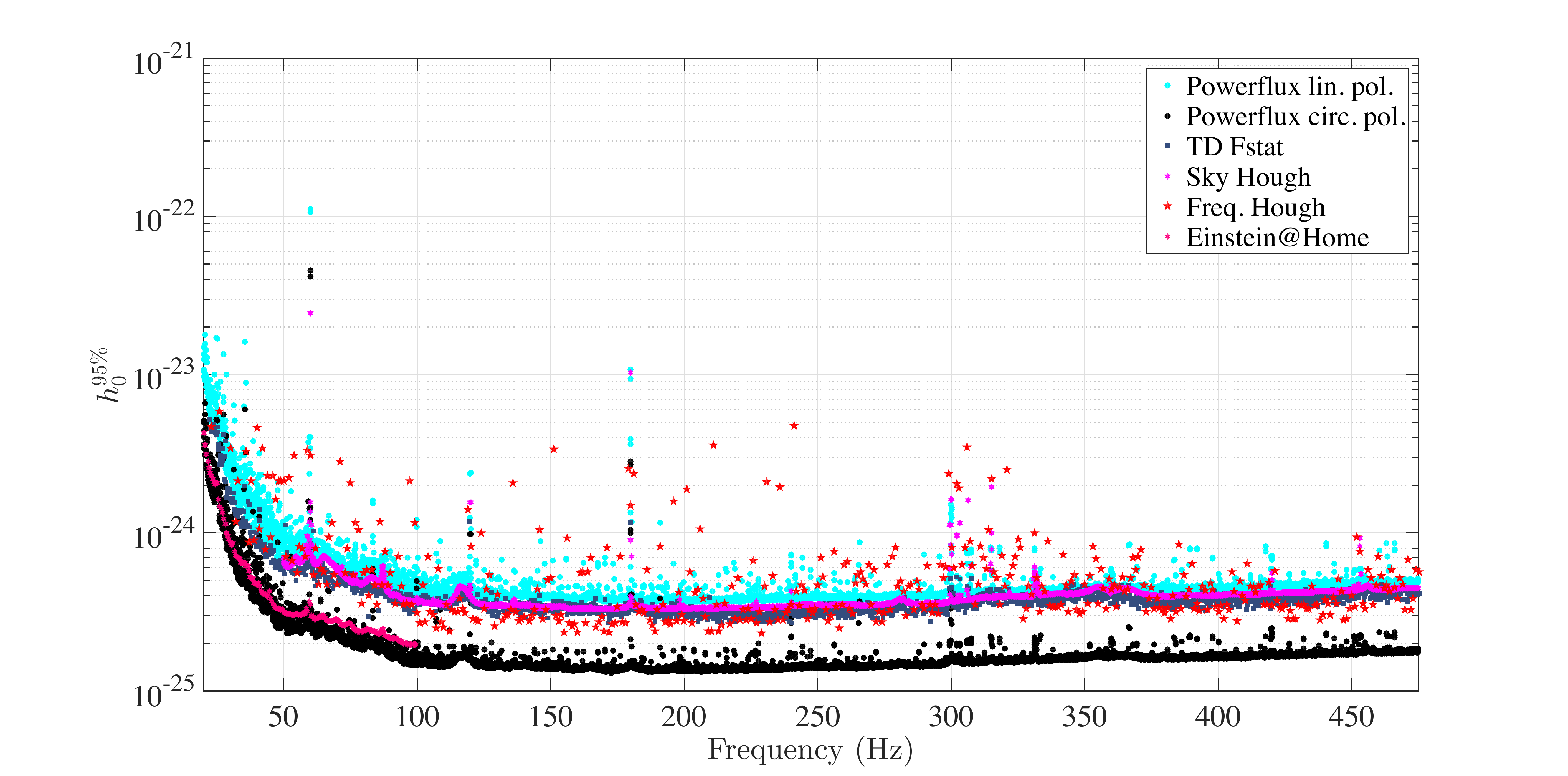}

  \caption{(Color online) Upper limit comparison for the four search pipelines used in this analysis. The curves represent the source strain amplitude $h_0$ at which 95$\%$ of simulated signals would be detected.
    Three of the pipelines (\fh, \sh, \td) present population-averaged limits over the
    full sky and source polarization, while one pipeline (\pf) presents strict all-sky limits for circular-polarization (most favorable orientation -- black) and linear-polarization (least favorable orientation -- cyan) sources. Converting the \pf\ upper limits to validated population-averaged upper limits would require extensive, band-dependent Monte Carlo simulations, but previous studies suggest that such limits would lie in a region similar to that of the other pipelines.    
    In addition, the population-averaged upper limits from the most recent Einstein@Home search are shown for comparison \cite{ref:eho1}. The Einstein@Home search explored the low frequencies, and a narrower spindown range using a much longer coherence length (210 hours).} 
  \label{fig:AllUls}
\end{figure*}

\section{\label{sec:conclusions}Conclusions}

We have performed the most sensitive all-sky searches to date for continuous gravitational waves
in the range 20-475\,Hz, using four independent search programs that apply a variety of
algorithmic approaches with different parameter choices and different approaches to
handling instrumental contaminations. The overall improvements in strain sensitivity come primarily
from the improved noise floors of the Advanced LIGO interferometers over previous LIGO and Virgo
interferometers, with reductions in upper limits of about a factor of 3 at frequencies above 100 Hz
and larger reductions at lower frequencies. 
We explored both positive and negative spindowns and found no credible
gravitational wave signals, allowing upper limits to be placed on possible source signal amplitudes.
Fig.~\ref{fig:AllUls} shows a summary of the strain amplitude upper limits obtained for the four
pipelines. Three of the pipelines (\fh, \sh, \td) present population-averaged limits over the
full sky and source polarization, while one pipeline (\pf) presents strict all-sky limits for circular-polarization and
linear-polarization sources.

At the highest frequencies we are sensitive to neutron stars with an equatorial
ellipticity as small as $\sci{8}{-7}$ and as far away as $1$\,kpc for favorable spin orientations.
The maximum ellipticity a neutron star can theoretically support is at least $\sci{1}{-5}$ according to \cite{crust_limit, crust_limit2}. Our results exclude such maximally deformed pulsars above $200$\,Hz pulsar rotation frequency ($400$\,Hz gravitational frequency) within $1$\,kpc.
Outliers from initial stages of each search method were followed up
systematically, but no candidates from any search survived scrutiny.

\section{Acknowledgments}

The authors gratefully acknowledge the support of the United States
National Science Foundation (NSF) for the construction and operation of the
LIGO Laboratory and Advanced LIGO as well as the Science and Technology Facilities Council (STFC) of the
United Kingdom, the Max-Planck-Society (MPS), and the State of
Niedersachsen/Germany for support of the construction of Advanced LIGO 
and construction and operation of the GEO600 detector. 
Additional support for Advanced LIGO was provided by the Australian Research Council.
The authors gratefully acknowledge the Italian Istituto Nazionale di Fisica Nucleare (INFN),  
the French Centre National de la Recherche Scientifique (CNRS) and
the Foundation for Fundamental Research on Matter supported by the Netherlands Organisation for Scientific Research, 
for the construction and operation of the Virgo detector
and the creation and support  of the EGO consortium. 
The authors also gratefully acknowledge research support from these agencies as well as by 
the Council of Scientific and Industrial Research of India, 
Department of Science and Technology, India,
Science \& Engineering Research Board (SERB), India,
Ministry of Human Resource Development, India,
the Spanish Ministerio de Econom\'ia y Competitividad,
the  Vicepresid\`encia i Conselleria d'Innovaci\'o, Recerca i Turisme and the Conselleria d'Educaci\'o i Universitat del Govern de les Illes Balears,
the National Science Centre of Poland,
the European Commission,
the Royal Society, 
the Scottish Funding Council, 
the Scottish Universities Physics Alliance, 
the Hungarian Scientific Research Fund (OTKA),
the Lyon Institute of Origins (LIO),
the National Research Foundation of Korea,
Industry Canada and the Province of Ontario through the Ministry of Economic Development and Innovation, 
the Natural Science and Engineering Research Council Canada,
Canadian Institute for Advanced Research,
the Brazilian Ministry of Science, Technology, and Innovation,
International Center for Theoretical Physics South American Institute for Fundamental Research (ICTP-SAIFR), 
Russian Foundation for Basic Research,
the Leverhulme Trust, 
the Research Corporation, 
Ministry of Science and Technology (MOST), Taiwan
and
the Kavli Foundation.
The authors gratefully acknowledge the support of the NSF, STFC, MPS, INFN, CNRS, PL-Grid and the
State of Niedersachsen/Germany for provision of computational resources.

This document has been assigned LIGO Laboratory document number \texttt{LIGO-P1700052-v19}.

\appendix
\section{\pf\ outlier tables}
\label{app:pfout}

\pf\ outliers are separated into three categories. Of the most interest are outliers in table \ref{tab:PowerFluxOutliers} spanning 2 or more segments that are outside a known comb of $0.25$~Hz lines. Outliers spanning only one segment are presented in table \ref{tab:PowerFluxOutliersShort}. Finally the table \ref{tab:PowerFluxOutliers25} lists outliers near $0.25$~Hz comb.

\begin{table*}[htbp]
\begin{center}
\scriptsize
\begin{tabular}{D{.}{.}{2}D{.}{.}{2}cD{.}{.}{5}D{.}{.}{4}D{.}{.}{4}D{.}{.}{4}l}\hline
\multicolumn{1}{c}{Idx} & \multicolumn{1}{c}{SNR}  & \multicolumn{1}{c}{Segment} &  \multicolumn{1}{c}{Frequency} & \multicolumn{1}{c}{Spindown} &  \multicolumn{1}{c}{$\RAJ$}  & \multicolumn{1}{c}{$\DECJ$} & Description \\
\multicolumn{1}{c}{}	&  \multicolumn{1}{c}{}	&  \multicolumn{1}{c}{}	& \multicolumn{1}{c}{Hz}	&  \multicolumn{1}{c}{nHz/s} & \multicolumn{1}{c}{degrees} & \multicolumn{1}{c}{degrees} & \\
\hline \hline
1 & 10532 & $[0,2]$ & 256.00854 & -1.475 & 255.225 & -85.537 & Extremely strong bin-centered line at 256.0 Hz\\
33 & 352 & $[1,2]$ & 52.80829 & 0.000 & 301.148 & -84.274 & Hardware injection ip5\\
36 & 269 & $[0,2]$ & 191.03127 & -8.663 & 351.292 & -33.643 & Hardware injection ip8\\
37 & 236 & $[1,2]$ & 265.57551 & -0.013 & 71.719 & -56.276 & Hardware injection ip0\\
39 & 227 & $[0,1]$ & 21.41061 & -0.500 & 230.715 & -1.630 & Sharp line in L1 near 21.41 Hz, H1 and L1 SNR inconsistent\\
46 & 172 & $[1,2]$ & 146.16942 & -6.775 & 356.992 & -65.953 & Hardware injection ip6\\
47 & 170 & $[0,1]$ & 31.11704 & -7.362 & 56.406 & -22.568 & Coincident combs with different morphology between H1 and L1\\
52 & 145 & $[1,2]$ & 108.85708 & 0.037 & 176.861 & -34.170 & Hardware injection ip3\\
56 & 110 & $[0,1]$ & 59.60507 & 0.113 & 283.407 & 68.377 & Coincident lines in spectrum, signal nearly stationary in detector frame\\
70 & 63 & $[1,2]$ & 99.96961 & 1.775 & 99.689 & 35.244 & Coincident regions between H1 and L1\\
72 & 52 & $[1,2]$ & 30.63391 & -6.688 & 357.777 & -40.140 & Both H1 and L1 spectra are contaminated\\
74 & 48 & $[1,2]$ & 412.00362 & -0.588 & 84.695 & -71.077 & Sharp bin-centered line in L1 at 412.0 Hz\\
78 & 39 & $[0,1]$ & 93.75912 & 0.062 & 279.035 & -13.320 & Sharp and coincident lines in H1 and L1, SNR inconsistent\\
85 & 33 & $[0,1]$ & 299.42508 & 1.175 & 317.303 & 46.331 & All SNR comes from large artifact in H1\\
95 & 23 & $[0,2]$ & 90.74396 & 0.250 & 86.019 & -18.203 & Coincident bin-centered lines at 90.75 Hz, 0.25 Hz comb\\
98 & 19 & $[0,2]$ & 33.59221 & 0.300 & 359.165 & -18.874 & Both H1 and L1 spectra are disturbed, H1 does not see anything\\
99 & 19 & $[1,2]$ & 306.01509 & -2.275 & 130.248 & 33.827 & Large artifact in L1, H1 does not see anything\\
100 & 19 & $[0,2]$ & 299.39436 & -0.287 & 116.867 & 73.521 & Large artifact in H1, L1 does not see anything\\
102 & 17 & $[0,2]$ & 307.31880 & -6.612 & 181.054 & -25.115 & Large artifact in L1, H1 does not see anything\\
109 & 16 & $[0,1]$ & 452.88717 & 0.238 & 82.204 & -67.295 & Large artifact in H1\\
122 & 14 & $[1,2]$ & 452.86305 & -2.350 & 198.764 & 49.143 & Large artifact in H1\\
128 & 13 & $[1,2]$ & 29.63552 & 0.312 & 107.619 & 1.373 & Coincident disturbances with different morphologies in H1 and L1\\
134 & 13 & $[1,2]$ & 265.75911 & -2.112 & 67.964 & -28.799 & Not confirmed by Einstein@Home followup\\
138 & 13 & $[0,1]$ & 178.60606 & -3.275 & 194.163 & 51.285 & Sharp line in L1 at 178.7 Hz is outside signal range\\
154 & 12 & $[1,2]$ & 404.79214 & -6.675 & 136.431 & 38.381 & Not confirmed by Einstein@Home followup\\

\hline
\end{tabular}
\caption[Outliers that passed detection pipeline]{Outliers that passed \pf\ detection pipeline spanning more than one segment and excluding those near a $0.25$-Hz comb. Only the highest-SNR outlier is shown for each 0.1-Hz frequency region. Outliers marked with ``line'' had strong narrowband disturbances identified near the outlier location. The ``Segment'' column reports the set of
contiguous segments of the data that produced the outlier, as
described in Section~\ref{sec:results}.
Frequencies are converted to epoch GPS $1130529362$.}
\label{tab:PowerFluxOutliers}
\end{center}
\end{table*}

\begin{table*}[htbp]
\begin{center}
\scriptsize
\begin{tabular}{D{.}{.}{2}D{.}{.}{2}cD{.}{.}{5}D{.}{.}{4}D{.}{.}{4}D{.}{.}{4}l}\hline
\multicolumn{1}{c}{Idx} & \multicolumn{1}{c}{SNR}  & \multicolumn{1}{c}{Segment} &  \multicolumn{1}{c}{Frequency} & \multicolumn{1}{c}{Spindown} &  \multicolumn{1}{c}{$\RAJ$}  & \multicolumn{1}{c}{$\DECJ$} & \multicolumn{1}{c}{Description} \\
\multicolumn{1}{c}{}	&  \multicolumn{1}{c}{}	&  \multicolumn{1}{c}{}	& \multicolumn{1}{c}{Hz}	&  \multicolumn{1}{c}{nHz/s} & \multicolumn{1}{c}{degrees} & \multicolumn{1}{c}{degrees} & \multicolumn{1}{c}{}\\
\hline \hline
3 & 5470 & $[1,1]$ & 35.70737 & -0.287 & 71.099 & -89.596 & Broadly coincident region between 35.7 and 35.71\\
22 & 933 & $[2,2]$ & 35.76455 & -0.700 & 241.377 & -7.873 & Broadly coincident region\\
31 & 417 & $[1,1]$ & 59.51547 & -2.112 & 282.450 & 16.555 & Coincident regions\\
40 & 223 & $[1,1]$ & 83.31080 & -2.000 & 184.935 & -16.484 & Coincident regions between H1 and L1\\
41 & 197 & $[0,0]$ & 100.00749 & 0.650 & 319.773 & -59.068 & Coincident regions between H1 and L1\\
43 & 185 & $[0,0]$ & 113.68611 & 1.062 & 26.888 & -57.341 & Strong line in H1, H1 and L1 SNR inconsistent\\
45 & 177 & $[2,2]$ & 31.42488 & -0.062 & 278.446 & -49.515 & Coincident regions between H1 and L1, is this really an injection\\
48 & 163 & $[2,2]$ & 46.95114 & -0.350 & 132.597 & -88.847 & Coincident sharp lines at 46.95\\
49 & 152 & $[2,2]$ & 38.47802 & -6.312 & 326.542 & -17.113 & Hardware injection ip12\\
57 & 108 & $[2,2]$ & 440.50273 & 1.762 & 65.270 & -37.191 & Sharp line in L1 at 440.5\\
62 & 86 & $[0,0]$ & 40.87517 & 0.375 & 236.742 & 79.741 & Elevated and disturbed spectrum, uneven sensitivity during run\\
66 & 70 & $[1,1]$ & 31.76233 & 0.613 & 64.921 & -9.503 & Coincident lines in H1 and L1\\
77 & 39 & $[1,1]$ & 39.76505 & -0.250 & 298.570 & 42.651 & Sharp and coincident lines in H1 and L1\\
82 & 37 & $[2,2]$ & 78.50513 & -0.688 & 299.793 & 45.680 & Sharp and coincident lines at 78.5 Hz, 0.25 Hz comb\\
83 & 34 & $[2,2]$ & 128.49647 & 0.962 & 323.700 & 86.937 & Outlier favors region next to the line at 128.5 Hz\\
87 & 32 & $[1,1]$ & 32.60162 & -0.912 & 221.578 & 10.228 & All SNR comes from disturbance in L1\\
89 & 29 & $[2,2]$ & 127.50310 & -0.900 & 135.660 & -85.458 & Coincident bin-centered lines at 127.5 Hz in both ifos\\
90 & 28 & $[1,1]$ & 47.68505 & -4.125 & 80.941 & -75.614 & H1 and L1 disagree on location, common overlapping artifact\\
92 & 24 & $[0,0]$ & 246.30806 & -0.800 & 257.315 & -88.956 & Coincident bin-centered lines at 246.3 Hz\\
96 & 22 & $[0,0]$ & 453.38509 & -7.400 & 213.641 & 21.911 & Large artifact in H1, L1 does not see anything\\
97 & 22 & $[0,0]$ & 70.92119 & 0.275 & 136.639 & -68.924 & Coincident artifacts in H1 and L1\\
101 & 17 & $[0,0]$ & 306.00102 & -6.513 & 106.860 & -14.138 & Small disturbance in L1\\
104 & 17 & $[2,2]$ & 26.36415 & -9.013 & 62.782 & -13.429 & Coincident combs, large artifact with different morphology in H1 and L1\\
105 & 16 & $[1,1]$ & 43.68056 & 0.087 & 114.317 & -3.107 & Large artifact in L1\\
106 & 16 & $[0,0]$ & 327.59196 & -8.162 & 23.524 & -50.419 & \\
107 & 16 & $[1,1]$ & 31.61544 & 0.238 & 17.372 & -10.319 & Different disturbances in H1 and L1 spectrum\\
108 & 16 & $[0,0]$ & 308.99815 & -3.925 & 289.216 & 47.738 & \\
110 & 15 & $[1,1]$ & 337.31874 & -0.675 & 244.827 & 41.178 & \\
111 & 15 & $[0,0]$ & 350.25984 & -8.188 & 313.477 & -34.804 & \\
112 & 15 & $[0,0]$ & 345.08507 & 0.900 & 136.135 & 40.606 & \\
113 & 15 & $[0,0]$ & 453.35293 & -5.725 & 143.703 & 64.705 & Appears to be associated with artifact at 453.40 Hz\\
114 & 14 & $[0,0]$ & 254.15359 & -9.938 & 204.993 & 43.892 & Weak sharp line in L1 at 241.1 Hz\\
115 & 14 & $[0,0]$ & 436.09611 & -6.750 & 346.514 & -36.245 & \\
116 & 14 & $[0,0]$ & 310.92629 & -3.975 & 359.968 & 14.226 & \\
117 & 14 & $[1,1]$ & 162.18514 & -1.587 & 101.520 & -3.877 & \\
118 & 14 & $[2,2]$ & 454.18958 & -4.138 & 254.665 & 0.668 & Sharp line in L1 at 454 Hz is outside signal range\\
119 & 14 & $[1,1]$ & 225.64096 & -2.725 & 234.719 & 31.199 & Large artifact in H1\\
120 & 14 & $[0,0]$ & 394.54298 & -2.238 & 147.069 & 51.380 & \\
121 & 14 & $[1,1]$ & 270.50488 & 1.025 & 192.977 & 26.380 & Sharp line in L1 at 270.5 is outside signal range\\
123 & 14 & $[2,2]$ & 213.68406 & 0.413 & 203.168 & 36.792 & \\
124 & 14 & $[1,1]$ & 293.36763 & -0.725 & 291.910 & -39.555 & \\
125 & 13 & $[1,1]$ & 336.94860 & -1.212 & 29.971 & -40.200 & \\
126 & 13 & $[1,1]$ & 318.76465 & -7.375 & 203.369 & 39.161 & Sharp line in L1 at 318.80 Hz is outside signal range\\
127 & 13 & $[1,1]$ & 97.00786 & -4.338 & 152.289 & 42.607 & Sharp line in L1 at 97.1 Hz is outside signal range\\
129 & 13 & $[2,2]$ & 446.76910 & -5.237 & 275.842 & 27.684 & \\
130 & 13 & $[2,2]$ & 415.31493 & -0.625 & 246.343 & 35.521 & Sharp line in L1 at 415.4 Hz is outside signal range\\
131 & 13 & $[1,1]$ & 289.16717 & -4.812 & 289.555 & 56.661 & Sharp line in L1 at 289.3 Hz is outside signal range\\
132 & 13 & $[2,2]$ & 435.04082 & 1.175 & 61.748 & -49.375 & \\
133 & 13 & $[0,0]$ & 353.38444 & -7.325 & 267.522 & 65.417 & \\
135 & 13 & $[0,0]$ & 247.50050 & -5.975 & 60.313 & -25.256 & Sharp line in L1 at 247.3 is outside signal range\\
136 & 13 & $[0,0]$ & 463.10329 & -1.700 & 30.527 & -20.284 & Sharp line in L1 at 463.2 Hz is outside signal range\\
137 & 13 & $[1,1]$ & 380.57045 & -9.338 & 244.202 & 45.262 & Sharp line in L1 at 480.6 Hz is outside signal range\\
139 & 13 & $[2,2]$ & 443.65961 & -1.988 & 82.034 & -43.630 & \\
140 & 13 & $[1,1]$ & 442.16364 & -6.525 & 80.319 & 31.835 & \\
141 & 13 & $[2,2]$ & 437.40225 & -9.825 & 139.879 & -37.221 & Lines in H1 and L1 outside signal range\\
142 & 13 & $[1,1]$ & 474.28332 & -8.562 & 354.078 & 39.661 & Sharp line in L1 at 474.3\\
143 & 13 & $[0,0]$ & 232.64393 & -7.500 & 244.442 & 30.182 & Sharp line in L1 at 232.8\\
144 & 13 & $[2,2]$ & 455.60513 & -9.125 & 264.100 & 12.723 & Identical slope in H1 and L1 spectrum\\
145 & 13 & $[2,2]$ & 321.96324 & -5.425 & 88.612 & -18.667 & Disturbed H1 spectrum\\
146 & 13 & $[0,0]$ & 377.36117 & -3.900 & 162.703 & 19.347 & \\
147 & 13 & $[1,1]$ & 463.81555 & -2.762 & 92.427 & -21.270 & Sharp line in L1 at 463.7 is outside signal range\\
148 & 12 & $[2,2]$ & 369.43296 & 1.113 & 318.118 & 43.385 & Sharp line in L1 at 369.5 is outside signal range\\
149 & 12 & $[0,0]$ & 348.24255 & -2.862 & 249.590 & 53.903 & Sharp line in L1 at 348.2 is outside signal range\\
150 & 12 & $[1,1]$ & 293.17864 & -1.713 & 1.204 & 9.380 & \\
151 & 12 & $[1,1]$ & 371.37093 & -9.562 & 17.812 & -38.934 & \\
152 & 12 & $[2,2]$ & 418.80178 & -9.588 & 180.042 & -37.122 & \\
153 & 12 & $[0,0]$ & 56.38793 & 0.812 & 11.265 & -26.818 & Rather large disturbance in L1 outside signal range\\
155 & 12 & $[1,1]$ & 389.58663 & 0.375 & 190.109 & 39.234 & \\
156 & 12 & $[1,1]$ & 381.17708 & -8.162 & 327.725 & -27.100 & Sharp line in L1 at 381.1 is outside signal range\\
157 & 12 & $[1,1]$ & 471.92300 & 0.800 & 351.848 & -37.756 & Sharp line in H1 at 472 Hz, sharp line in L1 471.9 Hz\\
158 & 12 & $[0,0]$ & 231.41586 & -6.038 & 151.842 & 51.699 & \\
159 & 12 & $[1,1]$ & 384.21151 & -5.938 & 205.820 & 36.442 & Sharp line in H1 at 384 Hz is far outside signal range\\
160 & 12 & $[2,2]$ & 359.81255 & -0.875 & 278.065 & 26.246 & Outlier favors sidelobes of 60 Hz harmonic\\
161 & 12 & $[0,0]$ & 302.90153 & 0.150 & 323.285 & -57.949 & \\

\hline
\end{tabular}
\caption[Outliers that passed detection pipeline]{Outliers that passed \pf\ detection pipeline spanning only one segment, excluding those near $0.25$-Hz comb. Only the highest-SNR outlier is shown for each 0.1-Hz frequency region. Outliers marked with ``line'' had strong narrowband disturbances identified near the outlier location. Segment column reports the set of
contiguous segments of the data that produced the outlier, as
described in \ref{sec:results}.
Frequencies are converted to epoch GPS $1130529362$.}
\label{tab:PowerFluxOutliersShort}
\end{center}
\end{table*}

\begin{table*}[htbp]
\begin{center}
\scriptsize
\begin{tabular}{D{.}{.}{2}D{.}{.}{2}cD{.}{.}{5}D{.}{.}{4}D{.}{.}{4}D{.}{.}{4}l}\hline
\multicolumn{1}{c}{Idx} & \multicolumn{1}{c}{SNR}  & \multicolumn{1}{c}{Segment} &  \multicolumn{1}{c}{Frequency} & \multicolumn{1}{c}{Spindown} &  \multicolumn{1}{c}{$\RAJ$}  & \multicolumn{1}{c}{$\DECJ$}  \\
\multicolumn{1}{c}{}	&  \multicolumn{1}{c}{}	&  \multicolumn{1}{c}{}	& \multicolumn{1}{c}{Hz}	&  \multicolumn{1}{c}{nHz/s} & \multicolumn{1}{c}{degrees} & \multicolumn{1}{c}{degrees} \\
\hline \hline
2 & 5603 & $[1,2]$ & 21.50025 & -0.150 & 151.422 & -79.107\\
4 & 4996 & $[1,2]$ & 34.49925 & 0.350 & 21.714 & 77.506\\
5 & 4938 & $[0,2]$ & 25.00029 & -0.413 & 234.953 & 13.114\\
7 & 4207 & $[0,2]$ & 29.50008 & -0.150 & 131.116 & -74.570\\
10 & 2921 & $[1,2]$ & 20.49974 & 0.113 & 309.323 & 79.014\\
11 & 2721 & $[0,2]$ & 33.24851 & -0.050 & 196.925 & 82.946\\
15 & 1593 & $[0,2]$ & 47.49882 & 0.463 & 23.877 & 80.095\\
17 & 1493 & $[1,2]$ & 46.49823 & 0.725 & 55.515 & 60.238\\
18 & 1238 & $[0,1]$ & 23.24905 & 0.263 & 62.224 & 5.482\\
19 & 1161 & $[0,2]$ & 31.25038 & -0.137 & 156.015 & -78.649\\
23 & 914 & $[0,2]$ & 54.50016 & 0.275 & 311.453 & 70.504\\
25 & 837 & $[0,2]$ & 22.24910 & -0.413 & 201.320 & 17.588\\
27 & 615 & $[0,2]$ & 58.50283 & -0.575 & 293.833 & -81.976\\
28 & 587 & $[0,2]$ & 57.49720 & 1.012 & 65.281 & 51.741\\
29 & 551 & $[0,2]$ & 41.50357 & -0.450 & 314.469 & -57.748\\
30 & 421 & $[0,2]$ & 43.24808 & -0.062 & 221.374 & 82.111\\
35 & 305 & $[0,1]$ & 48.24916 & -0.200 & 246.369 & 59.628\\
38 & 229 & $[0,2]$ & 20.24852 & 0.025 & 94.536 & -6.808\\
44 & 179 & $[0,1]$ & 27.24799 & -1.475 & 248.015 & -15.916\\
50 & 151 & $[0,2]$ & 50.24710 & -0.037 & 134.356 & 86.279\\
53 & 141 & $[1,2]$ & 42.99932 & 0.325 & 341.196 & 79.984\\
54 & 140 & $[1,2]$ & 34.24893 & 0.450 & 43.246 & 68.520\\
59 & 106 & $[0,1]$ & 66.75438 & 0.175 & 29.149 & 89.424\\
60 & 98 & $[0,2]$ & 38.99998 & -0.025 & 94.142 & -67.898\\
63 & 85 & $[0,1]$ & 85.50355 & -0.537 & 285.886 & -82.302\\
64 & 79 & $[1,2]$ & 40.00031 & -0.175 & 115.403 & -76.647\\
65 & 76 & $[0,2]$ & 45.00071 & 0.062 & 17.453 & -83.230\\
67 & 68 & $[1,2]$ & 39.25073 & -0.525 & 215.636 & -65.846\\
68 & 66 & $[0,2]$ & 98.49708 & 0.750 & 330.835 & 87.908\\
76 & 44 & $[1,2]$ & 21.24994 & -0.037 & 255.942 & 63.243\\
86 & 32 & $[0,2]$ & 88.75324 & -0.062 & 212.757 & -88.557\\
91 & 25 & $[0,2]$ & 89.75337 & 0.775 & 317.050 & 51.127\\
93 & 24 & $[0,2]$ & 91.75467 & -0.338 & 260.215 & 23.279\\
103 & 17 & $[0,2]$ & 97.75447 & -1.075 & 245.495 & -8.107\\
6 & 4508 & $[2,2]$ & 23.50090 & -0.250 & 240.850 & -80.885\\
8 & 3346 & $[2,2]$ & 42.49877 & 0.400 & 30.333 & 83.056\\
9 & 3071 & $[1,1]$ & 28.49919 & 0.250 & 35.734 & 80.126\\
12 & 2269 & $[2,2]$ & 36.50033 & 0.200 & 329.266 & 68.973\\
13 & 2218 & $[0,0]$ & 30.50049 & 0.062 & 66.712 & -70.484\\
14 & 1864 & $[0,0]$ & 33.49977 & 0.025 & 272.862 & 72.497\\
16 & 1581 & $[1,1]$ & 24.50010 & 0.163 & 327.225 & 69.131\\
20 & 1155 & $[1,1]$ & 31.49812 & -0.150 & 183.141 & 56.041\\
21 & 1112 & $[1,1]$ & 39.50029 & -0.400 & 182.468 & -69.888\\
24 & 881 & $[0,0]$ & 37.50214 & 0.050 & 308.365 & -78.677\\
26 & 652 & $[0,0]$ & 48.50039 & 0.087 & 283.523 & 66.911\\
32 & 414 & $[0,0]$ & 25.24979 & 0.312 & 48.291 & -15.967\\
34 & 339 & $[0,0]$ & 25.49812 & -0.200 & 163.735 & 61.435\\
42 & 194 & $[2,2]$ & 71.50111 & -0.438 & 261.002 & 49.485\\
51 & 148 & $[2,2]$ & 26.25072 & -0.275 & 212.451 & -78.398\\
55 & 122 & $[0,0]$ & 60.50055 & 0.087 & 281.309 & 63.866\\
58 & 107 & $[1,1]$ & 31.99943 & 0.250 & 6.435 & 78.791\\
61 & 97 & $[1,1]$ & 30.00010 & -0.200 & 147.604 & -74.370\\
69 & 65 & $[0,0]$ & 30.99823 & 0.087 & 76.765 & 69.482\\
71 & 57 & $[2,2]$ & 84.99812 & 0.525 & 292.900 & 84.017\\
73 & 52 & $[1,1]$ & 37.25095 & -0.338 & 218.785 & -80.271\\
75 & 44 & $[2,2]$ & 57.00140 & -0.463 & 175.217 & -84.532\\
79 & 39 & $[2,2]$ & 46.25171 & -0.713 & 240.550 & -62.392\\
80 & 38 & $[2,2]$ & 44.25014 & -0.250 & 132.213 & -77.498\\
81 & 37 & $[2,2]$ & 38.00200 & -0.762 & 250.147 & -18.662\\
84 & 33 & $[1,1]$ & 36.25184 & -0.400 & 270.188 & -69.467\\
88 & 31 & $[1,1]$ & 50.00082 & -0.325 & 148.358 & -81.767\\
94 & 23 & $[2,2]$ & 41.24653 & -0.100 & 150.901 & -22.875\\

\hline
\end{tabular}
\caption[Outliers near comb that passed detection pipeline]{\pf\ outliers below $100$~Hz found within $5$~mHz of $0.25$\,Hz comb. Only the highest-SNR outlier is shown for each 0.1-Hz frequency region. Segment column reports the set of
contiguous segments of the data that produced the outlier, as
described in \ref{sec:results}.
Frequencies are converted to epoch GPS $1130529362$.}
\label{tab:PowerFluxOutliers25}
\end{center}
\end{table*}

\section{\fh\ outlier tables}
\label{app:fhout}
\begin{figure}[htbp]
\begin{center}
\includegraphics[width=3.7in]{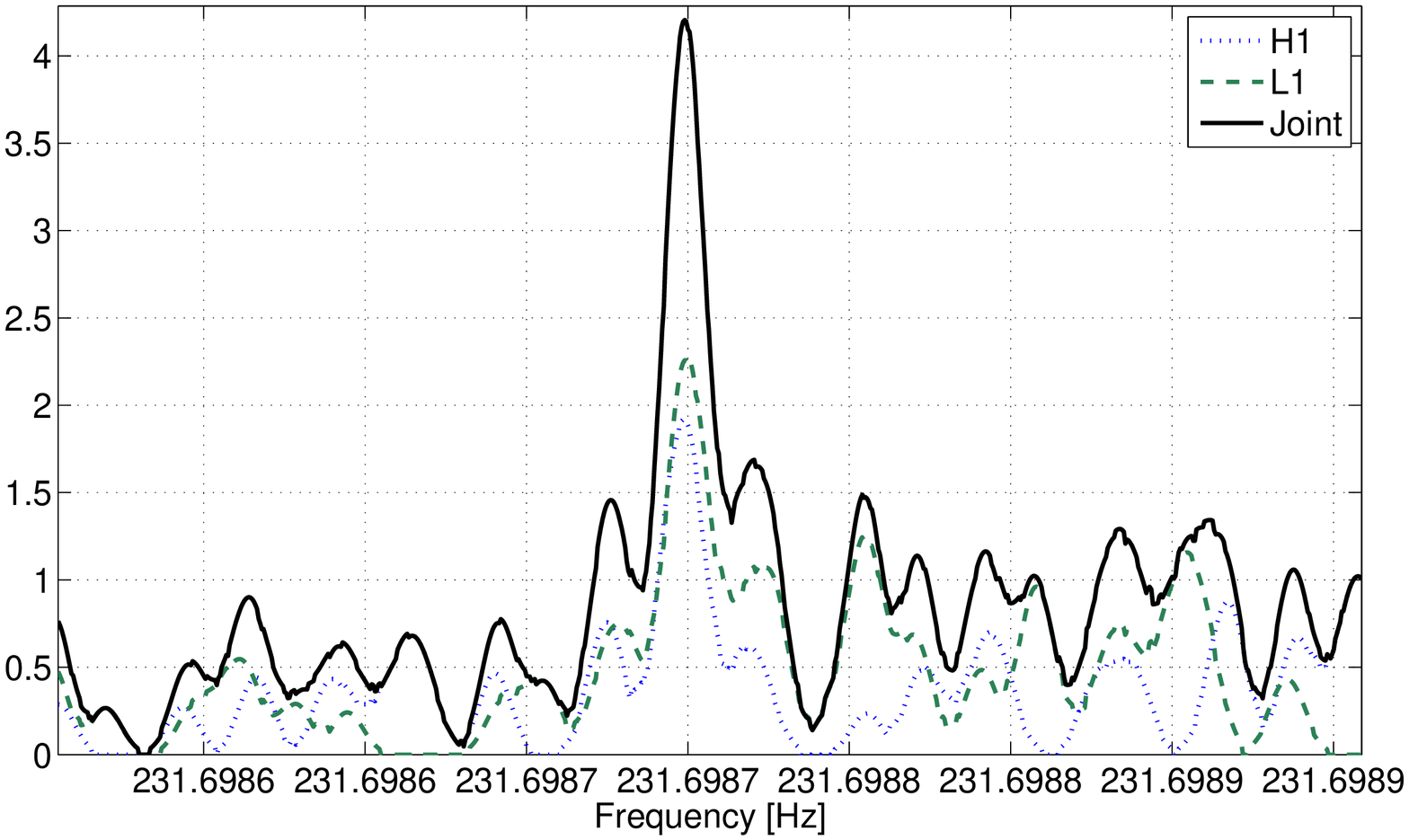}
\caption[FHPMproj]{ Peakmap projections for outlier 7 of the \fh\ search. The dotted blue line is for Hanford detector analysis, the green dashed line is for Livingston detector analysis and the continuous black line is for the joint search.}
\label{fig:outlier_0231_peakmap_projection}
\end{center}
\end{figure}

\begin{figure}[htbp]
\begin{center}
\includegraphics[width=3.7in]{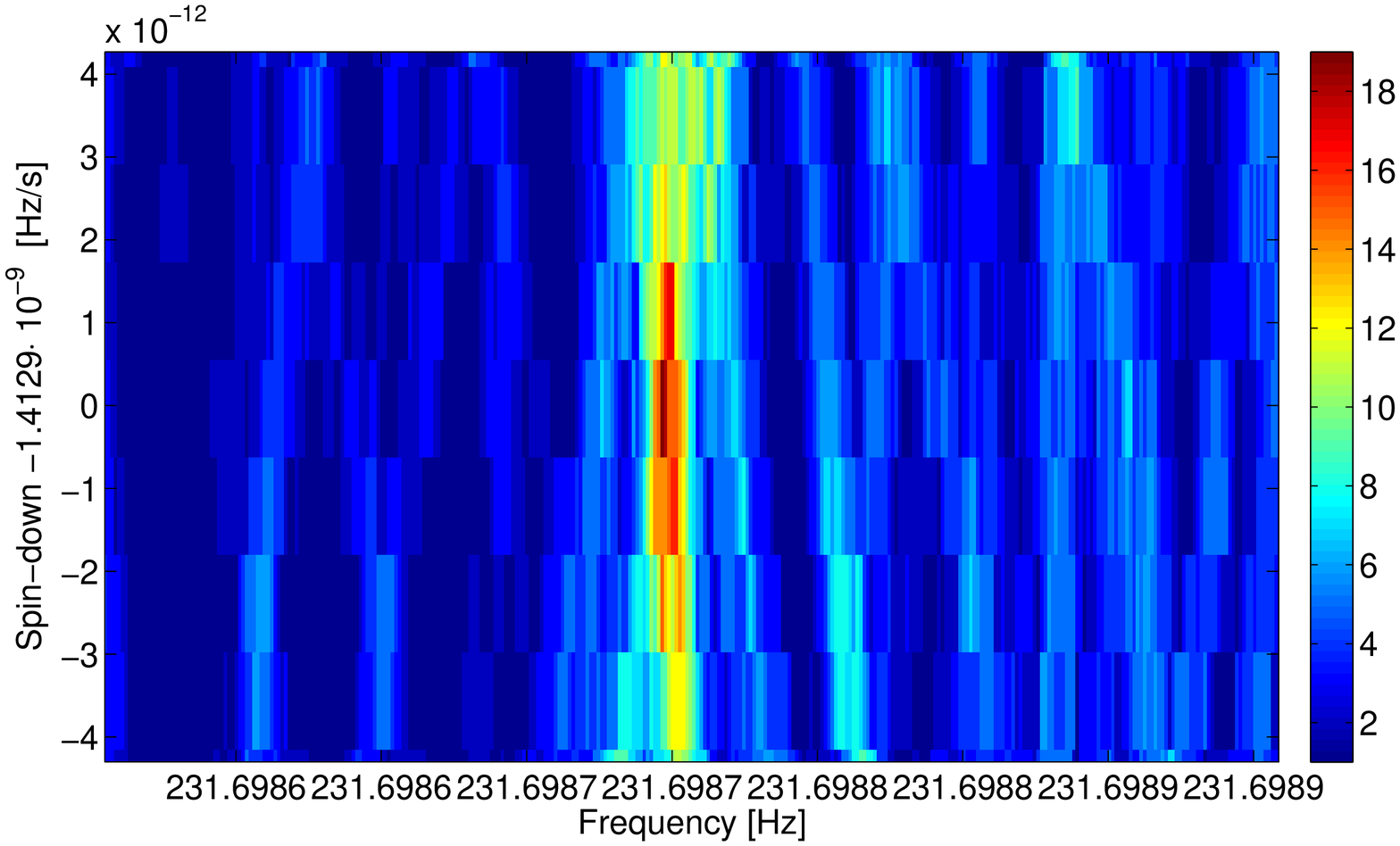}
\caption[FHHM]{Joint Hough map for outlier 7 of the \fh\ search.}
\label{fig:outlier_0231_hm_joint}
\end{center}
\end{figure}

In this section we describe in some detail the final outliers found in the \fh\ search and the analyses that have been carried on them. Tab. \ref{tab:fh_outliers} contains the list of the outliers and their main characteristic, including a brief comment.
\begin{table*}[htbp]
\begin{center}
\begin{tabular}{ccccccc}\hline
 Idx & Frequency [Hz] & Spin-down [Hz/s] & $\alpha$ [deg] & $\delta$ [deg] & CR & Description \\
\hline \hline \\
  1 & 19.3245 & $-9.2235\cdot 10^{-9}$ & 210.16 & -20.47 & 12.9 & Due to H1 alone \\
  2 & 27.8422 & $-8.2416\cdot 10^{-11}$ & 123.54 & -70.26 & 41.7 & Instrumental artifact mainly in L1 \\
  3 & 27.8425 & $-7.1820\cdot 10^{-11}$ & 76.67 & -74.73 & 63.2 & Instrumental artifact mainly in L1 \\
  4 & 59.6054 & $-7.8884\cdot 10^{-11}$ & 98.17 & -70.46 & 51.2 & Two nearby instrumental artifacts in H1 and L1 \\
  5 & 59.6053 & $-8.2416\cdot 10^{-11}$ & 263.16 & 62.68 & 39.9 & Two nearby instrumental artifacts in H1 and L1 \\
  6 & 217.4516 & $+1.8249\cdot 10^{-10}$ & 77.15 & -31.98 & 7.8 & Due to H1 alone \\
  7 & 231.6987 & $-1.4128\cdot 10^{-9}$  & 288.08 & 36.63 & 9.0 & Consistent among H1 and L1. \\
  8 & 269.8699 & $-5.7679\cdot 10^{-9}$ & 242.98 & 33.60 & 7.9 & Possibly transient disturbance in H1 \\
  9 & 281.5976 & $-2.5184\cdot 10^{-9}$ & 166.49 & 44.47 & 7.8 & Consistent but not highly significant in single IFOs \\
  10 & 289.8485 & $-6.9783\cdot 10^{-9}$ & 276.08 & 32.80 & 9.8 & Instrumental artifact in L1\\
  11 & 294.5292 & $-1.1774\cdot 10^{-10}$& 316.32 & 30.28 & 7.9 & Possibly transient disturbance in H1 \\
  12 & 304.8360 & $-5.1687\cdot 10^{-10}$ & 74.14 & -41.39 & 7.6 & Consistent among H1 and L1. \\
  13 & 393.3830 & $-7.2997\cdot 10^{-10}$ & 37.59 & -24.43 & 7.7 & Not consistent among single IFOs \\
  14 & 456.9495 & $-3.0612\cdot 10^{-10}$ & 248.33 & 44.97 & 8.0 & Consistent among H1 and L1. \\
 \hline
\end{tabular}
\caption[FH OUTLIERS]{Final outliers selected by the \fh\ pipeline. Each of them is identified by the frequency, the spin-down, the position in equatorial coordinates and the 
critical ratio computed on the corrected peakmap projection. Reference time is MJD$=$57328.59684 (GPS 1130509183.976).}
\label{tab:fh_outliers}
\end{center}
\end{table*}
Each of these outliers has been manually examined by looking at the details of the follow-up products, including the peakmaps and the Hough maps, and comparing single detector and joint results.
For all of these outliers a gravitational wave origin can be excluded. For outliers 2-5 and 10 in Table~\ref{tab:fh_outliers}, the presence of instrumental artifacts (of unknown origin) is 
clear. Outliers 1, 6, 8, 11 and 13 are not consistent between the two detectors. In particular, n. 8 and 11 are attributed to transient disturbance in the Hanford detector. Outlier 9 is consistent between detectors, but not highly significant in the two detectors. Finally, outliers 7, 12 and 14 were potentially more interesting: they are consistent among the two detectors, very significant also in the single-interferometer analysis, and the corresponding Hough maps look reasonable. As an example in Fig. \ref{fig:outlier_0231_peakmap_projection} we plot the corrected peakmap projections for outlier 7, and in Fig. \ref{fig:outlier_0231_hm_joint} the outlier joint Hough map.
For these outliers we have carried out a deeper follow-up using the method described in \cite{ref:ehfu,ref:eho1}, with a coherence time of 210 hours. In all cases the follow-up failed to 
yield a credible signal.
Hence none of the above outliers shows evidence of a true gravitational wave signal

\section{\sh\ outlier tables}
\label{app:shout}

Table~\ref{tab:sh_outliers} presents the parameters of the final 26 outliers from
the \sh\ search pipeline, along with comments on their likely causes.
None is a credible gravitational wave signal.

\begin{table*}[htbp]
\begin{center}
\scriptsize
\begin{tabular}{ccccccccccccccccc}\hline
Idx  & Frequency  & Spin-down & $\alpha$ & $\delta$  & $s_{\textrm{mean}}$ & $\#_{\textrm{cluster}}$ & $\#_{\textrm{L1}}$ & $\#_{\textrm{H1}}$  & $s^*_{\textrm{L1}}$  & $s^*_{\textrm{H1}}$  & $s_{\textrm{max}}$  & $s^*_{\textrm{multi}}$ & $s_{\textrm{theo}}$  & Description & \\
 &  [Hz] &  [nHz/s]&  [rad] &  [rad] &  &  &  &   &   &   &   &  &  & \\

\hline \hline \\
6 & 52.8084 & $-$0.0175 & $-$1.1478 & $-$1.4512 & 24.22 & 975 & 554 & 20 & 20.36 & 40.54 & 27.02 & 45.61 & 45.20  & 1 \\
9 & 53.8974 & 0.3693 & 0.4803 & 1.5115 & 6.31 & 14 & 6 & 3 & 5.66 & 10.47 & 6.50 & 21.20 & 11.90  & 2\\
14 & 57.0055 & $-$1.0014   & $-$2.2336 & $-$0.1685 & 9.84 & 14 & 7 & 2 & 15.84 & 9.17 & 11.43 & 14.28 & 15.37  & 3\\
24 & 62.4983 & 0.2480 & $-$1.0927 & 1.3699 & 14.64 & 1089 & 45 & 326 & 9.13 & 98.01 & 16.67 & 95.25 & 93.06 & 3\\
44 & 75.5055 & $-$0.9098  & $-$2.5598 & $-$1.1455 & 10.05 & 7 & 2 & 5 & 6.44 & 71.67 & 11.67 & 57.12 & 57.60  & 3\\
46 & 76.4988 & 0.2207 & $-$1.1572 & 1.2721 & 11.03 & 1090 & 34 & 282 & 7.59 & 76.60 & 11.49 & 68.14 & 67.41  & 3\\
47 & 77.4963 & 0.5741 & $-$0.1180 & 1.3766 & 9.54 & 959 & 70 & 121 & 6.28 & 76.87 & 10.61 & 70.23 & 66.58  & 3\\
49 & 78.4962 & 0.5587 & $-$0.2021 & 1.4096 & 9.65 & 308 & 13 & 129 & 6.38 & 75.89 & 10.29 & 69.25 & 64.34  & 3\\
52 & 79.9990 & 0.1808 & $-$1.2674 & 1.2599 & 10.05 & 615 & 59 & 175 & 6.59 & 107.63 & 10.90 & 97.21 & 91.77  & 3, 4\\
53 & 80.4994 & 0.0658 & $-$1.5722 & 1.2107 & 11.92 & 1254 & 97 & 138 & 7.41 & 68.79 & 13.27 & 60.94 & 60.89  & 3\\
60 & 84.4961 & 0.5454 & $-$0.4443 & 1.4369 & 8.95 & 1002 & 55 & 141 & 5.88 & 59.19 & 10.69 & 56.18 & 54.21  & 3\\
62 & 85.0013 & $-$0.2165 & 1.8264 & $-$1.2866 & 12.08 & 811 & 132 & 83 & 11.21 & 15.38 & 12.96 & 18.13 & 18.96  & 3\\
63 & 85.5012 & $-$0.1977 & 1.7877 & $-$1.2674 & 33.87 & 870 & 198 & 64 & 27.44 & 58.03 & 36.95 & 63.92 & 63.78  & 3\\
66 & 86.4989 & 0.1618 & $-$1.4494 & 1.2580 & 9.08 & 1357 & 147 & 125 & 6.38 & 57.24 & 10.49 & 54.70 & 53.67  & 3\\
76 & 98.4983 & 0.2876 & $-$1.2383 & 1.2866 & 10.16 & 1219 & 161 & 158 & 7.10 & 28.50 & 11.30 & 27.51 & 26.12  & 3\\
79 & 100.5032 & $-$0.5624 & 2.5925 & $-$1.3402 & 8.37 & 288 & 39 & 114 & 6.02 & 28 & 8.85 & 24.96 & 24.96  & 3\\
88 & 108.8573 & 0.0041 & $-$3.1371 & $-$0.5718 & 13.61 & 999 & 296 & 99 & 18.79 & 14.73 & 15.85 & 22.93 & 23.55  & 1 \\
92 & 112.5026 & $-$0.4739 & $-$1.8331 & 0.9775 & 6.52 & 68 & 2 & 55 & 5.98 & 11.70 & 6.78 & 11.72 & 12.84 & 3\\
104 & 127.4964 & 0.5926 & $-$0.9377 & 1.3368 & 6.36 & 877 & 204 & 58 & 7.04 & 7.64 & 6.94 & 9.08 & 10.39  & 3\\
105 & 127.9988 & 0.2101 & $-$1.3924 & 1.2317 & 70.70 & 1165 & 161 & 70 & 61.13 & 108 & 78.07 & 123.85 & 121.24  & 3, 4\\
107 & 128.4984 & 0.3602 & $-$1.1366 & 1.2478 & 6.94 & 971 & 223 & 82 & 8.40 & 7.35 & 7.60 & 9.50 & 11.07  & 3\\
112 & 146.1994 & $-$6.6167  & 0.0500 & $-$1.1133 & 16.08 & 1676 & 796 & 22 & 15.96 & 22.10 & 18.11 & 27.17 & 26.87  & 1 \\
136 & 191.0716 & $-$8.7553  & $-$0.2091 & $-$0.5534 & 22.83 & 988 & 262 & 107 & 30.61 & 22.80 & 25.66 & 37.00 & 37.76  & 1 \\
153 & 255.9995 & 0.0789 & $-$1.5493 & 1.1792 & 30.62 & 1082 & 203 & 43 & 20.10 & 105.72 & 33.03 & 96.46 & 86.90  & 3, 4\\
160 & 265.5736 & 0.3441 & 1.1995 & $-$0.9102 & 21.16 & 750 & 286 & 18 & 24.76 & 25.97 & 25.19 & 35.18 & 35.85  & 1 \\
193 & 452.8999 & $-$2.7816 & 2.5350 & $-$1.3173 & 10.01 & 222 & 4 & 109 & 6.68 & 88.07 & 10.04 & 61.91 & 28.21  & 5 \\
\hline
\end{tabular}
\caption[]{\sh\ outliers after  population and multi-interferometer consistency vetoes.
The table provides the frequency, spin-down and sky location of the cluster centers  found by the \sh\ search. $\#_{cluster}$ is the size of the cluster in terms of number of coincident pairs, $s_{max}$ and  $s_{mean}$ are  the maximum and mean value of the cluster significance, $\#_{L1}$ and $\#_{H1}$ are the number of different candidates producing coincidence pairs from the different data sets,  $s^*_{L1}$ and $s^*_{H1}$ are the maximum significance values obtained by analysing the data from H1 and L1 separately,
 $s^*_{multi}$  is maximum combined significance when the data from both detectors are analyzed jointly and $s_{theo}$ is the expected theoretical combined significance value.
Frequencies are converted to epoch GPS 1125972653. The outliers description codes mean the following:
1- hardware injection,
2- associated to unknown comb in H1 starting at 30.9430 Hz with 0.99816 Hz spacing,
3- miscellaneous combs with known or unknown sources at multiples of 0.5 Hz many of them due to blinking LEDs in timing system,
4- associated with the 8 Hz comb in H1 due to the OMC length dither,
and 5-spectral disturbance in H1. }
\label{tab:sh_outliers}
\end{center}
\end{table*}

\section{\td\ outlier tables}
\label{app:tdout}

Table~\ref{tab:TDFstat_outliers} presents the parameters of the final 49 outliers
from the \td\  pipeline, along with comments on their likely causes.
None is a credible gravitational wave signal.

\begin{table*}[htbp]
\begin{center}
\scriptsize
\begin{tabular}{D{.}{.}{2}D{.}{.}{6}D{.}{.}{5}D{.}{.}{4}D{.}{.}{4}l}\hline
\multicolumn{1}{c}{Idx} & \multicolumn{1}{c}{Frequency} & \multicolumn{1}{c}{Spindown} &  \multicolumn{1}{c}{$\RAJ$}  & \multicolumn{1}{c}{$\DECJ$} & Description \\
\multicolumn{1}{c}{}	& \multicolumn{1}{c}{Hz}	&  \multicolumn{1}{c}{nHz/s} & \multicolumn{1}{c}{degrees} & \multicolumn{1}{c}{degrees} &  \\
\hline \hline
1 & 12.49993 & 0.025 & 86.603 & -58.499 & 0.25 Hz comb \\
2 & 12.50002 & -0.002 & 90.000 & 66.561 & 0.25 Hz comb \\
3 & 31.76207 & -0.151 & 127.576 & 29.821 & PowerFlux idx 66 \\
4 & 36.60793 & -0.416 & 161.240 & -52.346 & Present only in L1\\
5 & 36.60841 & -0.075 & 100.874 & 43.930 & Present only in L1\\
6 & 13.49980 & 0.055 & 93.007 & 58.722 & 0.25 Hz comb \\
7 & 31.76213 & -0.029 & 121.201 & -22.659 & PowerFlux idx 66 \\
8 & 85.82924 & -0.782 & 122.451 & 42.487 & Present only in H1\\
9 & 85.82951 & 0.014 & 91.251 & -60.424 & Present only in H1\\
10 & 83.44610 & 0.356 & 57.433 & 74.096 & Instrumental artifact \\
11 & 83.44718 & -0.188 & 100.706 & -71.727 & Instrumental artifact \\
12 & 26.30940 & 0.120 & 65.866 & 28.651 & PowerFlux idx 104 \\
13 & 31.37813 & 0.076 & 86.489 & 60.660 & Hardware injection ip11
\\
14 & 31.51095 & -0.260 & 125.878 & -30.969 & Hardware injection ip11\\
15 & 13.49954 & 0.028 & 103.846 & -33.555 & 0.25 Hz comb \\
16 & 34.82444 & -0.178 & 125.214 & -37.390 & Present only in H1\\
17 & 34.82526 & 0.365 & 43.489 & 59.488 & Present only in H1\\
18 & 26.34245 & 0.188 & 101.305 & -41.914 & PowerFlux idx 104 \\
19 & 52.80802 & 0.272 & 75.131 & 46.185 & Hardware injection ip5
\\
20 & 52.80847 & -0.217 & 86.346 & -63.610 & Hardware injection ip5 \\
21 & 19.90049 & -0.049 & 120.105 & -66.933 & Transient disturbance \\
22 & 108.85713 & 0.016 & 172.077 & -30.650 & Hardware injection ip3 \\
23 & 108.85751 & -0.190 & 156.197 & 11.273 & Hardware injection ip3 \\
24 & 191.02943 & -8.248 & 19.785 & 8.689 & Hardware injection ip8 \\
25 & 191.02921 & -8.461 & 10.399 & -26.263 & Hardware injection ip8 \\
26 & 39.76317 & -0.250 & 116.156 & 35.030 & PowerFlux idx 77 \\
27 & 39.76228 & -0.111 & 120.585 & -49.208 & PowerFlux idx 77 \\
28 & 210.38111 & -0.719 & 111.146 & -73.636 & Present only in H1\\
29 & 146.16957 & -7.664 & 74.426 & 19.212 & Hardware injection ip6 \\
30 & 146.16861 & -6.847 & 9.592 & -64.231 & Hardware injection ip6 \\
31 & 46.94798 & 0.037 & 120.395 & 66.304 & PowerFlux idx 48 \\
32 & 46.94859 & 0.059 & 107.507 & -49.753 & PowerFlux idx 48 \\
33 & 93.89883 & 0.344 & 66.466 & 72.993 & Present only in H1\\
34 & 93.89971 & -0.084 & 94.143 & -67.379 & Present only in H1\\
35 & 33.33199 & 0.186 & 82.235 & 49.267 & Present at the beginning in H1 and in the middle in L1\\
36 & 33.33191 & 0.147 & 101.592 & -24.740 & Present at the beginning in H1 and in the middle in L1\\
37 & 28.94735 & -0.023 & 80.524 & -62.043 & Present only in H1\\
38 & 28.94644 & 0.217 & 56.600 & 61.593 & Present only in H1\\
39 & 140.25111 & -0.261 & 118.249 & 65.668 & 0.25 Hz comb \\
40 & 273.76680 & 0.022 & 279.363 & -0.225 & Present only in the first half of O1\\
41 & 66.75634 & 0.060 & 67.002 & 65.771 & Present only in H1 \\
42 & 286.11848 & 0.220 & 282.207 & 7.749 & Present only in the first half of O1\\
43 & 27.94841 & 0.150 & 79.648 & -46.995 & Present only in H1 \\
44 & 65.87802 & -0.007 & 95.671 & 68.015 & Present only in H1 \\
45 & 65.87734 & 0.237 & 82.021 & -39.834 & Present only in H1  \\
46 & 265.57578 & -0.077 & 71.932 & -57.608 & Hardware injection ip0\\
47 & 364.34662 & 0.322 & 281.187 & 22.135 & Present only in the first half of O1\\
48 & 43.74991 & -0.110 & 87.628 & -41.298 & 0.25 Hz comb \\
49 & 331.65142 & 0.296 & 301.174 & -56.957 & Present only in the first half of O1\\

\hline
\end{tabular}
\caption{\td\ pipeline outliers in the range of frequencies between 10 and 475 Hz.
  The columns provide the nominal frequencies and frequency derivatives, right ascensions and declinations
  found for the outliers, along with comments indicating the likely sources of the outliers.}
\label{tab:TDFstat_outliers}
\end{center}
\end{table*}

\newpage


\end{document}